\documentclass[12pt]{article}
\usepackage[pdftex,colorlinks=true,linkcolor=blue,citecolor=blue,urlcolor=blue]{hyperref}
\usepackage{amsmath,amsfonts,amssymb}
\usepackage{graphicx}
\usepackage{psfrag}
\usepackage{enumerate}
\usepackage{mathrsfs}
\usepackage{dsfont}

\newcommand{\RR}{\mathbb{R}} 
\newcommand{\ZZ}{\mathbb{Z}} 
\newcommand{\HH}{\mathbb{H}}

\newcommand{\Tr}{{\rm {Tr}}}

\newcommand{\be}{\begin{equation}}
\newcommand{\ee}{\end{equation}}
\newcommand{\beq}{\begin{eqnarray}}
\newcommand{\eeq}{\end{eqnarray}}

\def\[{\left [}
\def\]{\right ]}
\def\({\left (}
\def\){\right )}

\def\r2{\sqrt{2}}

\def\Tr{{\rm Tr}}

\newcommand{\nc}{\newcommand}
\nc{\rnc}{\renewcommand}
\nc{\bra}[1]{\langle#1|}
\nc{\ket}[1]{|#1\rangle}
\nc{\ketbra}[2]{|#1\rangle\!\langle#2|}
\nc{\braket}[2]{\langle#1|#2\rangle}
\nc{\proj}[1]{\left| #1\right\rangle\!\left\langle #1 \right|}
\rnc{\max}{\operatorname{max}}
\nc{\smfrac}[2]{\mbox{$\frac{#1}{#2}$}}
\nc{\tr}{\operatorname{tr}}
\nc{\ox}{\otimes}
\nc{\dg}{\dagger}

\nc{\choice}[2]{{\mathcal{C}^{#1}_{#2}}}
\def\Sadd{{S_{\text{sum}}}}

\newcommand{\bbibitem}[1]{\bibitem{#1}\marginpar{#1}}

\def\Label#1{\label{#1}%
  \smash{\hbox to0pt{\raise1ex\hbox{\tiny[#1]}\hss}}}
\def\noLabels{\let\Label=\label}
\def\nobbibitem{\let\bbibitem=\bibitem}

\numberwithin{equation}{section}

\begin{document}
\clearpage\thispagestyle{empty}

\begin{center}


{\Large \bf
Multiboundary Wormholes and  Holographic Entanglement}

\vspace{7mm}

Vijay Balasubramanian$^{a,b}$, Patrick Hayden$^{c}$, Alexander Maloney$^{d,e}$,  Donald Marolf$^{f}$, \\
Simon F. Ross$^{g}$  \\



\bigskip\centerline{$^a$\it David Rittenhouse Laboratories, University of Pennsylvania}
\smallskip\centerline{\it 209 S 33$^{\rm rd}$ Street, Philadelphia, PA 19104, USA}
\bigskip\centerline{$^b$\it CUNY Graduate Center, Initiative for the Theoretical Sciences}
\smallskip\centerline{\it 365 Fifth Avenue, New York, NY 10016, USA}
\bigskip\centerline{$^c$\it Department of Physics, Stanford University}
\smallskip\centerline{\it Palo Alto, CA 94305, USA}
\bigskip\centerline{$^d$\it Department of Physics, McGill University}
\smallskip\centerline{\it 3600 rue Universit\'{e}, Montreal H3A2T8, Canada}
\bigskip\centerline{$^e$\it Center for the Fundamental Laws of Nature, Harvard University}
\smallskip\centerline{\it Cambridge, MA 02138, USA}
\bigskip\centerline{$^f$\it Department of Physics, University of California,}
\smallskip\centerline{\it Santa Barbara, CA 93106, USA}
\bigskip\centerline{$^g$\it Centre for Particle Theory, Department of Mathematical Sciences}
\smallskip\centerline{\it Durham University, South Road, Durham DH1 3LE, UK}
\end{center}

\vspace{5mm}

\begin{abstract}
The AdS/CFT correspondence relates quantum entanglement between boundary Conformal Field Theories and geometric connections in the dual asymptotically Anti-de Sitter space-time. We consider entangled states in the $n-$fold tensor product of a 1+1 dimensional CFT Hilbert space defined by the Euclidean path integral over a Riemann surface with $n$ holes. In one region of moduli space, the dual bulk state is a black hole with $n$ asymptotically AdS${}_3$ regions connected by a common wormhole, while in  other regions the bulk fragments into disconnected components. We study the entanglement structure and compute the wave function explicitly in the puncture limit of the Riemann surface in terms of CFT $n$-point functions.  We also use AdS minimal surfaces to measure entanglement more generally. In some regions of the moduli space the entanglement is entirely multipartite, though not of the GHZ type.  However, even when the bulk is completely connected,  there are regions of the moduli space in which the entanglement is instead almost entirely bipartite: significant entanglement occurs only between pairs of CFTs.   We develop new tools to analyze intrinsically $n$-partite entanglement, and use these to show that for some wormholes with $n$ similar sized horizons there is intrinsic  entanglement between all $n$ parties, and that the distillable entanglement between the asymptotic regions is at least $(n+1)/2$ partite.
\end{abstract}

\setcounter{footnote}{0}
\newpage
\clearpage
\setcounter{page}{1}

\tableofcontents

%
\section{Introduction}
%

Quantum systems are fundamentally distinguished from classical ones by their capacity for entanglement, a property that gives rise to many of the most counterintuitive features of quantum mechanics.  A remarkable new role for entanglement has recently appeared in holographic descriptions of field theories. In the AdS/CFT correspondence, the entanglement structure of the quantum theory seems to be playing a central role in the emergence of a classical spacetime geometry.

The importance of entanglement in this context was first appreciated in the eternal black hole spacetime \cite{Maldacena:2001kr}. The eternal black hole has two asymptotic regions, connected by an Einstein-Rosen bridge.  The dual state is an entangled state on the two boundaries, which can be obtained by considering the Euclidean black hole geometry. The role of entanglement in other geometries has been studied in \cite{VanRaamsdonk:2010pw,Czech:2012be,Maldacena:2013xja}. Another relation between entanglement and geometry arises in the Ryu-Takayanagi (RT) formula relating the entropy of a reduced density matrix associated to a spatial subregion in  a static field theory state to the area of a minimal surface in AdS anchored on the boundary of the subregion \cite{Ryu:2006bv}. This entropy provides a measure of the amount of entanglement between the subregion and its complement. This relation has been extended to a covariant proposal \cite{Hubeny:2007xt}, and the RT formula has been related to the eternal black hole construction in \cite{Lewkowycz:2013nqa}. It has also  been extended to associate a ``differential entropy'' with the area of certain surfaces in AdS space that do not touch the  boundary \cite{Bianchi:2012ev,Balasubramanian:2013rqa,Balasubramanian:2013lsa,Myers:2014jia}.  From a different perspective, it has been been proposed that the general structure of the entanglement-based MERA ansatz for calculating ground state wavefunctions may provide a new understanding of the emergence of an additional dimension in the holographic description \cite{Swingle:2009bg}.  Finally, several recent efforts propose a derivation of linearized gravity from the dynamics of entanglement of the underlying quantum degrees of freedom, e.g. \cite{Nozaki:2013vta,Lashkari:2013koa, Faulkner:2013ica, Swingle:2014uza}.

All of these constructions are fundamentally bipartite.  For example, the thermofield double (TFD) is a state on two CFTs dual to a spacetime connecting two asymptotic regions.  Likewise, the Ryu-Takayangi formula describes entanglement between a boundary region and its complement in terms of a surface that divides the bulk space into two parts.  The extensions discussed in \cite{VanRaamsdonk:2010pw,Czech:2012be,Maldacena:2013xja,Bianchi:2012ev,Balasubramanian:2013rqa,Balasubramanian:2013lsa,Myers:2014jia} similarly describe entanglement between pairs of systems.

Entanglement, however, is an inherently multipartite concept. A system with many degrees of freedom can be entangled in a way that is not fully characterized by the entanglement between subsets of the degrees of freedom. An analogy can be drawn with quantum field theory, where many-point correlation functions cannot be inferred from lower-point correlations.      An example of intrinsically three-party entanglement is the GHZ state of three spins: $|{\rm GHZ}\rangle = (\left|\uparrow \uparrow \uparrow \right\rangle + \left|\downarrow \downarrow \downarrow \right\rangle ) / \sqrt{2}$. Tracing over any one of the three spins results in a classical mixture of the product states $\left|\uparrow \uparrow \right\rangle$ and $\left|\downarrow \downarrow\right\rangle$ even though the global pure state does not factorize. Nonlocal effects can be more pronounced in the multiparty setting as well: while any state of two spins can at best violate a Bell inequality on average~\cite{cirel1980quantum}, the GHZ state can with a single measurement~\cite{greenberger1990bell}.

We wish to study multipartite entanglement in the AdS/CFT correspondence.  We will focus on the AdS${}_3$/CFT${}_2$ case, though we expect similar considerations will apply in higher dimensions. The simplest entangled state is the TFD state, which lives in the Hilbert space ${\cal H}\otimes {\cal H}$, where ${\cal H}$ is the Hilbert space of a single CFT. The TFD state is defined by the Euclidean path integral on a finite cylinder, with the two states in ${\cal H}$ inserted on either end of the cylinder.  The inverse temperature of the state is the conformal modulus of the cylinder. A natural multiparty generalization is the state $|\Sigma \rangle$ in ${\cal H}^{\otimes n}$ given by the Euclidean path integral over a Riemann surface $\Sigma$ with $n$  boundaries.
The goal of this paper is to analyze the bulk connectivity and multiparty nature of entanglement in $|\Sigma \rangle$.

The state $|\Sigma\rangle$ depends on the conformal moduli of the Riemann surface $\Sigma$. For some values of the moduli, the dominant gravitational solution is a connected, multiboundary black hole: this geometry  has $n$ asymptotically AdS$_3$ regions connected  by a wormhole with the conformal geometry of $\Sigma$. This picture was initially suggested in \cite{Maldacena:2001kr} (see also \cite{Krasnov:2000zq, Krasnov:2003ye, Skenderis:2009ju}). In this case the moduli can be interpreted as $n$ black hole parameters (the mass, temperature or horizon area of the black hole) -- one for each asymptotic region -- along with a number of ``internal" moduli which encode the structure of the behind-the-horizon wormhole. In other regions of moduli space the dominant bulk solution is disconnected.  The transitions between these topologically distinct bulk solutions generalize the Hawking-Page phase transition.

The multipartite nature of the entanglement in $|\Sigma\rangle$  also depends on the moduli.
When all boundaries are connected in the bulk through a common wormhole, one might expect multipartite entanglement to play an important role.\footnote{One might interpret the arguments of \cite{Susskind:2014ira} to suggest that this entanglement should resemble that of GHZ states.  However, in a sense which will be made precise in section \ref{entang}, we will find that GHZ-like states give universally negligible contributions to the entanglement of $|\Sigma \rangle$. This outcome might be expected from the results of \cite{hayden2013holographic,Gharibyan:2013aha}.}
But the actual story is more complicated. In some regions of moduli space the entanglement is entirely multipartite (at leading order in the central charge), though never GHZ-like. In other regions it is entirely bipartite.  Both of these behaviours are possible even in parts of moduli space where the bulk geometry is completely connected!

We begin by reviewing relevant background material in section  \ref{review}.  We construct multiboundary black holes as quotients of AdS$_3$ and describe the moduli associated with the Riemann surface $\Sigma$ and the state $|\Sigma\rangle$, following \cite{Brill:1995jv,Aminneborg:1997pz,Brill:1998pr, Aminneborg:1998si} and \cite{Maldacena:2001kr, Krasnov:2000zq, Krasnov:2003ye, Skenderis:2009ju}.  We discuss bulk phase transitions for the states $|\Sigma \rangle$ and argue that the wormhole fragments into disconnected components in certain regions of moduli space. Black holes in the various asymptotic regions can undergo dramatic changes when a phase boundary is crossed.

In section \ref{cft} we argue that the wave function of $|\Sigma\rangle$ can be expressed as a sum of CFT $n$-point functions up to the action of some (complexified) conformal transformations which act separately on each individual boundary CFT. Thus $|\Sigma\rangle$ has much more structure than the simple TFD state. In the so-called puncture limit -- where $\Sigma$ becomes a Riemann surface with $n$ punctures -- we determine the leading conformal transformations. We then use factorization limits of the $n$-point functions to study disconnected phases, and identify limits of the fully-connected bulk phase where the entanglement becomes fully bipartite.

We examine the multipartite nature of the entanglement in section \ref{entang}.  Even in the puncture limit, the entropy of the reduced density matrices obtained by tracing over one or more CFTs is difficult to compute directly from $|\Sigma\rangle$. So instead we use the covariant HRT prescription \cite{Hubeny:2007xt}, which requires only that we find the area of certain bulk extremal surfaces. Even when the bulk describes a single connected wormhole, this analysis demonstrates that the amount of entanglement and its qualitative nature both depend on the moduli: there are  regimes where entanglement is largely bipartite and others where it is largely multipartite.

Unfortunately, there is no unique, agreed upon ``best'' measure of multipartite entanglement~\cite{vedral1997quantifying,bennett2000exact,barnum2001monotones}. (For reviews, see \cite{eisert2005multi,horodecki2009quantum}.) In the bipartite case, all pure states are asymptotically interconvertible into each other at a rate given by the ratio of their entanglement entropies using only ``local operations and classical communication''~\cite{bennett1996concentrating}. This establishes the entanglement entropy as the essentially unique measure of bipartite entanglement. But in the multipartite case there are inequivalent forms of entanglement, some of which can be interconverted but only irreversibly. For example, any multiparty entangled state can be prepared locally and then distributed using pairwise entanglement via teleportation, but this process cannot in general be run backwards. Other forms of multiparty entanglement cannot be interconverted at all \cite{dur2000three}.  Section~\ref{sec:more-parties} and section~\ref{intrinsic}  develop some methods to address these issues. We use these tools to show that for wormholes with $n$ equal sized horizons the distillable entanglement between the asymptotic regions is at least $(n+1)/2$ partite -- there is none for smaller subsystems and there is always some for larger subsystems.  We also show that some multiboundary black holes in which all the horizons are similarly sized must have intrinsically $n$-partite entanglement (when $n$ is even) in the sense that otherwise their HRT entropies computed from minimal surface areas would be significantly smaller. When $n$ is odd, we show that there is intrinsically $(n-1)$-partite entanglement.

As for the TFD state on a product of  two Hilbert spaces, many entanglement properties of $|\Sigma\rangle$ match those of a simple random state model.  In particular, the above results can be reproduced using only the fact that -- as in \cite{Page:1993df} -- if the system is divided into one large and multiple small subsystems, the small subsystems are entangled only with the large system and not with each other.  Interestingly, in many cases the entropy of a given asymptotic region is determined by geometric structures behind the horizon.

We conclude with a summary and closing comments in section \ref{conclude}.

%
\section{Multiboundary black holes in 3D}
\label{review}
%

Three dimensional general relativity has no local gravitons, allowing one to describe rich families of solutions analytically. This is in particular the case for black holes with multiple asymptotic regions, each with a geometry asymptotic to global AdS$_3$. Such spacetimes were constructed as quotients of (a subregion of) AdS$_3$ in \cite{Brill:1995jv,Aminneborg:1997pz,Brill:1998pr}.  See also \cite{Aminneborg:1998si} for the rotating case.  As noted in \cite{Maldacena:2001kr} and described in detail in \cite{Skenderis:2009ju} (see also \cite{Krasnov:2000zq, Krasnov:2003ye}), they are associated with a Euclidean path integral on a certain Riemann surface $\Sigma$ which provides a natural candidate for the dual CFT state $|\Sigma\rangle$.  With $n$ boundaries, the state lives in the Hilbert space ${\cal H}^{\otimes n}$, where ${\cal H}$ is the Hilbert space of a single CFT on the cylinder.

In section ~\ref{quotientreview} we review the bulk solutions, which are (in the non-rotating case) parameterized by a choice of Riemann surface $\Sigma$ with $n$ boundaries.  The surface $\Sigma$ is the spatial geometry of a constant time slice: each boundary of $\Sigma$ matches on to one of the asymptotic boundaries.  These solutions come in continuous families which are labelled by the moduli of the surface $\Sigma$.  In section \ref{EPI} we review the associated  Euclidean path integrals which define the state $|\Sigma\rangle$. We will be somewhat brief, and refer the reader to \cite{Krasnov:2000zq, Skenderis:2009ju} for a more detailed discussion.

In section~\ref{phases} we argue that, as the moduli are varied, there are phase transitions in which the bulk Lorentzian geometry changes topology.
In particular, as one crosses a phase boundary some asymptopia disconnect from the others in the Lorentz-signature bulk geometry.   In these disconnected phases the boundaries -- and thus the CFT copies -- are partitioned into subsets defined by their connectivity. In these phase transitions the Lorentzian geometry changes topology, but the Euclidean
geometry remains connected.
 The holographic construction of the wave function tells us that the mutual information between disconnected components vanishes at leading order in the central charge $c$.  This indicates that the wavefunction factorizes at this order.

Interestingly, our phase transitions changing Lorentz-signature connectivity lead to
sharp changes in the entropy, energy, and temperature of {\it each} copy of the CFT, even if the associated boundary remains connected to many others through a Lorentzian wormhole.
In particular, tuning the moduli to disconnect a single boundary from the 3-boundary wormhole leads not only to a small entropy for the disconnected CFT, but also to a sharp decrease in entropy for the two CFTs that remain connected.  In the bulk, the black hole changes significantly in each asymptotic region -- even in those where it does not disappear.

%
\subsection{Multiboundary solutions of 3D gravity}
%
\label{quotientreview}

AdS${}_3$ is the Lorentzian, maximally symmetric spacetime with constant negative curvature and isometry group $SO(2,2)\simeq SL(2,\RR)\times SL(2,\RR)$. A global coordinate system covering all of AdS$_3$ is
\begin{equation} \label{gads3}
{ds^2 \over \ell^2} = -\cosh^2 \chi d \tau^2 + d\chi^2 + \sinh^2 \chi d \phi^2.
\end{equation}
The spacetime has a conformal boundary at $\chi \to \infty$.  The induced metric on this boundary is conformally equivalent to the cylinder metric $d\sigma^2 = -d\tau^2 + d\phi^2$.
We will be interested in solutions which are locally, but not globally, AdS${}_3$.
 For Einstein gravity in three spacetime dimensions, all solutions to the equations of motion are  locally AdS${}_3$.
These locally AdS${}_3$ solutions will also be present in more complicated theories of AdS${}_3$ gravity, though other solutions may be present as well.

A locally AdS$_3$ solution can be constructed by quotienting AdS$_3$ by a discrete subgroup $\Gamma$ of its isometry group.
The prototypical example is the Banados-Teitelboim-Zanelli (BTZ) black hole \cite{Banados:1992wn}, which
is the quotient of AdS${}_3$ by the group generated by a single element $\gamma$. More complicated geometries are found by using a discrete group $\Gamma$ with multiple generators \cite{Brill:1995jv,Aminneborg:1997pz,Brill:1998pr,Aminneborg:1998si}.  We will restrict our attention to discrete groups which lie in the diagonal  $SL(2,\RR)$ subgroup of  the isometry group, as only in this case is the Euclidean continuation of the geometry real.
For this class of geometries, we can choose our global coordinates so that the action of $\Gamma$ maps the $\tau=0$ surface in \eqref{gads3} to itself.  The full action of $\Gamma$ on AdS$_3$ is uniquely determined by its action on this hyperbolic space.

The quotient by such a $\Gamma$ is naturally described in
the FRW coordinates on AdS$_3$:
\be\label{metric}
{ds^2 \over \ell^2} = -dt^2 + \cos^2 t \, d\Sigma^2,
\ee
where $d\Sigma^2$ is the unit negative curvature metric on hyperbolic space $\HH^2$.  The $t=0$ surface in \eqref{metric} can be taken to be the same as the $\tau=0$ surface in global coordinates \eqref{gads3}.  These coordinates do not cover the whole of AdS$_3$, but they have the advantage that the action of $\Gamma$ on the spacetime is simply an identification on the $\HH^2$.  In this coordinate patch, therefore, $\Gamma$ identifies space-like separated points.  The action of $\Gamma$ on the full AdS$_3$ spacetime is more complicated, as outside this coordinate patch it might identify timelike-separated points.  In order to obtain a space-time without closed timelike curves we will therefore need to remove certain portions of the full AdS$_3$ geometry before performing the quotient.
In the present work we will not need to discuss the full spacetime geometry in detail, so we will focus on the region inside the coordinate patch \eqref{metric}.

The entirety of this patch is present without excisions in the quotient. This fact has important implications for the causal structure of our wormholes.  Recall in particular that in global AdS${}_3$ the surface $t=\pi/2$ describes the past light cone of a point, and that the generators of this cone represent null geodesics launched from the boundary at what one may call $t=0$.  As a result, given any two spatial locations on the global AdS${}_3$ boundary, one can find timelike observers launched radially inward from those locations -- at times shortly to the past of $t=0$ -- that meet in the bulk before $t=\pi/2$.  This immediately implies that timelike observers who enter the quotient spacetime from distinct boundaries can meet inside our wormhole before reaching the singularity.  Our wormholes thus have direct operational meaning in the bulk quantum gravity theory; they lie within the class of wormholes discussed in \cite{Marolf:2012xe} and are distinct from the so-called long wormholes of \cite{Shenker:2013pqa,Shenker:2013yza}.

The action on AdS$_3$ preserves the time-reversal symmetry $t \to -t$, which will enable us to define a real Euclidean signature geometry which includes the above $t=0$ surface as a moment of (Euclidean) time symmetry.

Technically, discrete groups $\Gamma$ of this diagonal form are known as Fuchsian groups, and their action on hyperbolic space is well-studied mathematically.  We restrict our attention to the case where $\Gamma$ acts freely in order to avoid orbifold-type singularities.  This means that  $\Gamma$ is generated by a collection of hyperbolic elements -- elements which can, by conjugation, be put in the form $\left({\lambda~~0\atop~~0~~\lambda^{-1}}\right)$ in $SL(2,\RR)$. $\Gamma$ is then a Fuchsian group of the second kind. The quotient $\Sigma = \HH^2/\Gamma$ is a smooth Riemann surface, possibly with boundaries. A Riemann surface of genus $g$ with $n$ boundaries has $6g - 6 + 3n$ moduli, and the uniformization theorem tells us any such Riemann surface can be obtained by such a quotient construction. The group $\Gamma$ is isomorphic to the fundamental group $\pi_1(\Sigma)$. For each equivalence class of curves in $\pi_1(\Sigma)$ there is a unique minimum length geodesic; the length of this geodesic is just the trace $\Tr (\gamma)$ of the corresponding element in $\Gamma$.

The action of $\Gamma$ will have fixed points on the conformal boundary of $\HH^2$ at some discrete set of points, and the action of $\Gamma$ on AdS${}_3$ has corresponding fixed points on the conformal boundary at $t=0$. Removing from AdS$_3$ the causal future and past of these points yields the restricted space  $\widehat{\mathrm{AdS}}_3$ where the action of the quotient on the spacetime is free of pathologies. The spacetimes we consider are then $\widehat{\mathrm{AdS}}_3 / \Gamma$.

\begin{figure}
\centering
\includegraphics[keepaspectratio,width=0.5\linewidth]{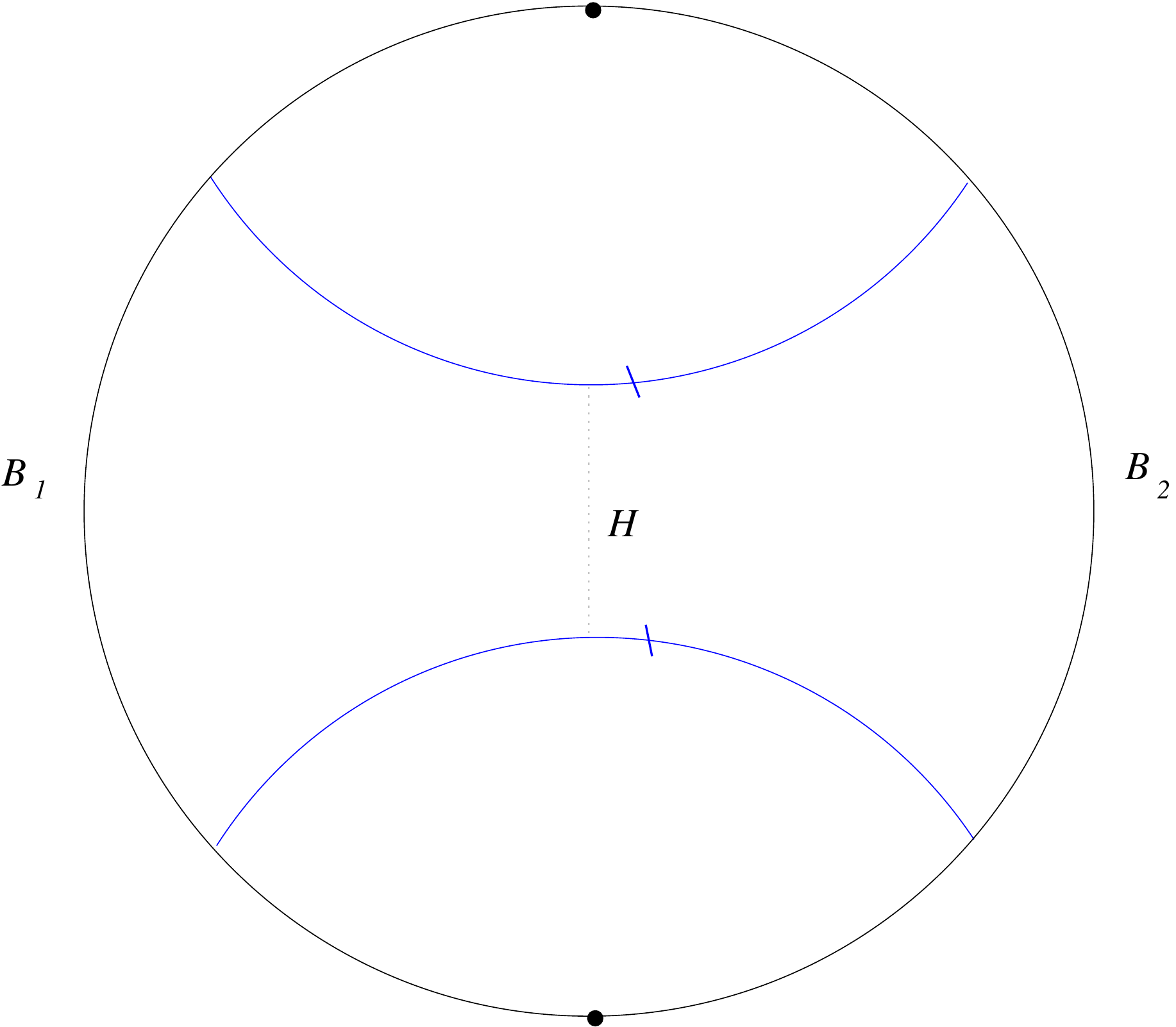}
\caption{The $t=0$ surface $\Sigma$ in non-rotating BTZ as a quotient of the Poincar\'e disc. The two marked geodesics (blue in the colour version, and located symmetrically above and below the center of the figure) are identified by the action of $\gamma$. The region between them provides a fundamental domain for the quotient. $B_1$, $B_2$ become the two circular boundaries of $\HH^2/\Gamma$. There is a minimal geodesic $H$, which coincides with the bifurcation surface of the BTZ event horizon; the length $L$ of this geodesic fully characterizes the geometry of $\Sigma$.}
\label{btz}
\end{figure}

\paragraph{Two boundaries:} For example, as noted above, when $\Gamma_\gamma =\ZZ$ is generated by a single element $\gamma$, the resulting spacetime is a BTZ black hole with two asymptotic regions.  The black hole is static (non-rotating) since we choose to preserve time-reversal symmetry as above.  The action of $\gamma$ identifies a pair of geodesics in the hyperbolic plane; by choosing appropriate coordinates in the Poincar\'e disc representation we can take it to act as shown in figure \ref{btz}. The region between the two geodesics defines a fundamental domain for the quotient.  The resulting surface $\Sigma$ is thus topologically a cylinder $S^1 \times \mathbb{R}$ and has two conformal boundaries. This surface is the analogue of the Einstein-Rosen bridge in BTZ. As we approach either boundary the proper size of the $S^1$ grows, and the geometry is asymptotically $\HH^2$.

There is a unique minimal-length closed geodesic on $\Sigma$, drawn as a dotted line in figure \ref{btz}. The proper length $L$ of this minimal geodesic labels the quotient geometries uniquely.\footnote{Genus zero is the unique exception to the general formula for the dimension of the parameter space given earlier; there is a one-parameter family of Riemann surfaces for $g=0, n=2$.}  The action of $\gamma$ has two fixed points on the boundary, marked as small disks in figure \ref{btz}. The restricted space $\widehat{\mathrm{AdS}}_3$ is defined by removing from AdS$_3$ the future and past of these two points and the BTZ spacetime is $\widehat{\mathrm{AdS}}_3 / \Gamma$.

The minimal geodesic $H$ is the bifurcation surface for the BTZ event horizon. In the spacetime it is the boundary of the intersection of both the causal future and the causal past of each boundary with the $t=0$ surface, and the region to the left (right) of this geodesic provides a Cauchy surface for the domain of outer communication $I^+( \mathcal I) \cap I^- (\mathcal I)$, where $\mathcal I$ is the left (right) asymptotic boundary. Its length is thus related to the temperature or mass of  the BTZ black hole. Indeed, for future reference we note that this black hole has temperature
\be
\label{TBTZ}
T = \beta^{-1} = \frac{L}{4 \pi^2 \ell},
\ee
where $L$ is the horizon length and  \eqref{TBTZ} defines the inverse-temperature $\beta$.  Here we have chosen the dimensionless notion of temperature associated with the dimensionless $\tau$ coordinate of \eqref{gads3} in each asymptotic region; we will also use the corresponding dimensionless notion of energy below.

The fact that the non-rotating black hole geometry is fully determined by the parameter $L$ makes it clear that the $t=0$ surface of any such black hole can be constructed by identifying a symmetric pair of geodesics as in figure \ref{btz}. Geodesics in the Poincar\'e disc are circle arcs meeting the boundary at right angles. In general, in the coordinates of \eqref{gads3}, a circle arc between points at $\phi = \alpha \pm \psi$ is given by
\be
\tanh \chi \cos (\phi - \alpha) = \cos \psi.
\ee
We will describe such a circle arc as being centered at $\alpha$, with opening angle $\psi$.

For the symmetric geodesics in figure \ref{btz} the centers are at $\alpha = \pi/2$, $\alpha' = - \pi/2$, and the opening angles are the same, $\psi = \psi'$, so we could also take $\psi$ as the parameter characterising the identification. This is related to the length $L$ of $H$ by
\be \label{length}
L = 2 \ell \tanh^{-1} (\cos \psi).
\ee

Note that there is a sense in which this solution contains {\it two} black holes, associated respectively with the right and left asymptotic regions.  As we will see, this point of view will be useful in giving a unified presentation of the $n$-boundary cases for all $n$.  We will therefore use this terminology below.  The interesting point, however, is that for $n=2$ one finds bulk solutions {\it only} when the right and left black holes are identical,\footnote{This is due to our restriction to purely gravitational solutions with time-reversal symmetry and no internal topology, and to the lack of 2+1 gravitational radiation.  Dynamical solutions involving matter fields can have  different left and right black holes, as can solutions with a wormhole of nontrivial topology linking the two asymptotic regions.} having the same mass $M$ and inverse temperature $\beta$.

\begin{figure}
\centering
\includegraphics[keepaspectratio,width=0.5\linewidth]{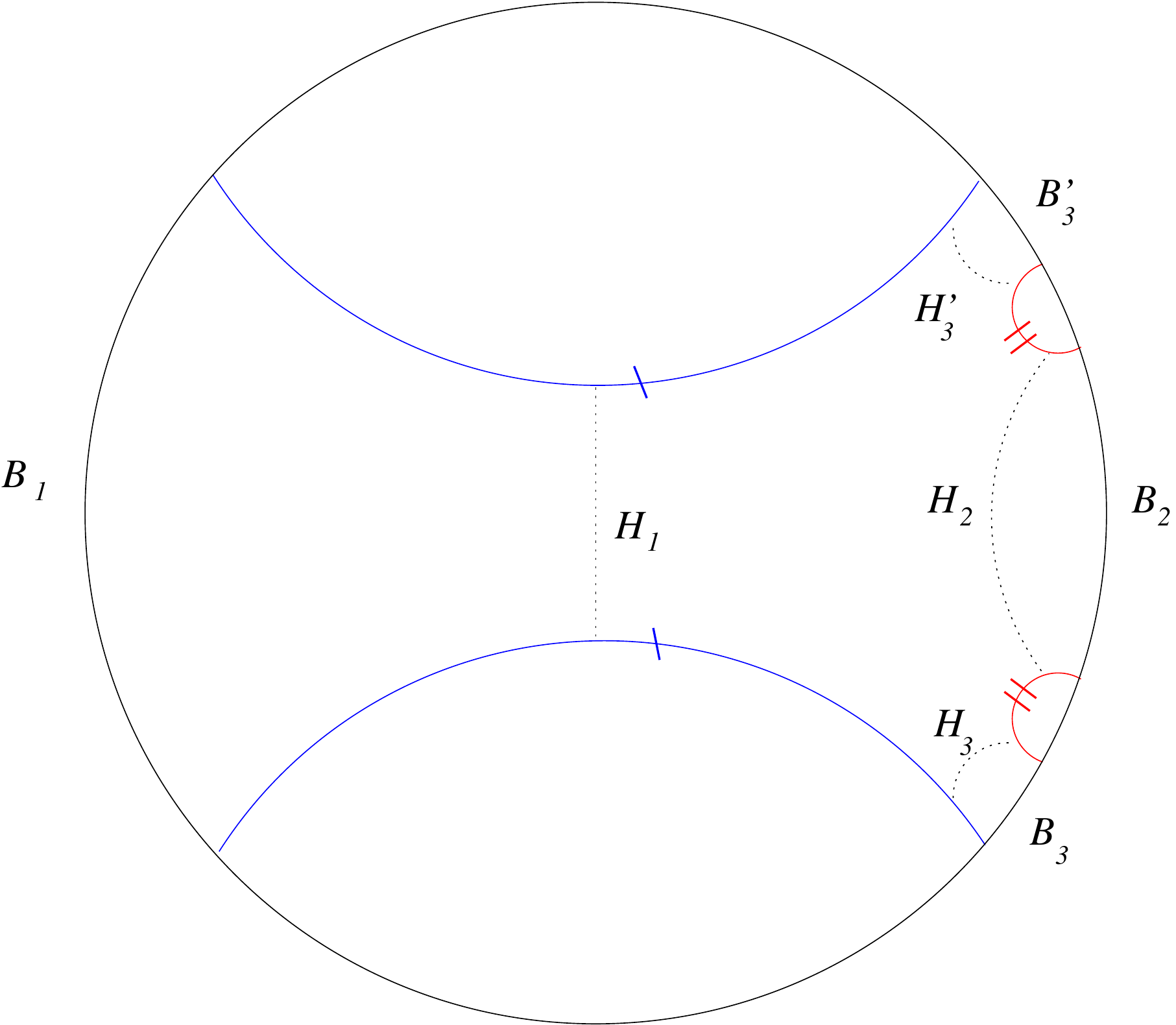}
\caption{The $t=0$ surface $\Sigma$ in the pair of pants wormhole as a quotient of the Poincar\'e disc. The pairs of labeled geodesics (blue and red in colour version) are identified by the action of $\Gamma$. The region of the Poincar\'e disc bounded by these geodesics provides a fundamental domain for the quotient. $B_1$, $B_2$ and $B_3 \cup B_3'$ become the desired three circular boundaries. There are corresponding minimal closed geodesics $H_1$, $H_2$ and $H_3 \cup H_3'$, each lying at a bifurcation surface associated with past and future event horizons for the corresponding asymptotic boundary. The lengths $L_a$ of these geodesic fully characterize the geometry of $\Sigma$.}
\label{pants}
\end{figure}

\paragraph{Three boundaries:} Another important case occurs when $\Gamma$ is generated by two elements that each identify pairs of geodesics as depicted in figure \ref{pants}.  Here $\HH^2/\Gamma$ has the topology of a pair of pants.  There are three circular boundaries: $B_1$  and $B_2$ are clearly circles, and the curves $B_3$ and $B_3'$ connect to give a third.  Note that the topology of $\Sigma$ is symmetric under permutations of the boundaries; it is only our presentation that treats boundary $3$ differently. The geometry of the $t=0$ slice is asymptotically $\HH^2$ at each of these boundaries, and the full spacetime will be asymptotic to global AdS$_3$ in each region. The spacetime in this example is thus a wormhole with three asymptotic boundaries.

The quotient has three independent minimal length geodesics, one in the homology class of each boundary, depicted as dotted lines in figure \ref{pants}. The proper lengths $L_a$ ($a = 1,2,3$) of these geodesics provide the three parameters labeling the geometry of $\Sigma$ and hence of our spacetime.

Any pair of pants geometry can be presented with a reflection symmetry as in figure \ref{pants}, with identifications between the geodesics having $\alpha_1 = \pi/2$, $\alpha_1' = - \pi/2$, $\psi_1 = \psi_1'$, and between the geodesics having $\alpha_2 = - \alpha_2'$, $\psi_2 = \psi_2'$. We can think of $\alpha_2$, $\psi_1$, $\psi_2$ as providing an alternative, more direct parametrization of the geometry of $\Sigma$. $L_1$ is related to $\psi_1$ as in \eqref{length}. The calculation of the other lengths is somewhat more involved, but conceptually straightforward.

The region on our $t=0$ surface outside any one of these minimal length geodesics is identical to a corresponding region in the $t=0$ hypersurface of BTZ. This BTZ geometry is determined only by the horizon length $L_a$. This is clear in the figure for $H_1$ and $H_2$, and we can choose a different representation of $\Sigma$ to make it similarly obvious for $H_3$. Thus, the region of $\Sigma$ outside the geodesic again provides a Cauchy surface for the domain of outer communication attached to the corresponding asymptotic boundary, and the geometry in this region is just as in BTZ. In particular, there are exact rotation and time translation Killing fields in this region.  The outgoing future and past light sheets from each minimal geodesic define future and past event horizons for observers at the corresponding asymptotic boundary.

Since the full geometry has no globally-defined Killing symmetries, the event horizons are not Killing horizons.  But no observation of the geometry near the horizon can detect the difference. There is also a finite piece of the spacetime behind the horizon which is exactly BTZ.  As a result, we may use the local horizon-generating Killing field near any horizon to define a notion of temperature given by \eqref{TBTZ}.     However, as we note in the next section,  the natural bulk state of quantum fields on this background determined by the Euclidean continuation is not thermal even in the region outside the horizons, so this ``temperature'' is a property of the classical geometry only. The bulk quantum fields are expected to evolve from their non-thermal initial state towards an appropriately coarse-grained thermality \cite{Louko:1998dj}.  But this evolution will not change the entanglement between the asymptotic regions, which are out of causal contact.

Perhaps the key difference from BTZ is the existence of a nontrivial region between the horizons $H_1,H_2,H_3$.  This region does not intersect the causal future or past of any asymptotic boundary.  It thus lies in what is known as the ``causal shadow'' region of the spacetime \cite{MVM}.  Understanding the description of this region in terms of the boundary field theories is a particularly interesting question for holography.

This region is also interesting mathematically. We have presented a geometry on the Riemann surface with three boundaries as a subregion of the hyperbolic plane with boundaries at  infinite proper distance. The hyperbolic plane is conformal to a hemisphere, so by doubling this geometry we can obtain a closed Riemann surface (the {\it Schottky double} of $\Sigma$) as a quotient of the sphere. This closed Riemann surface can also (and for many purposes more conveniently) be obtained as a quotient of the hyperbolic plane. (See \cite{Krasnov:2000zq, Skenderis:2009ju} for a more extensive discussion.)  The representation of the Schottky double as a quotient of the hyperbolic plane is obtained by working instead with a representation of $\Sigma$  where the boundaries are geodesics at finite distance. This geometry on the pair of pants is the geometry on the region bounded by the $H_a$ in figure \ref{pants}, but with some nontrivial relation between the parameters. That is, given a surface $\Sigma$ specified by some geodesic lengths $L_a$, we can conformally map $\Sigma$ to the region bounded by the $H_a$ with some geodesic lengths $L_a'(L_a)$. Unfortunately this relation between the parametrizations  is not known explicitly. We will work just with the representation of $\Sigma$ as the entire region shown in figure \ref{pants}, but the representation of the pair of pants as an interior region is useful for building more complex geometries.

\paragraph{More boundaries:} A genus $g$ Riemann surface with $n$ boundaries has a non-unique decomposition into $2g + n -2$ pairs of pants, by cutting the surface along sufficient closed cycles. This decomposition provides a convenient  parametrization of the moduli space of such Riemann surfaces, called the Fenchel-Nielsen parametrization, which consists of taking the 3 parameters $L_a'$ of each of the resulting pairs of pants in the representation as a finite region in the hyperbolic plane, setting equal the lengths of boundaries that are identified, and introducing a twist parameter $\theta$ on each identified boundary. We may use this decomposition to build up the general wormhole spacetime with $n$ boundaries by sewing together pairs of pants to build the conventional genus $g$ surface with $n$ boundaries and then attaching an exterior BTZ region to each of the boundaries $H_a$ to build our wormhole.

\begin{figure}
\centering
\includegraphics[keepaspectratio,width=0.5\linewidth]{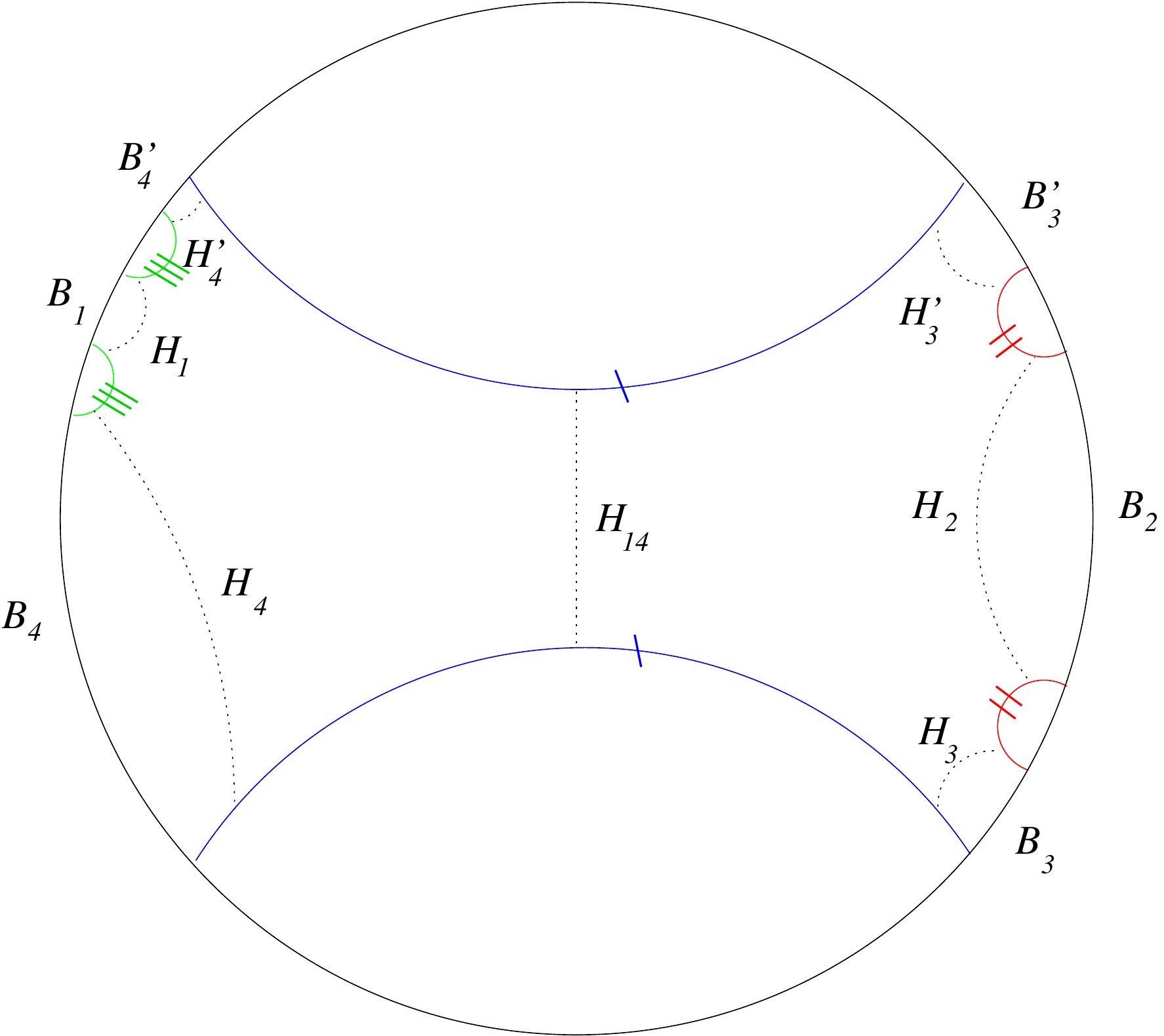}
\caption{The $t=0$ surface $\Sigma$ in the wormhole with four boundaries as a quotient of the Poincar\'e disc. The pairs of marked geodesics (blue, green and red in colour version) are identified by the action of $\Gamma$. The region they bound has four asymptotic boundaries and is a fundamental domain for $\HH^2/\Gamma$. It may be formed by sewing together two pairs of pants along $H_{14}$. The addition of a twist in the sewing would imply that we can no longer choose a reflection-symmetric representation of the geometry. While the first and second pair of identified geodesics are as before, the new identification introduces three new parameters; the centers $\alpha_3$, $\alpha_3'$ of the two identified geodesics are independent, although they can be taken to have the same opening angle, $\psi_3 = \psi_3'$. The identified geodesics are thus labeled by six parameters, corresponding to the moduli space of genus zero surfaces with four boundaries. In terms of geodesic lengths, we can take as independent parameters the lengths $L_a$ of the four horizons, and two additional moduli characterizing the geometry of the interior region.  These are naturally chosen to be the length $L_{14}$ of the minimal geodesic $H_{14}$ in the center and the twist $\theta_{14}$ applied along this geodesic (which is related to $\alpha_3 + \alpha_3'$). There are also similar geodesics $H_{13}$, $H_{12}$ corresponding to the different ways of splitting the surface with four boundaries into two pairs of pants.  But the lengths of $H_{13}$, $H_{12}$ are not independent; they are determined by the moduli above.}
\label{four}
\end{figure}

For simplicity we will restrict our discussion to genus zero surfaces; that is, we consider just spacetimes with multiple boundaries, without  introducing nontrivial topology behind the horizon.\footnote{Some comments on the higher genus case will appear in \cite{MRta}.}  The sewn-together pairs of pants then form a tree-like structure, with no closed loops.
It is reasonably straightforward to construct the full wormhole geometry as a quotient of the Poincar\'e disc; this is illustrated for the case of four boundaries in figure \ref{four}.

The surface $\Sigma$ will have $n$ real ``external" moduli $L_a$, $a=1,\dots,n$, which set the lengths of the horizons associated to each boundary.  In addition, $\Sigma$ will have $n-3$ complex ``internal" moduli which control the geometry of the Riemann surface once the horizon sizes are fixed.
We will denote these collectively as $\bf \tau_\alpha$. If we think of $\Sigma$ as constructed in terms of $n-2$ pairs of pants sewn together along some cuffs, then the ${\bf \tau}_\alpha$ are the length and twist parameters associated with each of the $n-2$ internal cuffs.  Of course, different pair-of-pants decompositions of $\Sigma$ will give different coordinates ${\bf \tau}_\alpha$ on the moduli space of $\Sigma$.
For example, in the $n=4$ case, we can think of $\Sigma$ as two pairs of pants, one with two external cuffs at horizons 1 and 4 and another with external cuffs a horizons 2 and 3.  The internal moduli are $L_{14}$, the length of the internal cuff of the first pair of pants (which of course equals $L_{23}$, the length of the internal cuff of the second pair of pants) along with a twist parameter $\theta_{14}$.
Or one could  instead choose the internal moduli to be $L_{13}$, the length an internal cuff of a pair of pants with external cuffs at horizons 1 and 3, along with an associated twist parameter.  This just gives different coordinates on the moduli space.

%
\subsection{Euclidean path integral}
%
\label{EPI}

We wish to interpret the above geometries holographically, as representing bulk saddle-point descriptions of dual CFT states on their conformal boundary. In \cite{Maldacena:2001kr} it was noted that related CFT states can be constructed by a path integral over the conformal boundary of the Euclidean bulk solution. This is a simple generalization of the BTZ case (see \cite{Skenderis:2009ju} for details).  Note, however, that the equivalence of the geometry near any asymptotic boundary with that of BTZ makes it natural to choose a CFT conformal frame in which the rotational symmetry $\partial_\phi$ of each CFT cylinder coincides for all time with the local rotational Killing field near the corresponding asymptotic boundary of the bulk black hole that evolves from $\Sigma$.  We refer to this choice as the BTZ frame below and caution the reader that it differs from the frame chosen in \cite{Skenderis:2009ju}.  Due to the conformal anomaly, this in particular leads to different expectation values for the CFT stress tensor.

On the bulk side the idea is to follow the usual Hartle-Hawking procedure and define a quantum state by a Euclidean gravitational path integral for which the wormhole spacetime describes a classical saddle-point. Since $\Sigma$ represents a moment of time symmetry in the Lorentzian spacetime it has zero extrinsic curvature.  Thus there is a corresponding Euclidean solution which also contains $\Sigma$ as a surface of zero extrinsic curvature. The Euclidean continuation can be described as a quotient of Euclidean AdS${}_3$ (i.e. $\HH^3$) by the same discrete group $\Gamma$.  The isometry group of Euclidean AdS${}_3$ is $SO(3,1)\simeq SL(2,\mathbb{C})$, and the Euclidean geometry will be the quotient of Euclidean AdS${}_3$ by $\Gamma$, now regarded as a subgroup of $SL(2, \RR) \subset SL(2,\mathbb{C})$. As in the Lorentzian case, we take the action of $\Gamma$ on $\HH^3$ to be determined by its action on $\Sigma$.  We now see the relevance of the restriction of $\Gamma$ to the diagonal subgroup; it is precisely for these groups that there is a real Euclidean geometry formed by quotienting $\HH^3$ by the same discrete group.

Less abstractly, the Euclidean geometry can be constructed by analytically continuing $t \to i t_E$ in \eqref{metric} to give the Euclidean metric
\be \label{emetric}
{ds^2\over\ell^2} = dt_E^2 + \cosh^2 t_E \, d\Sigma^2.
\ee
Here, as above, $d\Sigma^2$ is the constant negative curvature metric on the Riemann surface $\Sigma$.  The Euclidean geometry matches onto the Lorentzian geometry at the surface of vanishing extrinsic curvature $t_E=0$. In contrast to the Lorentzian case where \eqref{metric} only covered a part of the geometry, this coordinate system covers the whole of the Euclidean geometry. The geometry \eqref{emetric} for all $t_E$ is a solid handlebody whose boundary is the Schottky double of $\Sigma$.  This is not entirely apparent from \eqref{emetric}, which appears to have two asymptotic boundaries at $t_E\to \pm \infty$, each of which is a copy of $\Sigma$.  This, however, is a quirk of the coordinate system \eqref{metric} --  it can be shown that the boundaries of the two copies of $\Sigma$ are in fact glued to one another to obtain a single compact Riemann surface, the Schottky double of $\Sigma$.   We refer the reader to \cite{Krasnov:2000zq, Skenderis:2009ju} for further discussion of the Euclidean multi-wormhole geometry, and \cite{maskit1987kleinian} for a mathematical discussion of hyperbolic 3-manifolds of this type.

The suggestion of \cite{Maldacena:2001kr,Skenderis:2009ju} was to treat \eqref{emetric} for $t_E <0$ as a saddle point of the Euclidean path integral defining our Hartle-Hawking-like state on the $t=0$ surface of the Lorentzian geometry.  If this saddle dominates the bulk path integral, then via holography it also approximates the state $|\Sigma\rangle$ defined on the boundaries $B_a$ of $\Sigma$ by the field theory path integral over the asymptotic boundary of the bulk Euclidean geometry. In the coordinates \eqref{emetric}, this conformal boundary lies at $t_E \to -\infty$.  It follows that its conformal geometry is identical to that of $\Sigma$.  For simplicity we will describe the state $|\Sigma\rangle$ as being given by a path integral over $\Sigma$ itself, though the reader should understand that we do not imply any local identification of the CFT spacetime and any particular (e.g., $t=0$) surface in the bulk.

We conclude by noting that the full conformal boundary of \eqref{emetric} formed by joining the two surfaces at $t_E \rightarrow \pm \infty$ generally admits no $U(1)$ isometry.  As a result, there is no obvious sense in which Euclidean time is periodic, and there is no reason to expect precisely thermal behavior in any CFT copy or in any asymptotic region of the bulk.

%
\subsection{Bulk phases}
\label{phases}
%

We constructed the above CFT path integral to understand the dual description of the multiboundary black hole.   But we now argue that there are multiple saddle-points with the same boundary conditions, resulting in phase transitions  in which different saddle points exchange dominance  as the moduli of $\Sigma$ are varied (see also \cite{MRta}).  We will keep the discussion in this section rather brief, though in appendix \ref{mcg} we give a somewhat more mathematical discussion where these phase transitions are related to the mapping class group.  In a theory with fermions, there will be further interesting dependence on the choice of periodic or antiperiodic boundary conditions for the fermions on each circle. We relegate this discussion to appendix \ref{spins} and focus here and in the main text below on the simplest case with antiperiodic boundary conditions on each boundary.

To review the two-boundary case, consider the CFT state on $\mathcal{H}_1 \otimes \mathcal{H}_2$ described by the path integral over $\Sigma$ as shown in figure \ref{btz}.  This $\Sigma$ is conformal to a cylinder, and a nice way to determine the dominant bulk saddle is to study the CFT path integral for the norm of this state.  One then integrates over the torus obtained by gluing together two copies of $\Sigma$ at the corresponding boundaries.  We thus seek bulk saddles with torus boundaries.  Such solutions to the bulk equations of motion are solid tori with a single contractible cycle. For a fixed boundary condition (fixed conformal structure of the boundary torus) there is then an infinite family of bulk saddle points, characterized by choosing a homotopy class of the boundary torus to be identified with the contractible cycle in the bulk \cite{Maldacena:1998bw}.

Our torus has a $\mathbb{Z}_2$ symmetry exchanging the two copies of $\Sigma$ that we may think of as a Euclidean time-reflection, and which implies that our state is invariant under time-reversal. Two of the above saddles preserve the time-reversal symmetry: they are called Euclidean thermal AdS (when the original CFT spatial circle is contractible) and the Euclidean BTZ black hole (when the Euclidean time direction is contractible).   Here the term thermal is associated with having a Euclidean time translation symmetry, so that the period $\beta$ of this Euclidean time defines an inverse temperature. Taking the spatial circle to have period $2\pi$, there is a Hawking-Page
phase transition $\beta = 2\pi$, with Euclidean thermal AdS$_3$ dominating for small temperatures (large $\beta$) and Euclidean BTZ dominating for high temperatures.  At least one of these always dominates over those bulk geometries that break time-reversal symmetry.

Since Euclidean time-reversal symmetry implies that the $t=0$ surface has zero extrinsic curvature, we can use the bulk geometry on this surface as initial data to build a unique Lorentz-signature spacetime.  This spacetime is just the Wick rotation of the Euclidean saddle.  In this sense it is convenient to speak of our bulk path integral being dominated by the corresponding Lorentz-signature spacetime which in our case is either two copies of global AdS$_{3}$ or BTZ as implied by the terminology above.  In the global AdS${}_3$ case the quantum fields on the two copies combine to form a thermofield double state.

From the Lorentzian point of view, the thermal AdS and Euclidean BTZ saddles give contributions to the wave function which described different Lorentzian space-times.  The state is obtained by matching onto a Lorentzian signature geometry at a moment of time reflection symmetry (i.e. a surface of vanishing extrinsic curvature).  In  the thermal AdS saddle, the surface of vanishing extrinsic curvature is a pair of disconnected discs; this geometry constructs the thermofield double entangled state between a pair of disconnected AdS space times.  In the Euclidean BTZ saddle the surface of vanishing extrinsic curvature is a cylinder with two asymptotic regions; this geometry constructs the BTZ black hole  with two boundaries connected by a wormhole.

\begin{figure}
\centering
\includegraphics[keepaspectratio,width=0.3\linewidth]{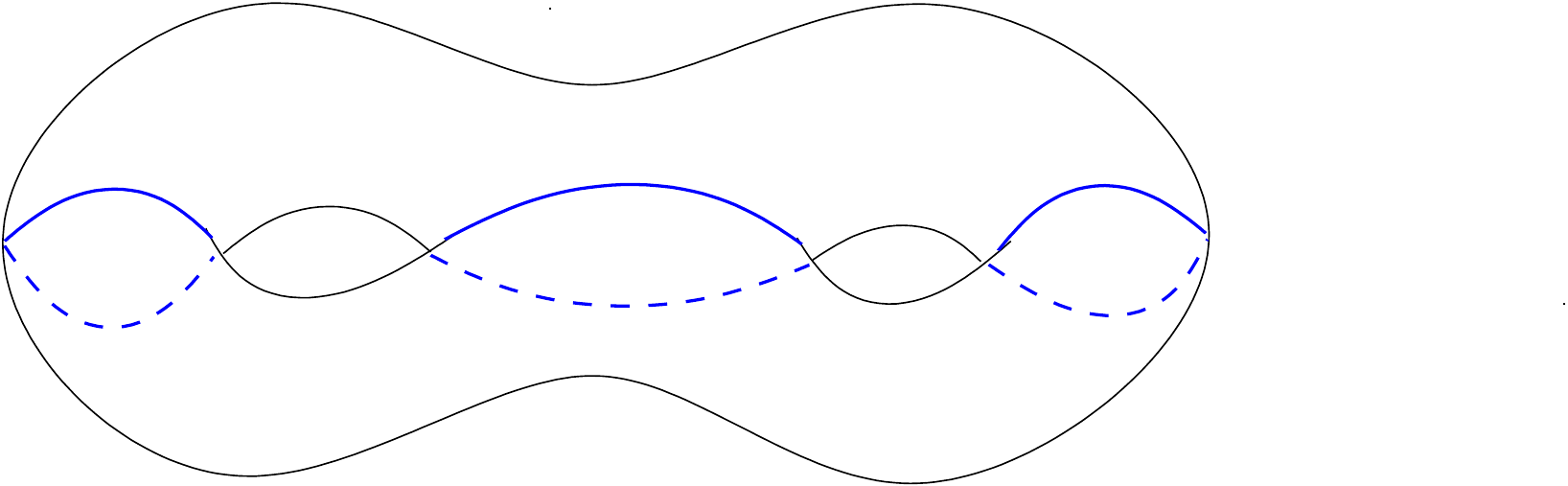}
\includegraphics[keepaspectratio,width=0.3\linewidth]{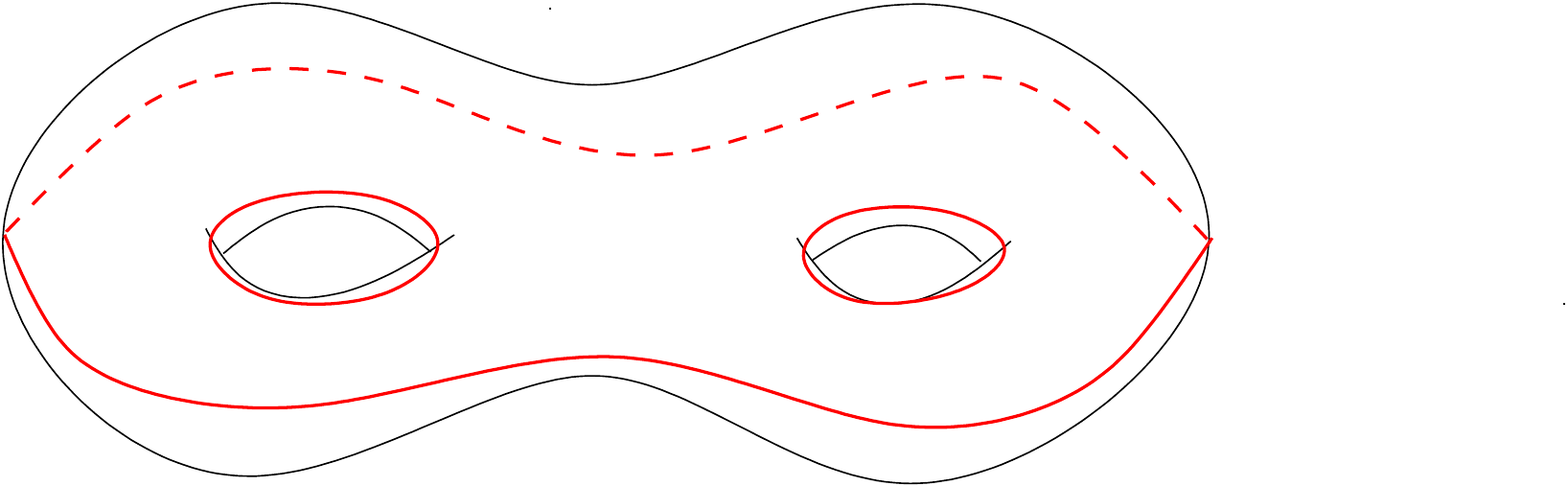}
\includegraphics[keepaspectratio,width=0.3\linewidth]{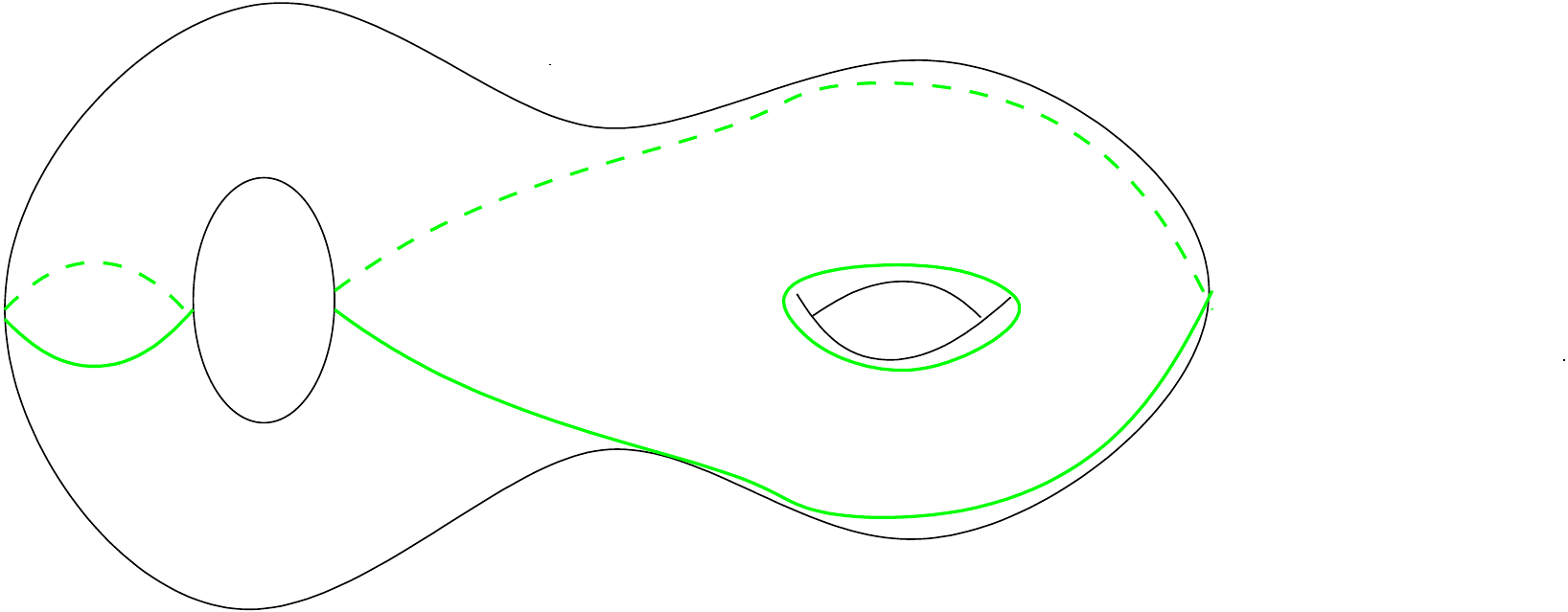}
\caption{Different decompositions of the genus two Riemann surface which preserve a $\mathbb Z_2$ reflection symmetry. In each case the boundary of the handlebody is decomposed into two pairs of pants. If we take the pair of pants $\Sigma$ to have the same geometry, the geometry of the resulting genus two surface is different in each case, but the different decompositions describe the same family of bulk handlebody geometries up to diffeomorphism. In the bulk handlebody, the cycles indicated in the left figure are contractible, while those indicated in the second are not. Slicing the bulk handlebody along the cycles in the left figure thus provides a bulk initial data surface consisting of three disconnected discs, corresponding to three copies of global AdS. The second slicing produces a bulk initial data surface for a three-boundary wormhole. The final slicing, along the cycles indicated in the right figure, yields a bulk initial data surface consisting of an annulus and a disconnected disc, corresponding to BTZ and a disconnected copy of global AdS.}
\label{genus2}
\end{figure}

The three-boundary case has a similar but richer structure. In the case where $\Sigma$ is a pair of pants, we can glue together two copies of $\Sigma$ to obtain a genus two Riemann surface.  The bulk solution of interest is a handlebody, the three-manifold constructed by filling in this Riemann surface. The Riemann surface has four elementary cycles, of which two are contractible in the bulk.
Different Euclidean saddle points can then be constructed by specifying which combinations of cycles are contractible in the bulk.
We are interested in those choices which have a $\mathbb Z_2$ time-reveral symmetry, so that they can be analytically continued to real Lorentzian geometries.
There are three different choices of geometry, shown in figure \ref{genus2}, which have this symmetry.  These choices correspond to three different ways one can construct a genus two Riemann surface by gluing together two pairs-of-pants.

As in the genus one case, each Euclidean geometry is regarded as preparing a state in the Lorentzian theory.
To understand the Lorentzian interpretation, we need to find the surface of vanishing extrinsic curvature in each case.
In the first case the surface of vanishing extrinsic curvature is three disconnected discs.
Thus this geometry constructs an entangled state in three different disconnected AdS space-times.
In the second case the surface is the pair of pants $\Sigma$.  So this geometry constructs a three-boundary (pair of pants) wormhole connecting all the boundaries.
In the third case the surface of zero extrinsic curvature is topologically a disc and a cylinder.  This geometry constructs
a BTZ black hole connecting one pair of boundaries which is entangled with one copy of global AdS$_3$.
In fact, this case yields three distinct saddles depending on which pair of boundaries are taken to be connected.

We would expect varying moduli (parametrized by the horizon lengths $L_a$) to induce transitions between these different phases. In the limit where all $L_a$ are large one expects the three-boundary wormhole to dominate; for all $L_a$ small one expects the 3-global-AdS$_3$ phase to dominate; while  for one $L_a$ small with the other two large we expect the global-AdS$_3$-plus-BTZ phase to dominate.

By construction our $\beta_a = \frac{4 \pi^2 \ell}{L_a}$ are the inverse temperatures of the black holes in the connected 3-boundary wormhole phase. But they will not generally be the inverse temperatures in the BTZ-plus-global AdS${}_3$ phase. This is because the $t=0$ geometry outside a given bulk BTZ horizon after disconnection is not generally the same as that in the associated asymptotic region of the fully-connected bulk wormhole of $\Sigma$.  This is clear from the fact that BTZ solutions require identical black holes on right and left, while the fully-connected wormhole allows all $\beta_a$ to be distinct.  Our phase transition will thus generically be associated with discontinuous changes in the geometry of the exterior black hole regions, even when the black hole does not completely disappear.  In fact, we will argue in section \ref{reduced} below that the black hole temperature changes by a factor of two even when e.g. $\beta_1 = \beta_2$ and the phase transition occurs due to varying $\beta_3$.

Further, while inverse temperatures are a useful way to characterize the geometries of the saddle-points that include black holes, we will see in Sec.~\ref{cft} that the quantum state of matter fields is generally non-thermal even when restricted to the exterior regions. In particular, while the bulk quantum fields on the global AdS${}_3$ retain some order-one entanglement with the other spaces, the associated density matrix on one such factor need not be precisely thermal. In the next section we will often focus on the puncture limit of small $L_a$; we will find that in this limit at least there is some thermal behaviour in the CFT state, with the contribution of high-energy states having a Boltzmann suppression determined by the $\beta_a$.

In addition to the above saddle-points obtained by various slicings of the handlebody, there will also be non-handlebody solutions in the bulk, as noted in \cite{Yin:2007at}. These could in principle produce additional phases. However, it was conjectured in \cite{Yin:2007at} that such non-handlebody saddles are always subleading in the path integral compared to the handlebodies. We will assume that this is true, and not consider the non-handlebody geometries further.

The phase transitions can in principle be calculated by evaluating the bulk action for the different saddles. But in special subspaces of the moduli space, we can avoid doing this explicit calculation. In the two-boundary case, the transition between two copies of thermal AdS$_3$ and BTZ occurs when the boundary torus is square, that is the length of the time and space circles are equal. This is a point of enhanced symmetry in the torus moduli space. Phase transitions in the moduli space of genus two surfaces should also occur at points of enhanced symmetry.  For example, the subspace where all the temperatures are equal includes the Bolza surface, the most symmetric genus two surface. There should then be a transition from the three AdS$_3$ saddle to the three-boundary wormhole saddle when we reach the Bolza surface. In appendix \ref{hp} we show that this occurs at $\beta$ of order one, as expected. By continuity, we can then argue that when the $\beta_a$ are of similar size and all small, the dominant saddle should be three copies of AdS$_3$, while when they are of similar size and all large, it should be the three-boundary wormhole.

For $n > 3$ boundaries there will be a similar structure but with more possibilities. Relevant phases include the $n$-boundary connected wormhole as well as any combination of disconnected components $i$ having respectively $m_i \ge 1$ boundaries with $\sum_i m_i = n$.

Understanding the details of this phase structure is interesting in its own right, but the important point for our present discussion is that the CFT states we consider do not always correspond to a single connected wormhole in the bulk. In particular, much of the discussion in the next section will be carried out in a puncture limit where the lengths $L_a$ are small so that -- at least for the simplest choice of spin structure -- fully connected bulk saddles will not dominate.  However, as in the two-boundary case, some of the qualitative features of the entanglement in the CFT state will be the same in any phase.

%
\section{CFT properties of the state}
\label{cft}
%

When there are $n$ asymptotic regions in the bulk Lorentzian spacetime, the dual CFT state $|\Sigma\rangle$ is an element of $\mathcal{H}_1 \otimes \mathcal{H}_2 \otimes \cdots \mathcal{H}_n$ where $\mathcal{H}_i$ is the Hilbert space of the CFT on a circle.    We are interested in a state $|\Sigma\rangle$ defined by the CFT path integral over a Riemann surface $\Sigma$ with $n$ boundaries. Holographically, this $\Sigma$ also gives the asymptotic boundary conditions of the corresponding bulk path integral, and in the particular case when the bulk is dominated by the fully-connected  wormhole we saw in section \ref{EPI} that $\Sigma$ is also the geometry of the bulk $t=0$ surface.

In the CFT path integral, if states $|\phi_1\rangle \in {\cal H}_1, \dots |\phi_n\rangle \in {\cal H}_n$ are used to fix boundary conditions on the respective boundaries of $\Sigma$, this path integral computes the amplitude $\langle \Sigma | \phi_1 \phi_2 \dots \phi_n\rangle$.  Working in a
basis of energy eigenfunctions $|i \rangle$ (of energy $E_i$) in each ${\cal H}_a$ we may write the state as
\be \label{nstate}
|\Sigma \rangle = \sum_{i_1 \ldots i_n} A_{i_1 \ldots i_n} |i_1 \rangle_1 \ldots  |i_n \rangle_n,
\ee
where the coefficients $A_{i_1 \ldots i_n}$ are functions of the $3n-6$ moduli of $\Sigma$, determined by the Euclidean path integral on $\Sigma$. For simplicity we again restrict attention to the case where $\Sigma$ has genus zero.

In the two-boundary case the $A_{i_1 i_2}$ can be determined explicitly. There $\Sigma$ is conformal to a flat cylinder, and the BTZ choice of conformal frame makes \eqref{nstate} invariant under both time-reversal and the natural rotational symmetry.  If its circumference is $2 \pi$, the cylinder has length $\beta/2$, where $\beta$ is the periodicity of the dimensionless bulk imaginary time coordinate corresponding asymptotically to the Wick rotation of $\tau$ in \eqref{gads3}.  Thus the CFT path integral produces the usual thermofield-double state
\be \label{tfd}
|\Sigma \rangle = \sum_i e^{-\beta E_i/2 \ell} |i\rangle_1 |i\rangle_2 \,
\ee
where $\beta$ coincides with the inverse temperature one would assign to both the BTZ and double-global-AdS${}_3$ bulk saddles.

In the multiboundary case the path integral over $\Sigma$ is more complicated and we will not generally be able to write the coefficients explicitly. In general the $A_{i_1\ldots i_n}$ will be non-vanishing even if the indices $i_a$ are unequal, unlike the two-boundary case.
Our state $|\Sigma\rangle$ is thus more complicated than a GHZ-like state, in that in the GHZ state
$|{\rm GHZ}\rangle = (\left|\uparrow \uparrow \uparrow \right\rangle + \left|\downarrow \downarrow \downarrow \right\rangle ) / \sqrt{2}$ only the diagonal elements $A_{\uparrow\uparrow\uparrow}$ and $A_{\downarrow \downarrow \downarrow}$ are non-vanishing.

To evaluate (\ref{nstate}) we will pursue the following general strategy.   First the Riemann surface $\Sigma$ can always be conformally mapped to a sphere with $n$ holes $h_a$.   Using the state-operator correspondence, we can map the states  $|i \rangle_a$ fixing the boundary conditions on these holes to local insertions of the corresponding operators $O_a$ at some point inside each hole; the  $A_{i_1  \ldots i_n}$ are thus related to the $n$-point correlation functions $\langle O_1 \dots O_n \rangle$.
The mapping from states to operators will introduce a factor depending on the size and shape of the hole $h_a$.  As a result, the $A_{i_1 \ldots i_n}$ are related to the $\langle O_1 \dots O_n \rangle$ by the action of operators $V_a$ acting on $\mathcal H_a$.   These $V_a$ are essentially complexified conformal transformations that depend only on the moduli of the Riemann surface ($L_a$, $\tau_\alpha$ in the notation of the previous section) and so are universal for all CFTs.   In this way, the coefficients $A_{i_1 \ldots i_n}$ can be written as a product of a universal piece that depends on the moduli of the Riemann surface (and hence on the parameters of the Lorentzian spacetime), and a piece that depends on the specific field content of the CFT via the $n$-point functions on the sphere.

We will be able to carry out this procedure explicitly when the  Riemann surface $\Sigma$ has genus zero and is close to the puncture limit ($L_a \rightarrow 0$). In this limit we will see that  we can choose a conformal frame where the shrinking holes are round; the moduli of $\Sigma$ then correspond to radii and locations of the holes. Using this picture we will show that, up to exponentially suppressed corrections and a convention-dependent rotation,
\begin{equation}
\label{puncture}
V_a = \exp(-\frac{1}{2} \hat \beta_a H_a), \ \ \ {\rm with} \ \ \hat \beta_a = \beta_a - \beta^0_a.
\end{equation}
Here $H_a$ is the Hamiltonian on boundary $a$, and $\beta_a= \frac{4\pi^2 \ell}{L_a}$ is the inverse temperature of the BTZ geometry in the exterior region of the surface $\Sigma$ (i.e. the region between the minimal geodesic whose length is $L_a$ and the boundary $a$).   The parameter $\beta^0_a$ depends only on the moduli that are held finite in this limit.\footnote{There is also a more general puncture limit where only some $L_a$ are taken small.  The corresponding $V_a$ again take the form \eqref{puncture}, but the $\beta_0^a$ can then also depend on the $L_b$ that remain finite.}  Thus, at least in the puncture limit, the $A_{i_1 \ldots i_n}$ involve a Boltzmann suppression similar in character to the two-boundary case.

A key simplification in the puncture limit is that the $V_a$ \eqref{puncture} are diagonal in the energy basis, so that the coefficients $A_{i_1 \ldots i_n}$ become simply proportional to the $n$-point functions. This simple relation holds only in the puncture limit; more generally we do not expect the energy basis to diagonalize the $V_a$.

In section \ref{3ptf} we carry out this puncture limit calculation in detail for the three-boundary case, and comment on the extension beyond the puncture limit. In section \ref{factorization} and \ref{reduced} we study  interesting limits in the general $n$-boundary moduli space. In particular, limits in which $n$-point functions factorize lead to corresponding factorization of the state \eqref{nstate}. A similar discussion at the level of density matrices will then allow us to make further observations regarding the entanglement structure of \eqref{nstate}.

%
\subsection{Three-boundary state}
\label{3ptf}
%

We now explore the relation between the CFT three-point function and the state on three boundaries. Because there are no internal moduli, this is the simplest example of the connection to $n$-point functions.  For three boundaries, the CFT state $|\Sigma \rangle$  is an element of $\mathcal H_1 \otimes \mathcal H_2 \otimes \mathcal H_3$, where $\mathcal H_a$ is the Hilbert space of the CFT on a circle. Introducing a basis of  energy eigenstates $|i \rangle_a$ on each boundary, it can in general be written
\be \label{3state}
|\Sigma \rangle = \sum_{ijk} A_{ijk} |i\rangle_1 |j\rangle_2 |k\rangle_3,
\ee
where $A_{ijk}$ is a set of coefficients whose structure we wish to determine. These coefficients are defined by a path integral over some pair of pants geometry, with the specified states on each boundary.

In section \ref{review} we parametrized the three-dimensional moduli space of pair-of-pants metrics by the lengths $L_1$, $L_2$, $L_3$ of the minimal geodesics. The $A_{ijk}$ will be functions of these parameters. In a theory with fermions, there will also be significant dependence on the choice of periodic or antiperiodic boundary conditions for the fermions on each circle; we relegate this discussion to appendix \ref{spins}.

\subsubsection{Mapping to the sphere}

As outlined above, we begin by conformally mapping our problem to a path integral over the round sphere with three holes $h_a$. We will do explicit calculations for small $L_a$. At least in some open set in which all $L_a$ are small, we will be able to choose a conformal frame in which the holes are round holes of circumference $2 \pi r_a$ centered at equally spaced points around the equator; the moduli of $\Sigma$ are mapped to the three radii $r_a$. Far from the puncture limit, such a description may not be possible because the conformal transformations mapping the individual holes to round circles can produce overlapping circles, though it should be possible to understand the description for general $L_a$ by a suitable analytic continuation.

We do not know  the explicit conformal map from $\Sigma$ to the sphere even for small $L_a$. But in the regime of small $L_a$, we can determine the radii $r_a$ as a function of the $L_a$ up to a universal factor. We do so by breaking the problem up into two overlapping regions: the exterior legs of $\Sigma$, which have the geometry of the $t=0$ surface in BTZ, and an interior region $I$, which is well-approximated by a universal geometry for small $L_a$.

\begin{figure}
\centering
\includegraphics[keepaspectratio,width=0.5\linewidth]{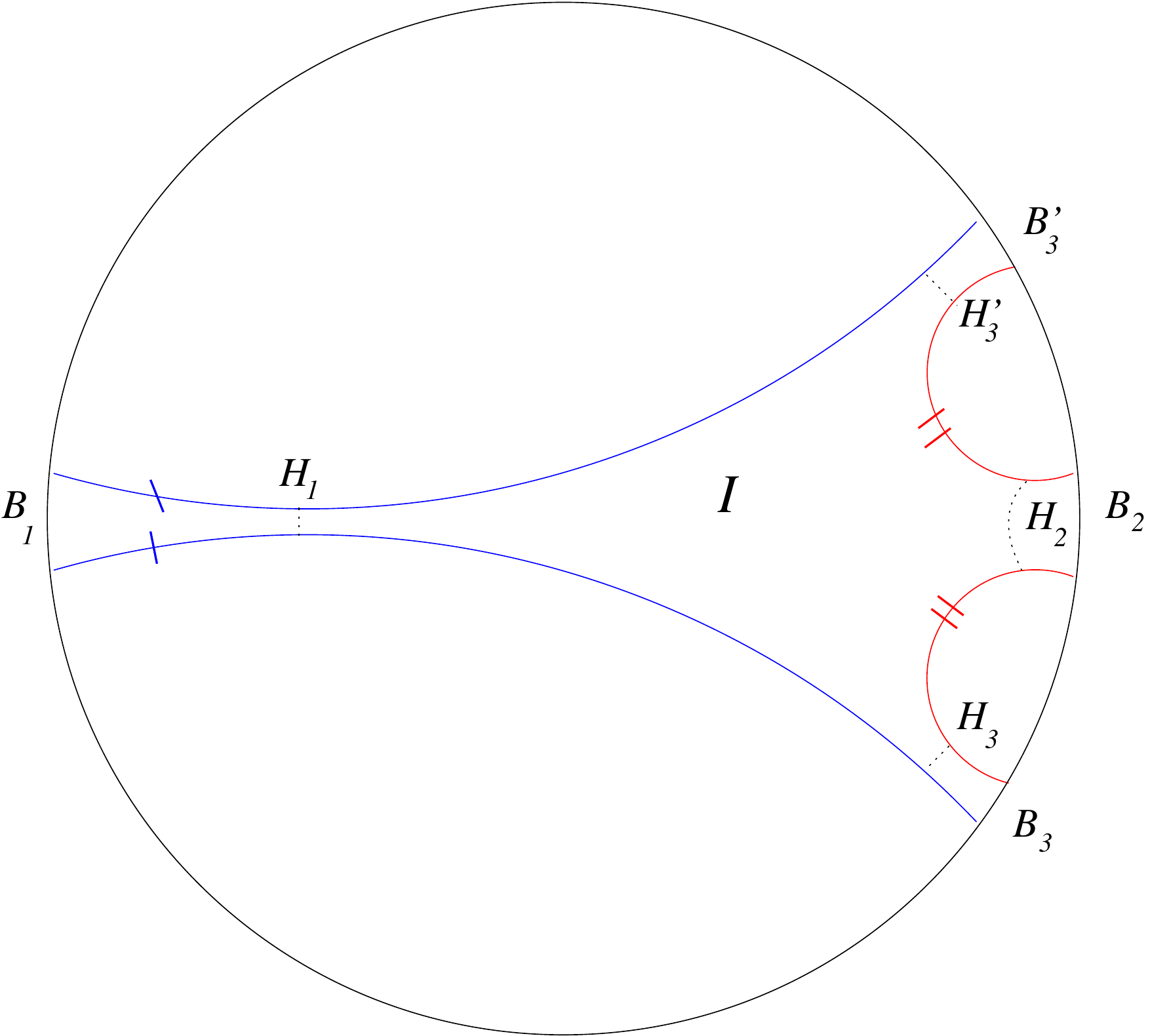}
\caption{The limit of small $L_a$ for the surface $\Sigma$, showing the region $I$ between the minimal geodesics. Here the identifications are drawn in a different presentation to make it obvious that all the minimal geodesics go off to infinite proper distance in the limit of small $L_a$.}
\label{sigma0}
\end{figure}

In the limit of vanishing $L_a$, the geometry of $\Sigma$ approaches a universal geometry $\Sigma_0$. This limit is illustrated in figure \ref{sigma0}, which redraws figure \ref{pants} to make it clear that in this limit all minimal geodesics recede to infinite proper distance from the interior region $I$, and the boundary segments $B_a$ shrink to points.\footnote{In the limit, $B_3$ and $B_3'$ approach points $\pm \phi_0$ on the boundary, which may be different for different values of $L_a$, but different values of $\phi_0$ are related by an $SL(2,\mathbb{R})$ transformation of the disc, so they define equivalent geometries, and there is a unique limiting geometry $\Sigma_0$.} For small $L_a$, the geometry in an interior region $I$ will be close to that of $\Sigma_0$, and we may  approximate the map from $I$ to the sphere by that from the corresponding region of $\Sigma_0$ to the sphere.  (This approximation breaks down before we reach the minimal geodesics $H_a$.)

The universal geometry $\Sigma_0$ is a hyperbolic metric with three infinitely long cuspy arms. There is some conformal map from this surface to a punctured sphere, sending the end of each cusp to one of the punctures, which we can take to be equally spaced on the equator of the sphere. We do not know the full map from $\Sigma_0$ to the sphere, but it is easy to give its form near one of these punctures.  As noted above, for any $\Sigma$  the geometry around each of the minimal geodesics is locally that of the $t=0$ surface for a BTZ black hole.    The $L_a \rightarrow 0$ limit gives the zero mass black hole which is described by the  hyperbolic cusp geometry
\begin{equation}
\label{cusp}
ds^2_{cusp} = \ell^2 \frac{d\rho^2}{\rho^2} + \rho^2 d \phi^2.
\end{equation}
Thus, the geometry near the cusp at the end of each arm of $\Sigma_0$  (see Fig.~\ref{sigma0}) will be isometric to (\ref{cusp}) in an open set containing the end of each arm.   In \eqref{cusp} the coordinate $\phi$ has period $2 \pi$ and $\rho \in (0, \infty)$ with $\rho \rightarrow 0$ at the tip of the cusp.

We may take the conformal map from $\Sigma_0$ to our punctured sphere to have the property that $\rho = constant$ curves near the tip of each cusp map to closed curves on the sphere that become round circles as $\rho \rightarrow 0$.  This is clearly the case when we conformally map the precise cusp geometry \eqref{cusp} to a disk (even when the disk is considered as a part of a sphere; i.e., with Ricci scalar $R_{disk} = 2$).  The conformal mapping from $\Sigma_0$ (which interpolates between three cusps) to a sphere is more complicated, but is fundamentally governed by the elliptic differential equation obtained from the Weyl transformation properties of the Ricci scalar. In particular, for $ds^2_{sphere} = \Omega^2 ds^2_\Sigma$ we require
\begin{equation}
\label{Rconf}
\Omega^2 R_{sphere} = R_{\Sigma} + 2 \nabla^2 \ln \Omega .
\end{equation}
Since this equation approaches that governing the above cusp-to-disk conformal map near the tip of each cusp, we may seek a solution of \eqref{Rconf} that approximates the cusp-to-disk map near the tip of each cusp.  Thus, near one of the punctures, the map from the region near the cusp to the punctured sphere will be
\begin{equation}
\label{cuspmap}
r \approx r_d e^{-\ell/\rho},
\end{equation}
where $r$ is the radial distance from the puncture on the sphere and $r_d$ represents the zero mode of $\Omega$ (i.e. the overall scale of the map).   This degree of freedom is not fixed by the above discussion in the cusp region and encodes our ignorance of the full map from $\Sigma_0$ to the sphere.

Corrections to the cusp-to-disk map implied by \eqref{Rconf} may further be computed perturbatively in the cusp region.  Perhaps the easiest way to understand the nature of these corrections is to work in the sphere conformal frame and to recall that the Green's function for Laplace's equation on a disk is of the form $r^m$ near the origin where $m$ is the angular momentum of the desired source and $r$ is the usual radial coordinate on the disk.  Effects that break rotational symmetry thus decay at least as $r$ as $r \rightarrow 0$. These corrections are thus exponentially suppressed for $\rho \ll \ell$.

For small $L_a$, the surface $\Sigma$ has an interior region $I$ which is well-approximated by a corresponding region of $\Sigma_0$, which extends down into the cusp regions where the approximate map \eqref{cuspmap} is valid. The conformal map to the sphere for this region should be well-approximated by the map from $\Sigma_0$ to the sphere. The remaining problem is then to map the regions of $\Sigma$ outside $I$ into these holes. This can be done using the fact that the geometry on each exterior arm of $\Sigma$ is just that of BTZ at $t=0$.  Recall that this description is valid even a finite distance inside the horizon.  In fact, as should be clear from figure \ref{sigma0},  in the limit of small $L_a$ it remains valid {\it very} far inside the horizon.  The metric in such a region is\footnote{The metric \eqref{btzgeom} is equivalent to that of \cite{Banados:1992wn} when written in terms of $r_{BTZ} = \sqrt{\rho^2 + \rho_a^2}$, but we have chosen the form \eqref{btzgeom} to make manifest the agreement with (\ref{cusp}) for $\rho \gg L_a$.}
\begin{equation}
\label{btzgeom}
ds^2_{btz} = \ell^2 \frac{d\rho^2}{\rho^2 + \rho_a^2} + \left(\rho^2 + \rho_a^2\right) d \phi^2
\end{equation}
with $\rho_a = L_a /2\pi$. Here we again take $\phi$ to have period $2\pi$ but now $\rho \in (-\infty, \infty)$.  Since $L_a$ is small, the merger of this arm with the others takes place essentially as in $\Sigma_0$.   The lack of free parameters in $\Sigma_0$ means that, in the coordinates of \eqref{cusp},\eqref{btzgeom} this merger must take place at some $\rho$ of order $\ell$.  Thus \eqref{btzgeom} is the exact metric on $\Sigma$ in the region $- \infty <  \rho \lesssim \ell$, where $\rho = -\infty$ corresponds to the exterior boundary and $\rho=0$ is the minimal geodesic.

The region covered by the BTZ coordinates \eqref{btzgeom} overlaps with $I$, so we can determine the conformal map to the sphere in the region $\rho \ll \ell$ by requiring agreement with \eqref{cuspmap} in the region $\rho_a \ll \rho \ll \ell$, where the metric \eqref{btzgeom}  is well approximated by \eqref{cusp}.  The map from $\Sigma$ to the sphere in this region should coincide to leading order in $\rho_a/\rho$ with that from $\Sigma_0$ to the sphere, and map the rotational Killing field near the $a$th boundary of $\Sigma$ to that of the sphere about the associated center $c_a$ to leading order in $\rho/\ell$. Using a radial coordinate around the center $c_a$ it must then take the form
\begin{equation}
r = r_d e^{\left(\ell /\rho_a\right) [\tan^{-1} (\rho/\rho_a) - \pi/2]}
\end{equation}
to leading order in $\frac{\rho}{\ell}$, $\frac{\rho_a}{\ell}$.
The asymptotic boundary $\rho = - \infty$ thus maps to
\begin{equation}
r_a = r_d e^{-\pi \ell/ \rho_a} = r_d e^{-\frac{1}{2} \beta_a},
\end{equation}
where $\beta_a = \frac{4 \pi^2 \ell}{L_a}$ is the inverse temperature associated to the BTZ geometry \eqref{btzgeom}. Recall that $r_d$ represents an overall scale in the  map from the limiting geometry $\Sigma_0$ to the sphere, and is hence independent of the moduli of $\Sigma$. In cases with more boundaries the analogous factors can depend on the internal moduli, though they are again independent of any horizon lengths which become small.

Thus in the puncture limit we can conformally map each hole to a round hole of radius $r_a$. Taking the state-operator map to be  that given by the path integral over the unit-radius disk in the plane, this implies that the operator $V_a$ is defined by the path integral over an annulus between the hole of radius $r_a$ and the unit circle.  This yields $V_a = e^{- \frac{1}{2} \hat \beta_a H_a}$ as claimed in \eqref{puncture}, with
\begin{equation}
\label{hb}
\hat \beta_a = \beta_a - 2 \ln r_d.
\end{equation}

Each $V_a$ in principle also includes a convention-dependent rotation associated with how one fixes the location of $\phi=0$ in each asymptotic region.  However, this rotation can be set to zero by noting that $\Sigma$ in figure \ref{pants} has a $\mathbb Z_2$ isometry that acts as reflection on each asymptotic boundary.  In the sphere conformal frame and with the punctures chosen to lie on the equator, this symmetry exchanges the northern and southern hemispheres.  In either frame the symmetry may be used to define a coordinate $\phi$ near each boundary/puncture such that it locally acts as $\phi \rightarrow - \phi$. In particular, the state-operator map at $c_1$ is related to that used at $c_2,c_3$ by a rotation of the equator about the poles.   This fixes compatible conventions on $\Sigma$ and the sphere that sets this extra rotation to zero.

\subsubsection{Relation to OPE coefficients}

We have shown above that to leading order in small $L_a$ our state takes the form
\be \label{gen3state}
|\Sigma \rangle = \sum_{ijk}  \langle O_i(c_1) O_j(c_2) O_k(c_3) \rangle V_1 V_2 V_3  |i\rangle_1 |j\rangle_2 |k\rangle_3,
\ee
where $\langle O_i(c_1) O_j(c_2) O_k(c_3) \rangle$ denotes the three-point function on the sphere with the $c_a$ equally spaced around the equator and the $V_a$ are given by \eqref{puncture} and \eqref{hb}.  Many readers will find it useful to rewrite \eqref{gen3state} in terms of the standard three-point function on the plane.  Since the operators are not generally scalars, the full expression on the plane takes the form
\begin{align}
\label{3pt}
\langle O_i(z_1) O_j(z_2) O_k(z_3) \rangle_{\mathbb{R}^2} = \hspace{12cm}  \cr  \frac{C_{ijk}}{(z_1-z_2)^{\Delta_{ij}/2} (z_1  -z_3)^{\Delta_{ik}/2} (z_2  -z_3)^{\Delta_{jk}/2} (\bar z_1- \bar z_2)^{\tilde  \Delta_{ij}/2} (\bar z_1  - \bar z_3)^{\tilde \Delta_{ik}/2} (\bar z_2  -\bar z_3)^{\tilde \Delta_{jk}/2}} ,
\end{align}
where $\Delta_{ij} = \Delta_i + \Delta_j - \Delta_k$ etc, the $\Delta_i, \tilde \Delta_i$ are the left- and right-moving conformal weights of the operators $O_i$, corresponding to $E_i \pm J_i$ in terms of the energies $E$ and the angular momenta $J$ of the states $|i\rangle$ in the CFT on the cylinder of circumference $2\pi$.  The $C_{ijk}$ are the OPE coefficients, and the $z_a$ are the images on the complex plane of the $c_a$ on the sphere.  On the plane, it is natural to take $z_1 = 1$, $z_2 = e^{2\pi i/3}$, $z_3 = e^{-2\pi i/3}$ and to define the state-operator map at each of $z_1, z_2, z_3$ to be related by translations of the plane\footnote{In contrast, the convention we chose on the sphere would induce state-operator maps at $z_1,z_2,z_3$ on the plane that are related by rotations about the origin.}.  This is reflected in the fact that, using the rotation invariance of the vacuum state to set $J_i +J_j + J_k=0$ for any nonzero three-point function,  \eqref{3pt} at our $z_1,z_2,z_3$ becomes
\be
\label{3pt}
\langle O_i(z_1) O_j(z_2) O_k(z_3) \rangle_{\mathbb{R}^2} = C_{ijk} e^{i \frac{2\pi}{3} J_j} e^{-i \frac{2\pi}{3} J_k} e^{-\ln 3 \left( E_i + E_j + E_k\right)}.
\ee
Here we have assumed that $J_a$ has integer eigenvalues corresponding to our focus on antiperiodic boundary conditions for fermions.  The more general situation involves branch cuts.

Using the same conventions on $\Sigma$ as above, with the coordinate $\phi$ being fixed by taking the reflection to be $\phi \rightarrow - \phi$, the factor of  $e^{i \frac{2\pi}{3} J_j - i \frac{2\pi}{3} J_k}$ in \eqref{3pt} precisely cancels the rotations in the $V_a$.  Thus, we can write the three-boundary state as
\be \label{gen3stateplane}
|\Sigma \rangle = \sum_{ijk}  C_{ijk} e^{-\frac{1}{2}\tilde \beta_1 H_1}e^{-\frac{1}{2}\tilde \beta_2 H_2}e^{-\frac{1}{2}\tilde \beta_3 H_3}    |i\rangle_1 |j\rangle_2 |k\rangle_3,
\ee
where
\be
\tilde \beta_a = \beta_a - 2 \ln r_d  - 2 \ln 3.
\ee
The three-boundary state in the puncture limit is thus determined by the CFT data -- the OPE coefficients $C_{ijk}$ -- together with universal factors which carry all moduli dependence. The latter give a Boltzmann suppression of a similar character to that in the thermofield double state.

\subsubsection{Beyond the puncture limit}

It is clear that similar asymptotic forms are obtained in other limits where some subset of the $L_a$'s vanish with the others held fixed.  We will also see later in section \ref{factorization} that analogous behavior occurs in the limit $L_3 \rightarrow \infty$ with $L_1, L_2 \rightarrow 0$.  But we may also relate the three-boundary state to the three-point function for more general $L_a$.

Even when the holes $h_a$ are not round, we can write the state as in \eqref{gen3state} with operators $V_a$ defined by a path integral over an annulus cut out of the complex plane whenever the $h_a$ are small enough that the required disks to not overlap on the sphere\footnote{We expect that this construction may be analytically continued as functions of the $L_a$ to cases where these disks would in fact overlap.  In this sense it can be used for general $L_a$.}. The outer boundary is a circle of unit radius, and the inner boundary is defined by conformally mapping $h_a$ to the plane.  We may now use the result that any annulus is conformal to an annulus with two round boundaries (see e.g. \cite{seppala11}) to write the operator $V_a$ in terms of some $\exp{-\frac{1}{2} \hat \beta H_a}$.  However, we must also take into account how this conformal transformation acts on the states that we attach to this operator.  States are of course functionals of field configurations on the boundary; i.e., functionals of field configurations $\Phi(\phi)$ written in terms of some coordinate $\phi$ on the $S^1$.  Changing the choice of this $\phi$ enacts a diffeomorphism of the circle, which is the diagonal subgroup of the conformal group on the cylinder that acts simultaneously on both left- and right-movers.  Since mapping our original annulus to, say, a round cylinder of circumference $2\pi$ will generally involve a Weyl factor $\Omega$ that varies along both boundaries, and since states are naturally attached to both the cylinder path integral and the path integral on $\Sigma$ using coordinates $\phi$ associated with proper distance along each boundary, we see that we must generally allow for the action of such a (real) $S^1$-diffeomorphism at each boundary.  In other words, for general $L_a$ any operator $V_a$ will take the form
\begin{equation}
\label{Va}
V_a = U(h_a) e^{-\frac{1}{2} \hat \beta H_a} \tilde U(h_a),
\end{equation}
where $U,\tilde U$ may be written
\begin{equation}
\label{UI}
U = \exp \left( i \sum_{n \ge 0}  a_{I,n} \left(L_n + L_{-n} - \bar L_n - \bar L_{-n} \right) \right),  \ \ \
\tilde U = \exp \left( i \sum_{n \ge 0}  \tilde a_{I,n} \left(L_n + L_{-n} - \bar L_n - \bar L_{-n} \right) \right)
\end{equation}
with $a_n, \tilde a_n$ real and $L_n, \bar L_n$ representing the standard Virasoro generators.  The notation $a_n, \tilde a_n$ should not be taken to imply any relation between these coefficients.  The form of a given $V_a$ is universal in the sense that the coefficients $a_n, \tilde a_n$, $\hat \beta$ are fully determined by the size and shape of the $a$th hole $h_a$, and thus by the moduli of $\Sigma$, and do not depend on the particular choice of CFT. Thus for general $L_a$ we may still write the state as in \eqref{gen3state} with the $V_a$ given by \eqref{Va}. Due to the $U, \tilde U$ factors
the $V_a$ are not generally diagonal in the energy basis.  But as above we find $U(h_a), \tilde U(h_a) \rightarrow \mathds{1}$ and $\hat \beta_a \approx \beta _a + \beta^0_a$ in any limit where $h_a$ shrinks to a point.  Here the $\beta^0_a$ are determined by those moduli which are held constant.

%
\subsection{More boundaries and factorization limits}
\label{factorization}
%

The above discussion of the three-boundary state can easily be extended to general $n > 3$.  The main novelty is that the surface $\Sigma$ then has internal moduli, which correspond to conformally invariant cross ratios in the CFT correlators. In the puncture limit $L_a \rightarrow 0$, holding the internal moduli $\tau_\alpha$ fixed, the treatment of the vanishing $L_a$ at each puncture proceeds as above and we find
\be
\label{4state}
|\Sigma \rangle = \sum_{i_1 \ldots i_n} C_{i_1 \ldots i_n}(\tau_\alpha) e^{-\frac{1}{2}\sum_a \beta_a H_{a}} |i_1 \rangle_1 \ldots  |i_n \rangle_n,
\ee
where the coefficients $C_{i_1 \ldots i_n}$ are related to the CFT $n$-point functions. The dependence of $|\Sigma \rangle$ on the 
$\tau_\alpha$
includes contributions from the dependence of the CFT $n$-point functions on the cross ratios, factors from the possible rotation in $V_a$, and the $\beta_a^0$ term in \eqref{puncture}.  We have absorbed both of these latter factors into the definition of $C_{i_1 \ldots i_n}$ since they become independent of $L_a$ in the puncture limit.

The most interesting aspect of the dependence on internal moduli is that the state exhibits factorization limits, which we now discuss in some detail. For simplicity, we restrict to the four boundary case and corresponding factorization limits of the four-point function. The surface $\Sigma$ is depicted in figure \ref{four} and the internal moduli can be taken to be $L_{14}$ and $\theta_{14}$.

In the limit $L_{14} \to 0$ for fixed $L_a$ (which need not be small), the surface $\Sigma$ splits into two spheres, each with two holes, that remain connected by a thin tube. This corresponds to an $s$-channel limit of the corresponding four-point function. For a CFT with a unique ground state, at leading order in this limit only the vacuum propagates in this channel and the four-point function is dominated by the contribution from the disconnected two-point functions.  The state then factorizes:
\be
\label{f1234}
|\Sigma \rangle \approx  |\Sigma_{14}\rangle|\Sigma_{23}\rangle,
\ee
where each factor is given by a path integral over an appropriate two-boundary manifold $\Sigma_{14}$ or $\Sigma_{23}$.  These two-boundary manifolds are constructed via a two-step procedure:  First cut the original 4-boundary $\Sigma$ along $L_{14} = L_{23}$ to obtain two disconnected pieces with three boundaries each.  Now cap off the $L_{14}$, $L_{23}$ boundary of each piece by sewing in a disk; i.e., by attaching the vacuum state at this boundary.

Each of the resulting surfaces $\Sigma_{14}$, $\Sigma_{23}$ is then topologically an annulus.  So as above we may write the state $|\Sigma_{14}\rangle$ in the form
\be
\label{alphastate}
|\Sigma_{14}\rangle =  U_1 U_4 \left( \sum_i e^{-\frac{1}{2}\hat \beta_{14} E_i} |i \rangle_1 |i \rangle_4 \right)
\ee
for some $\hat \beta_{14}$ and some unitary operators $U_1, U_4$ of the form \eqref{UI} acting respectively on boundaries $1$ and $4$.  These operators and the parameter $\hat \beta_{14}$ in \eqref{alphastate} are determined by the two remaining moduli $L_1, L_4$.  There is of course a corresponding expression for $|\Sigma_{23}\rangle$.

As in section \ref{3ptf}, taking the puncture limit\footnote{Here we imagine taking $L_1,L_4 \rightarrow 0$ but maintaining $L_1,L_4 \gg L_{14}$.} causes the length of each arm to diverge and the Weyl factor $\Omega$ governing the conformal transformation to become rotationally invariant near each boundary.  As a result, the operators $U_1, U_4$ become trivial.  Thus, in the puncture limit the state $|\Sigma_{14} \rangle$ reduces to a thermofield double state as one might expect.
However, the temperature $\hat \beta_{14}$ of this state is more surprising. We may also follow section \ref{3ptf} in deducing the asymptotic form of $\hat \beta_{14}$.  In the arm that includes each boundary,  we may again use the BTZ approximation \eqref{btzgeom} for $\rho < \rho_0$ to conclude that in mapping $\Sigma_{14}$ to a sphere with two round holes, the radii of the holes will be $r_{1} = r_0 e^{-\frac{1}{2}\beta_{1}}$, $r_{4} = r_0 e^{-\frac{1}{2}\beta_{4}}$.  We therefore find
\be
\label{btzbeta}
\hat \beta_{14} = \beta_1 + \beta_4 + \beta_0.
\ee
We will argue in section \ref{reduced} that in fact $\beta_0 =0$ and further that $\hat \beta_{14} = \beta_1 + \beta_4$ and even far from the puncture limit (at least when $\beta_1 = \beta_4$).  With appropriate conventions (again at least when $\beta_1 = \beta_4$) it will turn out that $U_1 = U_4 = \mathds{1}$ as well.

These moduli  $\beta_1, \beta_4$ are specifying the geometry of $\Sigma_{14}$; it has two exterior regions, which are $t=0$ surfaces for BTZ geometries at the inverse temperatures $\beta_1, \beta_4$. In the puncture limit, we find that the CFT path integral on $\Sigma_{14}$ defines a thermofield double state at the inverse temperature $\hat{\beta}_{14}$. Taking $\beta_0= 0$, this gives
\be
\label{BTZT}
T_{14} = T_1T_4/(T_1 + T_4),
\ee
where $T_1,T_4$ are the temperatures of the black holes in regions $1$ and $4$ of the bulk 4-boundary connected wormhole specified by our moduli.
This relation between the temperatures is somewhat surprising. One might have guessed that at least for $\beta_1 = \beta_4$ the temperature of the thermofield double would agree with those in the connected 4-boundary wormhole.  But this is not what we find.

It is interesting to compare this CFT analysis of factorization to the discussion of bulk phase transitions in section \ref{phases}.  In the bulk we expect a phase transition at some finite value of $L_{14}$: as we decrease $L_{14}$ at fixed (sufficiently large) values of the other parameters, we expect to pass from the regime dominated by the Lorentz-signature four-boundary wormhole to one dominated by two disconnected two-boundary wormholes.\footnote{Smaller values of $L_a$ may be dominated by four copies of global thermal AdS which pair up to form two thermofield double states, or perhaps by a single thermofield-double pair of global AdS${}_3$ spaces and a single BTZ spacetime.  The details depend on the choice of spin structures as in appendix \ref{spins}.} This bulk phase transition implies an approximate factorization of the state; the entropy of the reduced density matrix $\rho_{14}$ on $CFT_1 \otimes CFT_4$ changes from being proportional to the central charge (on the side represented by connection in the bulk) to $\mathcal O(1)$ (on the side represented by a disconnected bulk). But since the two copies of BTZ are still connected through the Euclidean geometry, this factorization is only approximate. There is still some CFT entanglement, which is reflected in quantum entanglement between bulk fields on the disconnected Lorentzian geometries. It is only in the limit of vanishing $L_{14}$ that the Euclidean separation between the two copies of BTZ diverges, and the state factorizes as seen above.

From this bulk perspective, we can interpret \eqref{BTZT} as determining the temperature of the bulk BTZ geometry which provides the saddle-point for the bulk path integral. It again feels somewhat surprising that the resulting BTZ temperature is related to the original moduli in this nontrivial way.
It would be very interesting to derive the temperature \eqref{BTZT} directly from the bulk path integral, perhaps using the technology of e.g. \cite{Faulkner:2013yia,Hartman:2013mia}.\footnote{See \cite{Louko:1998dj} for discussion of a related case in which the naive period of imaginary time also disagreed with the physical temperature of the Lorentz-signature black hole.}
Applying the same reasoning to the 3-boundary phase transition of section \ref{phases} from the connected 3-boundary wormhole to a BTZ-plus-global-AdS${}_3$ phase shows that the BTZ temperature is again related to the black hole temperatures of the connected wormhole by the analogue of \eqref{BTZT}.

In addition to the above $s$-channel limit, there are corresponding $t$- and $u$-channel limits. The $t$-channel limit $L_{12} \to 0$ corresponds to $L_{14} \to \infty$ with $\theta_{14}=0$. In this limit the state will factorize as $|\Sigma \rangle = |\Sigma_{12}\rangle |\Sigma_{34} \rangle$. The $u$-channel limit $L_{13} \to 0$ corresponds to $L_{14} \to \infty$ with $\theta_{14} = \pi$. in this limit the state will factorize as $|\Sigma \rangle = |\Sigma_{13}\rangle |\Sigma_{24} \rangle$. In each case, the factorized states take the same form as in \eqref{alphastate}.

%
\subsection{Reduced density matrices and factorization}
%
\label{reduced}

To further explore the structure of our CFT states it is useful to consider the reduced density matrices obtained by taking partial traces over subsets of the degrees of freedom. In our multiboundary context, a simple class of reduced density matrices is obtained by tracing over one or more of the boundaries. These reduced density matrices also have a simple interpretation in terms of the path integral, and again exhibit interesting factorization properties.

For example, starting with the three-boundary state \eqref{3state} and tracing over CFT${}_3$ leads to the density matrix
\be \label{rho3}
\rho_{12} = \sum_{iji'j'} \rho_{ij i'j'} |i \rangle_1 |j\rangle_2 \langle i' |_1 \langle j' |_2
\ee
with $\rho_{ij i'j'} = \sum_k A_{ijk} A^*_{i'j' k}$. As $A_{ijk}$ is given by a CFT path integral over the pair of pants, this density matrix is a corresponding path integral over the 4-boundary space defined by sewing together two identical pairs of pants along the pair of boundaries corresponding to CFT${}_3$. Note that we do not integrate over the parameter $L_3$ associated with that boundary; this becomes a modulus for the resulting Riemann surface. The density matrix $\rho_{ij i'j'}$ thus corresponds to a path integral over the sphere with four holes and so is related to a CFT four-point function. In contrast to the previous discussion of the four-point function this reduced density matrix still only depends on the three parameters labeling our original pair of pants geometry.  Thus the reduced density matrix involves only a three-dimensional slice in the six-dimensional moduli space of Riemann surfaces with four boundaries.

Nevertheless, we can again take an $s$-channel limit; this corresponds to $L_3 \to 0$, so that the two pairs of pants factorize along the identification.  Thus the density matrix \eqref{rho3} factorizes into the product of a bra-vector and a ket-vector and describes a pure state. From the point of view of the original state \eqref{3state}, this is because taking $L_3 \to 0$ sends the coefficients $A_{ijk}$ to zero except when $|k \rangle_3$ is the vacuum state ($k = \mathds{1}$). The nonzero coefficients $A_{ij\mathds{1}}$ are then determined by the path integral over a manifold conformal to an annulus as in section \ref{factorization}.  Indeed, \eqref{rho3} becomes
\be
\label{factorpure}
\rho_{12} \approx  |\Sigma_{12} \rangle\langle \Sigma_{12} |
\ee
with $|\Sigma_{12} \rangle$ given by \eqref{alphastate} (replacing $4$ by $2$). The corresponding puncture limit is described by \eqref{BTZT} and $U_1,U_2 \rightarrow \mathds{1}$ so that we then obtain a thermofield double state entangling boundaries $1$ and $2$.   What is happening here is that the limit $L_3 \to 0$ forces the entropy of the accessible degrees of freedom in the third Hilbert space to be small. The other two spaces are then mostly entangled with each other and very little with the third.

We can again compare with the dual bulk description, where we expect that reducing $L_3$ for fixed $L_1, L_2$ should lead to a phase transition at some finite value of $L_3$ where we pass from the three-boundary black hole to a two-boundary wormhole and a copy of global AdS.

The other interesting limit for the density matrix is $L_3 \to \infty$.  Since the twist is fixed to $\theta =0$, this corresponds to the $t$-channel limit for the four-point function.   This again gives a degeneration limit of the sphere with four holes, but where the boundaries are paired differently than in \eqref{factorpure}. Thus in this limit the density matrix $\rho_{12}$ factorizes into a density matrix for boundary 1 and a density matrix for boundary 2 with little correlation:\footnote{This particular factorization limit is special in that the above approximations are equally valid for all choices of spin structure. Since the shrinking geodesic separates two copies of boundary $1$ from two copies of boundary $2$, the sewing construction forces fermions on this cycle to have antiperiodic boundary conditions.  This may be seen for example from the fact that the 4-boundary path integral would vanish if periodic boundary conditions were assigned.  See appendix \ref{spins} for further discussion.}
\be \label{rho12}
\rho_{12} \approx  \rho_1 \otimes \rho_2.
\ee
Reasoning as before, the density matrix $\rho_1$ will be given by a bulk path integral over a two-boundary Riemann surface whose geometry depends just on $L_1$, which gives for general $L_1$ a result of the form
\be
\label{rho1}
\rho_1 = U_1 e^{-\hat \beta_1 (L_1)H_1} U_1^{-1},
\ee
where $U_1$ has the form \eqref{UI}.   There will be a similar result for $\rho_2$.  The factorization of $\rho_{12}$ implies that the 3-boundary state for large $L_3$ takes the approximate form
\be
\label{L3large}
|\Sigma \rangle \approx U_1 U_2 e^{-\frac{1}{2} \hat \beta_1 H_1} e^{- \frac{1}{2}\hat \beta_2 H_2}  \sum_{ij} |i\rangle_1 |i\rangle_{31} |j\rangle_2 |j\rangle_{32} \otimes |\gamma \rangle_{33},
\ee
where as usual $U_1,U_2 \rightarrow \mathds{1}$ as $L_1,L_2 \rightarrow 0$.  In writing \eqref{L3large} we have factored the CFT${}_3$ Hilbert space ${\cal H}_3$ into a product of three Hilbert spaces ${\cal H}_3 \approx {\cal H}_{31} \otimes {\cal H}_{32} \otimes {\cal H}_{33}$ and introduced bases $\{|i\rangle_{31}\}$, $\{|j\rangle_{32}\}$ for ${\cal H}_{31}, {\cal H}_{32}$ (which are not energy eigenstates) as well as a state $|\gamma\rangle_{33} \in {\cal H}_{33}$. By leaving arbitrary the precise choices of these factor spaces and bases (as well as the choice of $|\gamma \rangle_{33}$) we have been able to remove the operator $U_3$ that one might have expected from \eqref{gen3state} and \eqref{Va}.

The above $t$-channel factorization property is again easy to understand from elementary considerations in the overall Hilbert space. In this limit the space of accessible degrees of freedom in the third Hilbert space is much larger than for the other two, so in direct analogy with the classic results of Page \cite{Page:1993df}, each of the remaining Hilbert spaces is strongly entangled with the large space.  This leaves no room for significant entanglement between the small Hilbert spaces and the reduced density matrix $\rho_{12}$ approximately factorizes into $\rho_1 \otimes \rho_2$.

Unlike the previous factorization limits, this limit is not associated with a phase transition in the bulk geometry; the connected 3-boundary wormhole remains the dominant bulk saddle even though the CFT state takes the approximate product form \eqref{L3large}, where there is little entanglement between boundaries $1$ and $2$. We will see in section \ref{entang} that it is instead associated with a change in minimal surface in the calculation of the entanglement entropies from the bulk.

Thus, tripartite entanglement is not a necessary condition to connect three asymptotic regions in the bulk. Indeed, any notion of intrinsically tripartite entanglement will by definition vanish for  \eqref{L3large} and we may say that the entanglement in this factorization limit is entirely bipartite.  This expressly rules out any conjecture that GHZ-like states always play an essential role in the CFT description of connected multiboundary wormholes.  Note that the approximation in \eqref{L3large} relies on large $L_3$ but not large $c$.  In particular, even corrections at order $c^0$  are exponentially suppressed at large $L_3$.

We could now consider our usual analysis of the puncture limit to determine $\hat \beta_1$ for small $L_1$. But in the present context we can do much better by recalling that the bulk solution is the connected 3-boundary wormhole described in the conformal frame that makes the boundary stress tensor rotationally symmetric -- and thus time-independent -- on each boundary, at leading order in the central charge $c$.  However, other than a pure rotation, any non-trivial conformal transformation $U_1$ in \eqref{rho1} would add boundary gravitons, deforming the boundary stress tensor calculated in \eqref{rho1} away from spherical symmetry and adding non-trivial time-dependence.  Since pure rotations commute with the Hamiltonian, we may therefore take $U_1 =\mathds{1}$ in \eqref{rho1}. By appropriate choices of phase for the states $|i\rangle_{31}$,  $|j\rangle_{32}$ we may  take $U_1 =\mathds{1}$,  $U_2 = \mathds{1}$ in \eqref{L3large} as well.   Noting that the expected stress tensor must match that of the bulk black hole also fixes $\hat \beta_1 = \beta_1$, and similarly $\hat \beta_2 = \beta_2$. These results hold for all $L_1,L_2$ and not just in the limits $L_1,L_2 \rightarrow 0$. Furthermore, since $U_1,U_2,\hat \beta_1, \hat \beta_2$ are determined entirely by conformal transformations independent of the central charge, these results are exact; they hold for any finite central charge as well as in the limit $c \rightarrow \infty$. As a result, while we began only with the assumption that stress tensor expectation values are thermal at leading order in large $c$, thermality of the the full reduced density matrix at this large $L_3$ limit is an exact result at all finite $c$.

This argument may also be used to determine the Boltzmann factor $\hat \beta_{14}$ and conformal transformations $U_1,U_4$ in the two-boundary states $|\Sigma_{14} \rangle$, $|\Sigma_{23} \rangle$ from \eqref{f1234}\eqref{alphastate} that describe the approximate factorization of the 4-boundary state $|\Sigma_{1234} \rangle$ at small $L_{14}$..  The key observation is that path integral computing the density matrix
\begin{equation}
\rho_{12} = \tr_{{\cal H}_3} \left[ |\Sigma_{123} \rangle \langle \Sigma_{123} | \right]
\end{equation}
defined by tracing the 3-boundary state $|\Sigma_{123} \rangle$ over  ${\cal H}_3$ coincides with that defining $|\Sigma_{1234} \rangle$.  Here we make appropriate restrictions on the moduli of $\Sigma_{1234}$ and reinterpret bra vectors on two of the four boundaries as kets. Explicitly, we may identify 1 and 4 in $|\Sigma_{1234} \rangle$ with the bra and ket copies of 1 in $\rho_{12}$, and 2 and 3 in $|\Sigma_{1234} \rangle$ with the bra and ket copies of 2 in $\rho_{12}$. Then the limit $L_3 \to \infty$ corresponds to the $s$-channel limit of section  \ref{factorization}, and the path integral which gives $\rho_1$ in \eqref{rho1} coincides with the one for \eqref{alphastate} when $\beta_1 = \beta_4$.  Comparing \eqref{rho1} and \eqref{alphastate} then requires
\be
\label{hb14}
\hat \beta_{14} = 2 \beta_1 = \beta_ 1 + \beta_4
\ee
for any $\beta_1 = \beta_4$ and forces $U_1,U_4$ to be pure rotations. Thus $U_1 = \mathds{1} = U_4$ for appropriate choices  of the origin $\phi=0$ of the angular coordinate on each boundary.  In particular, this must occur when we choose $\phi=0$ on boundaries $1$ and $4$ to represent points related by the $\mathbb{Z}_2$ reflection symmetry of $\Sigma_{14}$ that exchanges boundaries $1$ and $4$ while mapping the $L_{14}$ boundary to itself with two fixed points. It is tempting to speculate that \eqref{hb14} might hold even when $\beta_1, \beta_4$ are distinct.

For equal $\beta_1, \beta_4$ the relation \eqref{hb14} must in particular hold in the puncture limit where it must agree with \eqref{btzbeta}.  This implies that $\beta_0$ vanishes in \eqref{btzbeta}, since $\beta_0$ is independent of $\beta_1, \beta_4$ in this limit.  It would clearly be of interest to check this result directly by finding the desired conformal transformation.  That a consistent value of $\beta_0$ can be found at all constitutes a useful check on our arguments, and in particular verifies the absence of a factor of $1/2$ in \eqref{btzbeta}, so that $\hat \beta_{14}$ is the sum of $\beta_1$ and $\beta_4$ rather than the average.

Returning to the discussion of tracing over boundaries, we note that reduced density matrices obtained by tracing over more than one boundary are related to CFT path integrals on surfaces of higher genus.\footnote{We could also have encountered path integrals over higher genus surfaces simply by considering black holes with topology behind the horizon.  We have avoided this further complication for simplicity of the discussion, though the topic will be addressed in \cite{MRta}.} For example, taking the three-boundary state \eqref{3state} and tracing over two of the boundaries leads to the density matrix $\rho = \sum_{i i'} \rho_{i i'} |i \rangle \langle i'|$ with
\be
\rho_{i i'} = \sum_{jk} A_{ijk} A^*_{i' jk},
\ee
which corresponds to a path integral on the torus with two holes. This density matrix is a function of three moduli, so it corresponds to a three-dimensional subspace of the associated moduli space. The absence of Dehn twists implies that this is a slice of the moduli space where the torus is rectangular, so that the A and B cycles are orthogonal. In the puncture limit, this density matrix is related to the CFT two-point function on the torus.

%
\section{Multipartite holographic entanglement}
\label{entang}
%

We saw in section \ref{cft} that the $n$-boundary state factorizes in various limits. Some limits are associated with phase transitions in the bulk geometry, so that factorization occurs in regions of moduli space where the dominant bulk saddle-point describes a disconnected Lorentz-signature spacetime. Notably, however, the $L_3 \rightarrow \infty$ limit of section \ref{reduced} leads to
a CFT state of the approximate product form \eqref{L3large}, with correspondingly negligible entanglement between CFT${}_1$ and CFT${}_2$,  in a region where the connected 3-boundary wormhole remains the dominant bulk saddle.

This section will further probe the entanglement structure of the $n$-boundary states by using the Hubeny-Rangamani-Takayanagi (HRT) generalization \cite{Hubeny:2007xt}  of the Ryu-Takayanagi proposal \cite{Ryu:2006bv} to compute entropies holographically.  We confine ourselves to regions of moduli space for which the dominant Lorentz-signature bulk phase is the connected $n$-boundary black hole, so these results are largely complementary to the preceding discussion of factorization limits.  In some cases this may require imposing periodic boundary conditions on fermions as discussed in appendix \ref{spins}.  Due to our emphasis on bulk methods below, this section will describe entropy as being associated with a given boundary $B_a$ rather than a given CFT copy CFT${}_a$ as in section \ref{cft}.

We will see that even when the geometry remains connected, there are phase transitions corresponding to changes in the HRT extremal surface.  Such transitions indicate changes in the  leading-order entanglement structure at large central charge.  In particular, in the three-boundary case we find a phase transition associated with the large $L_3$ factorization limit, where the ${\cal O}(c)$ component of the mutual information between boundaries $1$ and $2$ vanishes for $L_3 \ge L_1 + L_2$.  The relation of the phase transition to the factorization limit in the CFT is similar to that for the changes of bulk geometry: the phase transition occurs at finite $L_3$, long before we get to the factorization limit described in section \ref{reduced}. Thus we learn that essentially bipartite entanglement (at leading order in the central charge) suffices to generate a 3-boundary connected wormhole over a sizeable part of the moduli space.  Conversely, we will also find points in moduli space where the entanglement is fully tripartite.  Intriguingly, at the level we are able to probe in this work we find that our entanglement results always coincide with properties of appropriate random states defined by sets of constraints associated with minimal geodesics.

For much of section \ref{entang}, we will be somewhat cavalier in calling correlation as diagnosed by the mutual information and its tripartite analog ``entanglement''. However, in quantum information theory, entanglement refers very specifically to correlation that cannot be generated without quantum mechanical interaction~\cite{werner1989quantum,bennett1996mixed}. While there is reason to believe that these measures do diagnose proper entanglement in holographic theories~\cite{hayden2013holographic}, no proof exists. Therefore, to solidify our claims, we repeat some of our analysis in section \ref{sec:entang-measures} quantifying entanglement more rigorously. Doing so leads to similar conclusions.

We begin by reviewing the holographic calculation of entanglement entropy in section \ref{hrt}.   Sections~\ref{p3} and \ref{depend} then study phase transitions between different minimal surfaces as a function of the bulk parameters and show that the bipartite vs. multipartite nature of the dual state depends strongly (and perhaps surprisingly) on the moduli of $\Sigma$.   We also use these calculations to fix the parameters $\hat \beta_1$, $\hat \beta_2$ in \eqref{L3large}, as well as $\beta_0$ in \eqref{btzbeta}. Section~\ref{sec:entang-measures} addresses correlation vs. entanglement, and section \ref{sec:more-parties} discusses $k$-party entanglement in the $n$-boundary state for larger $k, n$ and shows that the {\it distillable} entanglement between the asymptotic regions is at least $(n+1)/2$-partite.
Section~\ref{intrinsic} develops new tools to characterize $n$-partite entanglement and uses these tools to show that for wormholes with $n$ similar sized horizons there is $n$-partite entanglement for even $n$ and $n-1$-partite entanglement for odd $n$.  Section~\ref{random} then notes that, so far as the above probes can tell, the nature of the entanglement of the state is consistent with what one would expect for suitably generic states.  We also discuss extensions of this random-state model beyond the regime investigated thus far.

\subsection{Holographic calculation of entanglement entropy}
\label{hrt}

For time-independent states invariant under time-reversal, the entropy of spatial subregions may be studied holographically using the Ryu-Takayanagi proposal \cite{Ryu:2006bv}.  This prescription states that, given a spatial subregion $A$ in a constant-time surface of the CFT, the associated reduced density matrix $\rho_A$ obtained by tracing over degrees of freedom in the complement of $A$ has an entropy $S_A$ given by the area (i.e., the co-dimension two notion of volume, so in fact a length on AdS${}_3$) of a minimal surface $\gamma_A$ in the associated bulk constant-time surface whose boundary coincides with that of $A$ ($\partial \gamma_A = \partial A$) and for which $\gamma_A$ is homologous to $A$.  The symmetries imply that the bulk is static, so there is a unique bulk surface orthogonal to the time-translation which contains $A$.  This proposal has been extensively tested and explored in the literature, and a strong connection to Euclidean quantum gravity arguments was made in \cite{Lewkowycz:2013nqa}.

This construction must be generalized for time-dependent states such as our \eqref{nstate}.  The HRT prescription \cite{Hubeny:2007xt} replaces the minimal surface $\gamma_A$ in the spatial slice with an extremal surface $\mathcal E_A$ in the bulk spacetime which is homologous to $A$.  As evidence in its favor, this generalization has been shown to satisfy strong subadditivity \cite{Wall:2012uf}.

In addition, while it may be difficult to extend the argument of \cite{Lewkowycz:2013nqa} to the case of fully general time-dependence, the extension to subregions of the time-reflection invariant $t=0$ surface in states like the ones we consider
appears straightforward.  Since each state \eqref{nstate} is defined by a Euclidean path integral, replica partition functions may be defined in the usual way for any subregion $A$ of the $t=0$ slice.  Following the rest of the argument of \cite{Lewkowycz:2013nqa}, and in particular using the assumption that replica symmetry is unbroken, then leads one to consider conical singularities in the bulk $t=0$ surface.  Since our system is invariant under time-reversal, the Euclidean bulk $t=0$ surface also appears in the bulk Lorentzian geometry.  Since this surface is positive definite, any homology class contains a minimal surface.  And since time-reversal invariance sets the extrinsic curvature of our $t=0$ surface to zero, this minimal surface is an HRT extremal surface.  So at least for regions $A$ lying at $t=0$, we arrive at the HRT proposal so long as there are no further extremal surfaces with smaller area\footnote{While we have no sharp argument to forbid such surfaces, we note that any one such surface must spontaneously break the time-reversal symmetry.  In addition, under the assumptions of \cite{Wall:2012uf} they must be spacelike separated from the $t=0$ extremal surfaces and also, in an appropriate sense, farther from the boundary.  It seems unlikely that such surfaces exist in our geometries. We will henceforth assume that they do not.}.  In principle one should be able to move $A$ away from $t=0$ by an appropriate analytic continuation of the Euclidean results, though we will have no need to pursue this complication below; studying the behavior of sub-region entropy under Lorentzian evolution is an interesting problem for the future.

We shall restrict attention here to the case where $A$ consists of one or more of the boundaries in their entirety.  Unitarity then implies that the entanglement entropy is independent of time
and the above argument applies directly.  Thus we compute the entropy $S(B_1 \cup  \ldots  \cup B_k)$ of the reduced density matrix on some collection of asymptotic boundaries.  For brevity we write this entropy as simply $S( B_1 \ldots B_k)$.  Since $A$ is a closed manifold, the corresponding bulk minimal surface $\gamma_A$ must be a collection of closed geodesics in $\Sigma$ which together are homologous to $B_1 \cup  \ldots  \cup B_k$.  One natural candidate is the union $H_1 \cup  \ldots  \cup H_k$ of the horizons associated with the boundaries over which we do not trace, though there will also  be others involving internal geodesics and other horizons. One alternative that always exists is the union $H_{k+1} \cup \ldots \cup H_n$ of the horizons in the regions over which we trace. Since the overall state \eqref{nstate} is pure, we always have $S(B_1 \ldots B_k) = S(B_{k+1} \ldots B_n)$, and $S(B_1 \ldots B_n) = 0$.

From this basic data we can construct quantities such as the mutual information,
\begin{equation}
I(A : B) = S(A) + S(B) - S(A B)
\end{equation}
the triple information,
\begin{eqnarray}
I_3(A : B : C)
&=& I(A : B) + I(A : C) - I(A : B   C) \label{eqn:I3}
\\ &=& S(A) + S(B) + S(C) - S(AB) - S(BC) - S(AC) + S(ABC) \nonumber
\end{eqnarray}
and higher-order generalizations. The mutual information is nonnegative and equal to zero if and only if the state factorizes. $I_3$, on the other hand, can take either sign in general, but is always nonnegative in holographic theories~\cite{hayden2013holographic}. As such, its magnitude measures how much additional information there is in $BC$ about $A$ that is not present in $B$ and $C$ separately.

Note that in the subsequent discussion we will find that in some regimes of parameters, various mutual informations vanish. What this really means is that at the leading order in central charge probed by the HRT formula the mutual information is vanishing; we do not expect the value to be precisely zero, but rather some value of order one, which is not visible in the holographic analysis.

\subsection{Phases of entanglement for three boundaries}
\label{p3}

As we vary the parameters of $\Sigma$, there can be transitions where the relative area of two candidate minimal surfaces changes sign, so that control over the HRT entanglement passes from one to the other. These phase transitions reflect changes in the entanglement structure of the dual CFT state at leading order in the central charge.

Consider the three-boundary case, with $L_i$ parametrizing the size of the horizon of region $i$.     As described above, the HRT minimal surfaces that compute entanglement are unions of these horizons.   In this case, there is a transition at $L_3 = L_1 + L_2$; if $L_1 + L_2 < L_3$, the minimal surface for $B_1 \cup B_2$ and $B_3$ is $H_1 \cup H_2$, so
\be
S(B_1   B_2) = S(B_3) =  \frac{\pi}{2 G} (L_1 + L_2).
\ee
But for $L_1 + L_2 > L_3$, the minimal surface for $B_1 \cup B_2$ and $B_3$ is $H_3$, so
\be
\label{3small}
S(B_1   B_2) = S(B_3) =  \frac{\pi}{2 G} L_3.
\ee
It is natural to associate this phase transition with the factorization limit $L_3 \to \infty$, since for $L_1 + L_2 < L_3$ the holographic mutual information satisfies
\be
I(B_1 : B_2) = S(B_1) + S(B_2) - S(B_1   B_2) = \frac{\pi}{2G} ( L_1 + L_2 - (L_1 + L_2) ) =0.
\ee
Thus there is no entanglement of boundaries 1 and 2 at leading order in central charge whenever they are small subsystems, not just in the strict decoupling limit. Taking $L_1 \leq L_2$, there is also a further phase transition when $L_2 > L_1 + L_3$, where the minimal surface for $B_2$ changes from $H_2$ to $H_1 \cup H_3$. Thus, setting $L_{tot} = L_1 + L_2 + L_3$ and choosing $L_1 \leq L_2$, the mutual information is
\be \label{eqn:mi-piecewise}
I(B_1 : B_2 ) =
	\frac{\pi}{2G} \times \begin{cases}
		0  & \text{if } 2( L_1 + L_2 ) \leq  L_{tot} \\
		2(L_1 + L_2) - L_{tot} & \text{if } 2( L_1 + L_2 ) > L_{tot}
				\text{ and } 2 L_2 \leq L_{tot} \\
		2 L_1 & \text{if } 2 L_2 > L_{tot}.
	\end{cases}
\ee
In particular, the mutual information vanishes (at leading order in large central charge) until the horizons for black holes 1 and 2 are sufficiently large and then rises with slope 2 as a function of $L_2$ until saturating at $2 L_1$.

When $L_3 > L_1 + L_2$, the entanglement seems almost entirely bipartite, between 1 and 3 and between 2 and 3, though we will argue this more carefully in section \ref{sec:entang-measures}. In the extreme limit where $L_3 \to \infty$, we saw in the earlier field theory analysis that the entanglement becomes purely bipartite, and the reduced density matrix on tracing over 3 factorizes. Here we see that there is a sizeable region in the parameter space where bipartite entanglement dominates.  This strengthens the conclusion of section \ref{reduced}: a large multipartite entanglement component is not required to generate a branched womhole connecting multiple asymtptotic regions.

Of course, the relation to the CFT three-point function suggests that the entanglement for finite $L_3$ is not purely bipartite; we expect a subdominant tripartite component. But this can be compared to the situation in different geometric phases: When the bulk saddle is three copies of AdS, the holographic entanglement entropy calculation of course says that the mutual information between any two copies vanishes.  The actual CFT state will have some order one entanglement between the two copies, but this subleading entanglement does not appear to generate a geometric connection. Similarly, here the subdominant contribution presumably plays no important role in generating the connected geometry. Thus it appears to be enough that $B_1$ and $B_2$ have large entanglement with different parts of the $B_3$ Hilbert space for the bulk dual to have a geometric connection.

On the other hand, when $L_3 < L_1 + L_2$, there seems to be more than purely bipartite entanglement. This is an example of a general phenomenon; the extent to which our CFT state involves multipartite entanglement depends on the moduli of $\Sigma$. To discuss the multipartite nature of the entanglement in detail, we will turn below to cases with more boundaries, where this can be more clearly diagnosed. But first we make some further remarks on this three-boundary example.

We note that the leading-order behaviour \eqref{eqn:mi-piecewise} matches the mutual information for typical pure states chosen according to the unitarily invariant measure on $\overline{\mathcal{H}}_1 \otimes \overline{\mathcal{H}}_2 \otimes \overline{\mathcal{H}}_3$, where $\overline{\mathcal{H}}_a$ is a finite-dimensional Hilbert space of dimension $e^{\pi L_a/2G}$. This may be seen by using the results of \cite{Page:1993df} to compute the entropies $S(B_1)$, $S(B_2)$, $S(B_1B_2)$ and assembling these building blocks to find \eqref{eqn:mi-piecewise}. Thus, for the three-boundary case, the nature of the entanglement between subsystems is what we would expect for a ``random'' entangled state. Let us stress that this is not to say that the state \emph{is} generic; the state dual to the three-boundary wormhole is very special. It is just the very coarse overall entanglement structure that appears to be generic. We will discuss this issue further in section \ref{random}.

It is also interesting to note that the fact that $S(B_3) = \frac{\pi}{2 G} (L_1 + L_2)$ implies that knowledge of the reduced density matrix $\rho_3$ is telling us about the geometry beyond the horizon $H_3$ for this asymptotic region.  Indeed, in general the HRT surface corresponding to a region $A$ in the boundary lies outside the causal wedge associated to $A$ in the bulk spacetime \cite{Wall:2012uf}. These black holes provide a particularly striking example of this behaviour. In the BTZ case, the entropy of the reduced density matrix on one side is related to the area of the horizon. Here, it is related either to the area of the horizon $H_3$, or to the sum of the areas of the horizons $H_1$ and $H_2$.\footnote{Note that the minimal surface will always lie somewhere in the shadow region, or perhaps on its boundary.  This places the surface in the part of the space that is causally inaccessible from any of the boundaries and corresponds to the fact that the amount of entanglement cannot be affected by local quantum operations on any of the entangled systems. Acting with local operators on the boundaries can change the geometry in the regions that are causally accessible from the boundary, so it is natural that the minimal surface that calculates the entanglement entropy must lie outside these influenceable regions \cite{Wall:2012uf,MVM}.}

%
\subsection{Phases of entanglement for four boundaries}
%
\label{depend}

We now explore the phases of holographic entanglement in the four boundary case, for which the triple mutual information provides a useful probe of truly multipartite correlations.

Let us first consider the limit of small $L_a$, where we can assume that the length of any internal geodesic is larger than that of any horizon.  Thus, as in the three-boundary case, the only relevant minimal surfaces are those built from the horizons.  We again consider only the connected 4-boundary wormhole in the bulk\footnote{We expect that -- at least by making appropriate choices of spin structure (see appendix \ref{spins}) and tuning moduli -- one can arrange for this bulk phase to dominate the path integrals of section \ref{EPI} somewhere in each of the regimes discussed below.  But this remains to be shown in detail.  If not, our analysis applies to whatever CFT state is in fact dual to these bulk spacetimes.}.

There are then a number of different regions in the parameter space of such four-boundary black holes, distinguished by the behaviour of the mutual informations and triple informations. For example, if $L_1 + L_2 < L_3 + L_4$, the smallest minimal surface for $S(B_1   B_2) = S(B_3   B_4)$ will be $H_1 \cup H_2$, and the mutual information between boundaries 1 and 2 will vanish:
\be
I(B_1 : B_2) = S(B_1) + S(B_2) - S(B_1   B_2) = 0
\ee
as in the three-boundary case. Alternatively, if $L_1 + L_2 > L_3 + L_4$, the  smallest minimal surface for $S(B_1   B_2) = S(B_3   B_4)$ will be $H_3 \cup H_4$, and then  it is the mutual information $I(B_3 : B_4)$ which vanishes. If  $L_1 + L_2 = L_3 + L_4$, both these mutual informations will vanish. There are three different such divisions of the set of boundaries into pairs, so there will be sixteen different regions of the parameter space, labeled by the different mutual informations which are nonzero. When all of the lengths are equal, $L_1 = L_2 = L_3 = L_4 = L$, all of the bipartite mutual informations vanish. This suggests that the entanglement in this case does not involve bipartite entanglement between the pairs of boundaries, but is truly multipartite. Indeed, we find
\be
\label{triple}
I_3(B_2 : B_3 : B_4)
= I(B_2 : B_3) + I(B_2 : B_4) - I(B_2: B_3   B_4)
= -2L \cdot \frac{\pi}{2G}
 < 0,
\ee
which is in magnitude twice the entropy of an individual boundary, so $I_3$ is the most negative it can possibly be. (By strong subadditivity, $I_3(B_a : B_b : B_c) \geq -2 S(B_a)$.)

There is also an interesting division of the parameter space associated with the triple informations. If any one of the lengths is greater than the sum of the others, all of the triple informations vanish. To see this, consider without loss of generality $L_1 > L_2 + L_3 + L_4$. Then the minimal surface for $B_a \cup B_b $ will be  $H_a \cup H_b$ for $a,b = 2,3,4$, and the minimal surface for $B_2 \cup B_3 \cup B_4$ will be $H_2 \cup H_3 \cup H_4$.  Thus all mutual informations between boundaries 2, 3 and 4 vanish: in particular $I(B_2 : B_3) = 0$, $I(B_2 : B_4) = 0$, and $I(B_2 : B_3   B_4) = S(B_2) + S(B_3   B_4) - S(B_2   B_3   B_4) = 0$, so $I_3(B_2:B_3:B_4) = 0$.

Moreover, although the mutual informations involving $B_1$ are nonzero, the triple mutual informations vanish: for example,
\begin{eqnarray}
I(B_1 : B_2 : B_3) &=& S(B_1) + S(B_2) + S(B_3) - S(B_1   B_2) - S(B_1   B_3)
\\ &&- S(B_2   B_3) + S(B_1   B_2   B_3), \nonumber
\end{eqnarray}
but $S(B_1) = S(B_2) + S(B_3) + S(B_4)$, $S(B_1   B_2) = S(B_3) + S(B_4)$, and $S(B_1   B_3) = S(B_2) + S(B_4)$, and $S(B_1   B_2   B_3) = S(B_4)$, while $S(B_2   B_3) = S(B_2) + S(B_3)$, so
\begin{eqnarray}
I(B_1 : B_2 : B_3) &=& S(B_2) + S(B_3) + S(B_4) + S(B_2) + S(B_3) \\ &&- S(B_3) - S(B_4) - S(B_2) - S(B_4)\nonumber \\ && - S(B_2) - S(B_3) + S(B_4) = 0 . \nonumber
\end{eqnarray}
Similarly for $I(B_1 : B_2 : B_4)$ and $I(B_1 : B_3 : B_4)$. Thus, in this extreme part of the parameter space, all the triple mutual informations vanish. There are four such regimes, which are subsets of four of the sixteen regimes identified in the previous analysis of the mutual informations.

Thus, in the parameter space with four boundaries:
\begin{itemize}
\item
There is a special point where all pairwise mutual informations vanish (always only to leading order) but at which the triple information is as negative as it can be. At that point, there is very little correlation between pairs of subsystems but there is correlation between any collection of three subsystems.
\item There are regions where all the triple informations vanish, but that nonetheless have a large pairwise mutual information, which we interpret as indicating that the entanglement is essentially bipartite.
\end{itemize}

As a final example, again set all the lengths $L_a$ to be equal to a fixed $L$ and reintroduce $L_{14}$. As $L_{14} \rightarrow 0$, all other structures in figure 3 are pushed to the boundary, where the conformal factor diverges.  Holding fixed the $L_a$ then forces both $L_{12}$ and $L_{13}$ to become large, ensuring that of the internal minimal surfaces, only $L_{14}$ can ever be small enough to be relevant in any holographic entropy calculation.
One then finds
\beq
S(B_a) &=& \smfrac{\pi}{2G} L \label{eqn:L14-one} \\
S(B_1 B_4) &=& S(B_2 B_3) = \smfrac{\pi}{2G} L_{14} \\
S(B_a B_b ) &=& \smfrac{\pi}{2G} 2L \quad \text{for all } (a,b) \not\in \{(1,4),(2,3)\} \\
S(B_a B_b B_c ) &=& \smfrac{\pi}{2G} L \label{eqn:L14-four}
\eeq
so that
\beq
I(B_1:B_4) &=& I(B_2:B_3) = \smfrac{\pi}{2G}( 2L - L_{14})
	\quad \text{as calculated earlier, and} \\
I(B_a : B_b) &=& 0 \quad \text{for all } (a,b) \not\in \{(1,4),(2,3)\} \\
I_3(B_a : B_b : B_c ) &=& - \smfrac{\pi}{2G} L_{14}.
\eeq
We see that when $L_{14} = 2L$, there is no mutual information between any pair of
boundaries and the triple informations are equal to $-\smfrac{\pi}{2G}2L$. Shrinking $L_{14}$ linearly interpolates toward the other extreme, where there is only pairwise mutual information:  In the factorization limit $L_{14} \rightarrow 0$, the triple informations vanish, leaving only pairwise maximal mutual information between $B_1$ and $B_4$ and between $B_2$ and $B_3$. We may forbid a bulk phase transition in this limit to a pair of two-sided BTZ black holes by taking opposite spin structures for $B_1$ and $B_4$ (and thus also opposite spin structures for $B_2$ and $B_3$) as discussed in appendix \ref{spins}.

%
\subsection{Entanglement versus classical correlation}
%
\label{sec:entang-measures}

The mutual information $I$ quantifies bipartite correlations and, in holographic theories at least, the triple information $I_3$ provides a way of quantifying the extent to which correlations are intrinsically tripartite. None of these measures, however, distinguishes between correlation and entanglement. For pure states, there is no difference between the two, but a nontrivial distinction arises for mixed states. For example, consider the state $(\proj{\uparrow \uparrow} + \proj{\downarrow \downarrow})/2$. While correlated, it is the statistical mixture of two product states so should not be considered entangled. Formally, a state is said to be entangled only if it cannot be decomposed as a convex combination of product states~\cite{werner1989quantum,bennett1996mixed}.
In the holographic context, it is commonly assumed that bulk connectivity is related to true entanglement.  However, we know of no definitive analysis of the role that might be played by mere correlation.  We thus take care to distinguish the two below.

To begin, let us return to the 3-boundary black hole with $L_1 + L_2 < L_3$ for which we found that to leading order, $I(B_1:B_2) = 0$, $I(B_1:B_3) = 2S(B_1)$ and $I(B_2:B_3)=2S(B_2)$. From this, we provisionally drew the conclusion that the entanglement was essentially all bipartite between the pairs $(B_1, B_3)$ and $(B_2, B_3)$.
The extreme version of this situation is the $L_3 \rightarrow \infty$ limit discussed in section \ref{phases} in which the $B_1 B_2$ density operator factorizes. In that case, $B_1$ and $B_2$ are indeed each entangled only separately with $B_3$. When $L_3$ is only slightly larger than $L_1 + L_2$, however, we expect that  $I(B_1:B_2)$ will be $\mathcal O(1)$ so the state will not factorize in general.

In what quantitative sense, then, can we say that most of the entanglement is bipartite? A stringent operational definition of entanglement is to ask for the maximum rate at which near-perfect Bell pairs can be extracted from many copies of a given state using only local operations and classical communication (LOCC), a quantity known as the distillable entanglement, $E_D$~\cite{bennett1996mixed}. When a bipartite state is pure, $E_D$ and the entanglement entropy are one and the same: $E_D(A:B) = S(A)$. For mixed states, however, the entanglement entropy is generally just an upper bound: $E_D(A:B) \leq \min\{ S(A), S(B) \}$. In the case under consideration, $S(B_1) = \smfrac{\pi}{2G} L_1$ but there is very little mutual information $I(B_1:B_2)$ so we expect $E_D(B_1:B_2)$ to be small.  A famous coding theorem in quantum information theory, on the other hand, gives the following ``hashing'' bound~\cite{devetak2005distillation}:
\beq
E_D(A:B)
&\geq& S(A) - S(AB).
\eeq
Applying this inequality to the three boundary black hole gives
\beq
E_D(B_1:B_3) \label{eqn:hashing-bound}
&\geq& S(B_3) - S(B_1 B_3) \\
&=& S(B_1 B_2) - S(B_2) \\
&=& S(B_1) - I(B_1 : B_2).
\eeq
So, when $L_1 + L_2 < L_3$, we get $E_D(B_1:B_3) \geq S(B_1) - \mathcal O(1)$ and, similarly $E_D(B_2:B_3) \geq S(B_2) - \mathcal O(1)$, confirming that $B_1$ and $B_2$ are each nearly as entangled as they can be with $B_3$. (In fact, there is a single LOCC distillation procedure that will simultaneously extract both the $B_1:B_3$ and the $B_2:B_3$ entanglement at the specified rates in this case~\cite{horodecki2007quantum}.)

So far, the distillable entanglement has simply confirmed the more heuristic mutual information-based analysis. However, states can be highly correlated and even entangled without being distillable. Applying the lower bound again, we find that in general, for $L_1 \leq L_2$,
\beq
E_D(B_1 : B_2 )
&\geq& \max\{ 0, S( B_2 ) - S( B_1 B_2 ) \} \\
&=& \nonumber
	\frac{\pi}{2G} \times
	\begin{cases}
		0 & \text{if } L_2 \leq L_3 \\
		L_1 + 2 L_2 - L_{tot} & \text{if }  L_2 > L_3 \text{ and } 2 L_2 \leq L_{tot} \\
		L_1 & \text{if } 2 L_2 > L_{tot},
	\end{cases}
\eeq
where $L_{tot} = L_1 + L_2 + L_3.$
Like the mutual information, this lower bound on $E_D$ is piecewise linear as a function of $L_2$. But it starts its increase from $0$ at a later point, when $L_2 \geq L_3$ (or equivalently $L_1 + 2 L_2 = L_{tot}$), nonetheless saturating at its maximum value  $\smfrac{\pi}{2G} L_1$ when $2 L_2 = L_{tot}$, just as does the mutual information. When $E_D$ is positive, we can conclude unequivocally that there is bipartite entanglement between $B_1$ and $B_2$. Otherwise, we can't be sure.  Thus we cannot exclude the possibility that $B_1$ remains unentangled with $B_2$ in the regime close to, but on the $I(B_1:B_2) >0$ side of, the HRT phase transition.

In the analysis above, for reasons of convenience and conservatism, we have focused on the distillable entanglement.  It is the smallest of the many inequivalent mixed state entanglement measures~\cite{horodecki2000limits}. Another measure that might arguably be more relevant to holography is the entanglement of formation $E_F$~\cite{bennett1996mixed}, which is related to the minimal rate of Bell pairs required to produce near-perfect copies of the state in question using only LOCC~\cite{hayden2001asymptotic}. In general, $E_F \geq E_D$ and the gap can be very large: the entanglement of formation can be near maximal even when the mutual information is very small~\cite{hayden2006aspects}.  So in the example above, it is possible that $E_F(B_1:B_2)$ remains large even after the HRT phase transition leads to vanishing $I(B_1:B_2)$. Since $E_F(B_1:B_2)$ is defined as the minimum average entanglement entropy of the pure states in any convex decomposition of the mixed state $\rho_{1 2}$, this
would be an indication that it is impossible to describe the connected 3-boundary Lorentzian wormhole without the use of entanglement even when the correlations between the $B_1$ and $B_2$ asymptotic regions are very weak. Unfortunately, HRT calculations alone can't be used to verify if that is the case because getting good lower bounds on $E_F$ requires more detailed information about the structure of the state.

As in section \ref{depend}, we close by considering the 4-boundary case with all horizon lengths equal to a fixed $L$ and parameters chosen so that of the internal minimal surfaces only $L_{14}$ is small enough to play a role in HRT calculations. Using equations \eqref{eqn:L14-one}-\eqref{eqn:L14-four}, it is easy to see that
\be
E_D(B_1:B_3 B_4) = \frac{\pi}{2G} L, \quad
E_D(B_1:B_3) = 0\quad \text{and} \quad
E_D(B_1:B_4) \geq \frac{\pi}{2G} ( L - L_{14} ),
\ee
with equality holding in the first case because $E_D$ can never exceed entanglement entropy and in the second because it can't exceed half the mutual information~\cite{christandl2004squashed}. So the distillable entanglement between $B_1$ and $B_3B_4$ is maximal, with the length of the internal minimal surface $L_{14}$ controlling how much of the entanglement can be distilled using $B_1$ and $B_4$ alone.  This is consistent with what we expect in the factorization limit $L_{14} \rightarrow 0$,   but indicates that the distillable entanglement between $B_1$ and $B_4$ is significant even above any phase transition where the bulk geometry disconnects into a pair of two-sided BTZ black holes.

%
\subsection{Entanglement structure as a function of $n$}
%
\label{sec:more-parties}

Although the entanglement structure becomes more complicated as $n$ increases, similar ideas can be used to develop a sense of the relative importance of bipartite, tripartite and higher order correlations in any given situation. Consider a black hole with $n$ boundaries $B_a$ and horizon lengths $L_a$ sufficiently small that we can ignore internal minimal surfaces.

Since $I_3$ was a useful diagnostic for the 4-boundary case, it is tempting to evaluate its
higher order generalization
\be
I_k(B_1: B_2: \cdots : B_k) = \sum_{\sigma} (-1)^\sigma S(\sigma),
\ee
where the sum is over subsets of the arguments $\{ B_1, B_2, \ldots , B_k\}$. Interpreting these functions as measures of multipartite correlation only makes sense, however, when each $I_k$ is either always nonnegative or always nonpositive. An earlier computer search revealed that $I_4$ can be made to take both positive and negative values in holographic theories by choosing appropriate combinations of intervals in a single CFT~\cite{hayden2013holographic}. One might hope that such complications do not occur in this case since each $B_a$ is an entire CFT. But in fact it is even easier to find configurations in which $I_4$ takes either sign. When $L_a = L$ for all $a$, one finds that $I_4 = \smfrac{\pi}{2G} L$ for $n=5$ but $I_4 = - \smfrac{\pi}{2G} 2L$ for $n=6$.

With the naive approach ruled out, let us proceed instead using the more trustworthy mutual information and distillable entanglement functions.
Set $L_{tot} = \sum_{a=1}^n L_a$ to be the total length of all the horizons and for convenience write $L_Z = \sum_{z \in Z} L_z$ for the sum of the horizons indexed by set $Z$.
If $X, Y \subseteq \{ 1, \ldots, n \}$ such that $L_X \leq L_Y$ then
\be
\label{manyBMI}
I(B_X : B_Y ) =
	\frac{\pi}{2G} \times
	\begin{cases}
		0  & \text{if } 2( L_X + L_Y ) \leq  L_{tot} \\
		2(L_X + L_Y) - L_{tot} & \text{if } 2( L_X + L_Y ) > L_{tot}
				\text{ and } 2 L_Y \leq L_{tot} \\
		2 L_X & \text{if } 2 L_Y > L_{tot}.
	\end{cases}
\ee
This equation has exactly the same form as \eqref{eqn:mi-piecewise} and reflects the fact that the entanglement structure is similar to that of a generic random state.
Likewise, we can consider the distillable entanglement between collections of subsystems. Performing the same substitution gives
\beq
E_D(B_X : B_Y )
&\geq& \max\{ 0, S( B_Y ) - S( B_X B_Y ) \} \\
&=& \nonumber
	\frac{\pi}{2G} \times
	\begin{cases}
		0 & \text{if } L_X + 2 L_Y \leq L_{tot} \\
		L_X + 2 L_Y - L_{tot} & \text{if } L_X + 2 L_Y > L_{tot} \text{ and } 2 L_Y \leq L_{tot} \\
		L_X & \text{if } 2 L_Y > L_{tot}.
	\end{cases}
\eeq
Combining these bounds on $I$ and $E_D$ does permit rigorous conclusions about multipartite entanglement.

For example, set $L_a = L$ for all $a$ for simplicity. Then the bounds demonstrate that when $|X|+|Y| \leq n / 2$, the mutual information between $B_X$ and $B_Y$ is small. Since $E_D \leq I/ 2$~\cite{christandl2004squashed}, this means that the distillable entanglement of any subsystem of size $n/2$ or smaller is negligible. Conversely, if $|X| + 2|Y| > n$, then there \emph{will} be distillable entanglement. In particular, substituting $|X|=1$ tells us that this will be the case for any subsystem of size $|X|+|Y|$ strictly larger than $\smfrac{n+1}{2}$. In this precise sense, the distillable entanglement of the state is at least \smfrac{n+1}{2}-partite: there is none for subsystems of size smaller than the threshold and there always is for subsystems larger.

Earlier, we used the triple information $I_3$ to infer that, in a 4-boundary black hole with all horizon lengths equal, all correlation was at least tripartite.  But that argument was really at the level of correlation rather than entanglement. Nonetheless, the conclusion is consistent with our analysis here, which indicates that all the distillable entanglement is at least $\lceil \smfrac{4+1}{2} \rceil = 3$-partite.

Finally, we note that the fact that the entropy of a pair of boundaries exceeds that of a single boundary means that GHZ-like $n$-party entanglement is unimportant.\footnote{For $n=4$ one can also see this from \eqref{triple}, whose sign is opposite to that of the GHZ${}_4$ state as is always the case for holographic $I_3$ \cite{hayden2013holographic}.}

\subsection{Intrinsically $n$-partite entanglement}
\label{intrinsic}

Earlier, we identified the large-$L_3$ limit of a three-boundary black hole as an example of a system whose entanglement was not intrinsically tripartite because the state could secretly be factorized into a pair of bipartite entangled states, one between $B_1$ and $B_3$ and the other between $B_2$ and $B_3$. In this section we will generalize that idea to define intrinsically $k$-partite entanglement and then show that some $n$-boundary black holes have definitely have some $n$-partite entanglement.  If they didn't, then some of their HRT entropies would have to be significantly smaller than they are. However, we warn the reader that the argument below is not strong enough to quantitatively compare this $n$-party entanglement to $k$-party entanglement with $k < n$.

Our starting point will be that, for sufficiently small moduli $L_x$  (with the index labeling boundaries now $x = 1,\dots, n$), the entropies calculated using the HRT formula resemble those of random quantum states in the sense that
\be \label{eqn:generic-defn}
S( \cup_{x \in X} B_x )
= \min \left[
	\sum_{x \in X} S(B_x), \sum_{x \not\in X} S(B_x)
	\right]
\ee
to leading order in the central charge. We will again focus for simplicity on the case in which all the $S(B_x)$ are the same, so that the right hand side the formula is simply $\min(|X|,n-|X|) S(B_1)$. From now on, we will say that states satisfying \eqref{eqn:generic-defn} are \emph{entanglement-generic}. Our objective will be to use \eqref{eqn:generic-defn} alone  to demonstrate the existence of $n$-partite entanglement when $n$ is even, and $(n-1)$-partite entanglement when $n$ is odd.

\begin{figure}
\centering
\includegraphics[keepaspectratio,width=0.8\linewidth]{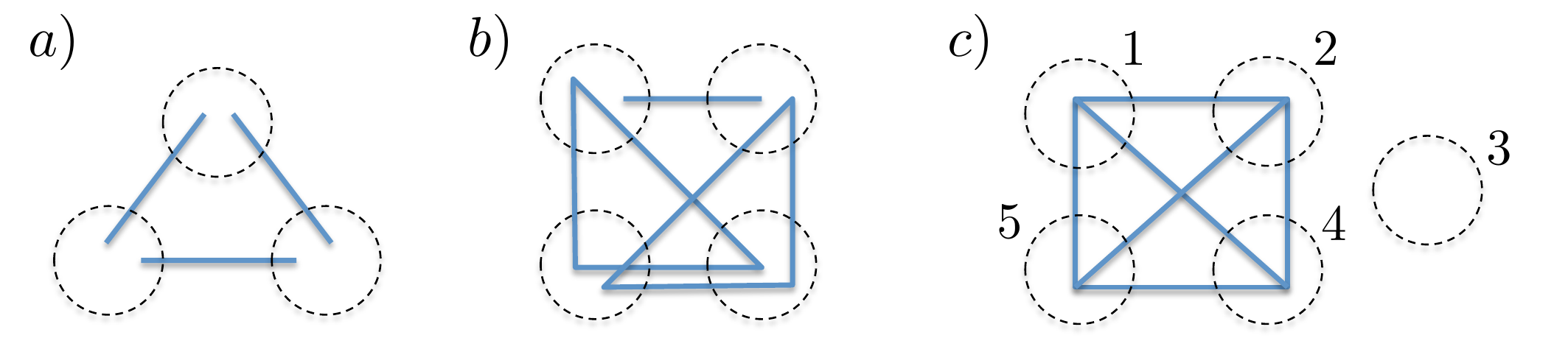}
\caption{a) A $2$-producible state on $B_1B_2B_3$. Each subsystem is indicated by a dashed circle, with component entangled states drawn as graphs, in this case lines, between the systems entangled by the state. Observe that if each $2$-party state is a Bell state, then the overall tripartite state will have generic entropy behaviour. b) A $3$-producible state composed of pure states entangled between pairs and triples of systems. c) A $4$-producible state. If the state on $B_1 B_2 B_4 B_5$ is generic, then the version of this state symmetrized over all 5 systems will have generic entanglement entropies.}
\label{fig:kprod}
\end{figure}

To get a sense of the difficulties inherent in drawing conclusions from boundary entanglement entropies alone, consider figure \ref{fig:kprod}a, which depicts a tripartite quantum state which is 2-producible, meaning that it is the tensor product of bipartite entangled states~\cite{guhne2005multipartite} . Unlike the $L_3 \rightarrow \infty$ example considered above, no 2-party reduced density matrix constructed from \ref{fig:kprod}a  will factorize. But among all tripartite states, the $2$-producible ones are still very special and lack intrinsically tripartite entanglement. Nonetheless, choosing each bipartite factor to be maximally entangled is sufficient to make the states entanglement-generic. (The impossibility of detecting intrinsically tripartite entanglement in a pure tripartite state using entropies alone is reflected in the fact that the triple information $I_3$ will always be zero for such states.)
Also, it is worth keeping in mind that, while $2$-producible states contain only pairwise entanglement, starting with enough pairwise entanglement, one could make any tripartite state using only local operations and classical communication. The strategy would just be to manufacture the desired state entirely in one factor and then teleport the appropriate pieces to the other two. Figure \ref{fig:kprod}b depicts a 3-producible state: it is the tensor product of states which are themselves entangled between at most three subsystems. If each of the three factors is chosen to be a pure state chosen at random from the Haar measure, the resulting state will not be entanglement-generic this time: the entropy of the top two subsystems will be strictly less than the sum by an amount equal to twice the entanglement entropy of the bipartite factor.

Figure \ref{fig:kprod}c provides a final illuminating example. As drawn, it looks highly atypical: system $B_3$ factorizes from the rest. But suppose the $B_1B_2B_4B_5$ portion is selected according to the Haar measure and that all five $B_x$ have the same dimension. Then the resulting state will with high probability be entanglement-generic. Because $S(B_3)$ will be zero, it doesn't cause any problems; the entropy of any two out of the five subsystems will be the sum of their individual entropies. While the example treats $B_3$ very specially, it is straightforward to symmetrize it such that all $S(B_x)$ will be the same. Simply take a tensor product of five similar states, each one singling out a different $B_x$ for factorization. The crucial fact that makes this trick work is that in our context \eqref{eqn:generic-defn} implies that the entropy of any $X$ is determined by a collection involving at most half the subsystems. But five is an odd number, so that ``at most half the subsystems" actually means two out of five and a Haar random state on four factors is generic on any two of its factors. As such, analogous constructions of entanglement-generic $(n-1)$-producible states exist whenever $n$ is odd.  Therefore, we will at best be able to show that odd-$n$ entanglement-generic systems cannot be $(n-2)$-producible, meaning they are intrinsically $(n-1)$-entangled.

Now consider the same construction in even dimensions. The entropy of a fixed subset in the symmetrized model is the same as the entropy of a random subset in the unsymmetrized model. So suppose that $B_1 B_2\cdots B_{n-1}$ is in a random pure state factorizing with the state of $B_n$. A random subset of $B_1 B_2 \cdots B_n$ of size $n/2$ will contain $B_n$ with probability $1/2$. If it does, the entropy, instead of being $nS(B_1)/2$, will be $(n/2 - 1)S(B_1)$. Otherwise, the entropy is as it should be for an entanglement-generic state. The expected deficit will therefore be $S(B_1)/2$. In the symmetrized model, the entropy of a system of size $n/2$ will therefore be $S(B_1)/2$ smaller than would be required to be entanglement generic. Our analysis below will show that this is optimal to within a constant factor: every $(n-1)$-producible state will exhibit an average entropy deficit on subsystems consisting of $n/2$ factors of at least an $n$-independent constant times $\smfrac{1}{n} \sum_x S(B_x)$.

Before proceeding, let us introduce some definitions and notation in the interest of being more precise. Let $[n] = \{1,\ldots,n\}$ and $\choice{X}{j}$ consist of the $j$-element subsets of $X$. In the rest of the section, $B_x$ will refer specifically to a Hilbert space. A state $\ket{\psi} \in \otimes_{x=1}^n B_x$ is said to be $k$-producible~\cite{guhne2005multipartite} if there exist Hilbert spaces $B_x^K$ and isometries $U_x : B_x \rightarrow \otimes_{K \in \choice{[n]}{k}} B_x^K$ such that
\be
\otimes _{x \in [n]} U_x \ket{\psi}
= \otimes_{K \in \choice{[n]}{k}} \ket{\psi^K}
\ee
for some $\ket{\psi^K} \in \otimes_{x \in K} B_x^K$.\footnote{This definition is a slight generalization of the usual notion, which doesn't explicitly allow for the isometries $U_x$.}
While this definition is transparent and completely sufficient, it will be very convenient below to be able to refer to $B_x^K$ for $x \not\in K$ with the understanding that $B_x^K$ is trivial in that case. So we will instead work with the equivalent definition that $\ket{\psi^K} \in \otimes_{x \in [n]} B_x^K$, with $B_x^K = \mathbb{C}$ for $x \not\in K$.

For any Hilbert spaces $B_x$, write $B_X = \otimes_{x \in X} B_x$. Our goal will be to quantify deviations from being entanglement-generic, which means we will be comparing $S(B_X)$ to $\Sadd(B_X) = \sum_{x \in X} S(B_x)$. Our main result will be a lower bound on the difference $S_{\rm{sum}}(B_X) - S(B_X)$ in terms of $|X|$. The main idea will be to generalize the observation made above about $n$-partite states composed of generic $(n-1)$-producible factors. The challenge is to prove a lower bound that works \emph{only} assuming that the state is $k$-producible and nothing else about its internal structure. (As noted above, however, we assume in this section that the entropies $S(B_x)$ are identical for all subsystems $x \in [n]$. The bound holds regardless but that assumption simplifies its interpretation.)

The derivation of the bound is quite technical so we defer the argument to appendix \ref{bound}, with the precise statement appearing as \eqref{eqn:multi-lb}.    The result may be expressed in terms of a fractional entropy deficit $\delta(j,k,n)$:
\be
\label{eqn:avbound}
\overline{[ S_{\rm{sum}}(B_X) ]}_{X \in \choice{[n]}{j}} - \overline{[ S(B_X) ]}_{X \in \choice{[n]}{j}}
\geq \delta(j,k,n) \overline{[ S(B_x)]}_{x \in [n]},
\ee
where the notation $\overline{[ \cdot ]}_{X \in \choice{[n]}{j}}$ denotes the average over $j$-element subsets.  This average is of course trivial for $S_{\rm{sum}}(X)$, as we have assumed that all $X \in \choice{[n]}{j}$ give the same $S_{\rm{sum}}(X)$.   Equation \eqref{eqn:multi-lb} shows that $\delta(j,k,n)$ is usually positive for allowed $j,k,n$ but does not give immediate insight into its magnitude so we have evaluated it on a computer.

\begin{figure}
\centering
\includegraphics[keepaspectratio,width=1.0\linewidth]{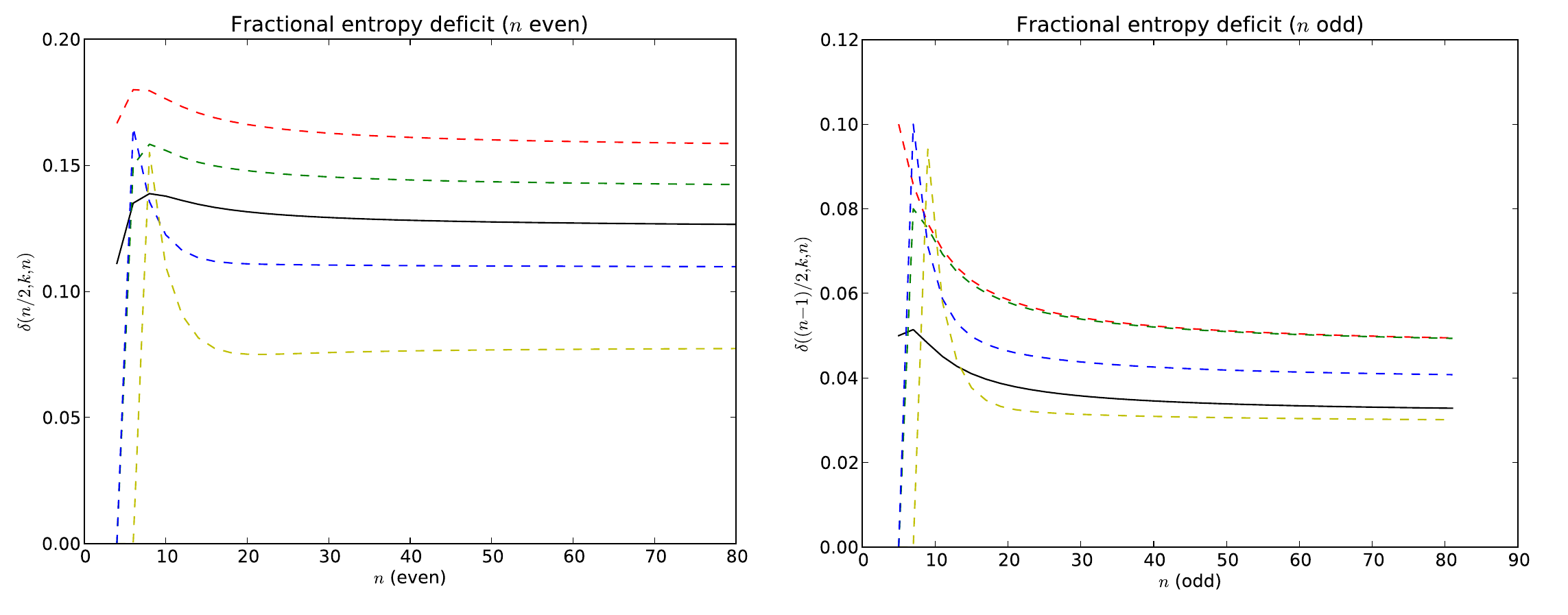}
\caption{Lower bounds on the fractional entropy deficits as a function of $n$. The plot on the left depicts $\delta(n/2,k,n)$ for $n$ even. The solid black line corresponds to $k=n-1$ and the dashed red, green, blue and yellow lines to $k=n-2$, $k=n-3$, $k=n-4$ and $k=n-5$, respectively. In all cases, $\delta$ converges to a constant, with the strongest bound actually occurring for $k=n-2$. The plot on the right depicts $\delta((n-1)/2,k,n)$ for $n$ odd. In this case, the deficit for $k=n-1$ is precisely zero so the solid black line corresponds instead to $k=n-2$. The dashed red, green, blue and yellow lines in turn represent $k=n-3$ through $k=n-6$. Once again, the deficits converge to constants at large $n$. By definition, the actual entropy deficits can only increase as $k$ decreases so these plots illustrate that our bounds could be strengthened for small $k$: the dashed yellow line, which in both plots corresponds to the smallest value of $k$, is also in both plots the weakest bound.}
\label{fig:entropy-deficits}
\end{figure}

Consider first $(n-1)$-producible states when $n$ is even. The strongest test is when $j = n/2$.  As shown in figure \ref{fig:entropy-deficits}, $\delta$ appears to be bounded below by a constant independent of $n$: there will always be an appreciable fractional entropy deficit. The smallest deficit, $2/9$, occurs when $n=4$, with the values up to $n=100$ being as high as $0.138$ and converging to a value in the vicinity of $1/8$. Since the symmetrized $(n-1)$-producible state constructed in the introduction to this section had an entropy deficit of $1/2$, the lower bound is within a constant factor of being optimal.

When $n$ is odd, the fractional entropy deficit for $(n-1)$-producible states is always 0, reflecting our earlier observation that it is possible to construct $(n-1)$-producible entanglement-generic states. There is an appreciable fractional entropy deficit for $k=n-2$, however. For $n=5$, one finds a value of $1/20$, for example, and it appears to converge to a value in the vicinity of $0.03$.

Studying the bound as a function of $k$ is unfortunately not as satisfying. Since being $k$-producible becomes more and more stringent as $k$ decreases, one would expect the bound to likewise get stronger. This happens up to a point, as is visible in figure \ref{fig:entropy-deficits}. The bound generally strengthens in going $k=n-1$ (or $k=n-2$ when $n$ is odd) to $k=n-2$ ($k=n-3$ when $n$ is odd) but thereafter gets weaker as $k$ decreases. That isn't a surprise since the bound was designed to demonstrate the existence of intrinsically $n$-partite entanglement. Nonetheless, it would be interesting to improve the analysis so as to extract good bounds for all $k$.

In summary, we have confirmed that  when all $L_x$ are equal, and when the HRT surfaces are unions of horizons, the state $\ket{\Sigma}$ is $n$-partite entangled for even $n$ and $(n-1)$-partite entangled for odd $n$.  If it weren't then the average entanglement entropies evaluated using the HRT formula of subsystems of size $n/2$  would be smaller by an amount proportional to the central charge and independent of $n$ than their actual values.  Our assumption about the HRT surfaces will certainly hold in the picture limit.

\subsection{The random state model}
\label{random}

Our CFT states have an interesting entanglement structure that depends on the choice of moduli.  While the details may seem complicated, we noted in section \ref{p3} that the leading-order bipartite mutual information \eqref{eqn:mi-piecewise} for the 3-boundary wormhole is precisely that found in the large $L_a/G$ limit for a random state on the Hilbert space $\overline{\mathcal{H}}_1 \otimes \overline{\mathcal{H}}_2 \otimes \overline{\mathcal{H}}_3$, where $\overline{\mathcal{H}}_a$ is a finite-dimensional Hilbert space of dimension $e^{\pi L_a/2G}$.  Such Hilbert spaces provide natural finite-dimensional models of our CFTs in the connected 3-boundary wormhole phase where the leading-order entropy of each black hole is just ${\pi L_a/2G}$.  For this reason we may think of the states  in $\overline{\mathcal{H}}_a$ as all having the same energy, namely that corresponding to the $a^{\text{th}}$ black hole.  In the puncture limit we might read off this conclusion directly from the Boltzmann factors in the CFT wavefunction.

In just the same way we may again use \cite{Page:1993df} to see that our general mutual information formula \eqref{manyBMI} for an arbitrary number of boundaries -- and thus also the triple mutual information \eqref{triple} for arbitrary $L_a$ -- agree with corresponding random state models at large $L_a/G$ so long as the internal geodesics can be ignored.  The point is again that any subsystem with less than half the total dimension has nearly maximal entropy, while purity of the full state requires the entropy of larger subsystems to be equal to that of their complement.

This result seems rather natural for a strongly interacting CFT with a high density of states.  It indicates that the $n$-point functions have rather little structure when viewed at a sufficiently coarse-grained level.  Of course, the actual $n$-point functions in any given CFT are definite objects and, in the same way, the $n$-boundary states are rather specific states.  This is all in direct analogy with the two-boundary case where the resulting thermofield-double state might be approximated by the fully entangled microcanonical ensemble state

\begin{equation}
\label{micro}
|\psi \rangle_{micro} = \sum_{E \in (E_1, E_2)} |E\rangle |E\rangle
\end{equation}
defined by an appropriate range of energies.  The entropy of each factor, and indeed the full reduced density matrix on each factor, agrees to leading order with that of a random state chosen with respect to the unitarily invariant measure on the appropriate ${\cal \bar H}_1 \otimes {\cal\bar H}_2$.  Nonetheless it is clear that the actual state
\eqref{micro} requires very specific correlations between the energies of the two systems that will not be reproduced by generic states. See e.g. \cite{Marolf:2012xe,Marolf:2013dba,Balasubramanian:2014gla} for further discussion of this point.

Taking this model seriously makes predictions for both the distillable entanglement $E_D$ and the entanglement of formation $E_F$. The most striking of these governs the behaviour of $E_F$ in the three-boundary case. When $L_1 + L_2 > L_3$, we have seen that $I(B_1:B_2) \propto L_1 + L_2 - L_3$ so that the correlation between $B_1$ and $B_2$ shrinks as $L_3$ increases. In the associated random state model, however, $E_F$ does not shrink, instead staying essentially constant and maximal right until $L_1 + L_2 = L_3$, at which point it drops abruptly to a negligible level~\cite{hayden2006aspects}. If the same is true for the entanglement of formation of the CFT states, then as long as $I(B_1:B_2)$ is of order the central charge, any description of the state $\rho_{12}$ as a mixture of pure states would involve only nearly maximally entangled states. This would be particularly interesting if those maximally entangled states could be interpreted holographically as different connected two-boundary spacetimes. The consequences for $E_D$ of the random state model are less striking, just that the lower bounds calculated in sections \ref{sec:entang-measures} and \ref{sec:more-parties} are tight modulo violations of the so-called additivity conjecture~\cite{hayden2006aspects}, which are believed to be small for the random state model~\cite{belinschi2013almost}.

One would very much like to test these predictions -- or indeed those associated with any other entanglement measure -- via further bulk calculations. It would be particularly interesting to study cases with a large number of boundaries $n$.  While the entanglement properties predicted by the simplest random state model will be invariant under arbitrary permutations of the boundaries, any choice of $\Sigma$ will break this symmetry.  Working in the puncture limit we may note that placing any $n$ points on the sphere to define an $n$-point function will induce a notion of proximity, so that some subsets of points may be said to be closer together than others.  One would very much like to understand the extent to which this manifests itself in entanglement properties of the CFT state and leads to discrepancies with the random state model.

Now, as we have already seen, even at small $n$ the state becomes somewhat less generic in regions of moduli space where the internal geodesics become relevant.  These internal geodesics effectively function as constraints that restrict the structure of the state.  In the presence of such constraints one might construct a natural random state model by first decomposing the relevant $n$-boundary manifold $\Sigma$ into pairs of pants by cutting $\Sigma$ along some set of internal geodesics associated with the moduli $L_{ab}$.  Considered separately, the state on each pair of pants would then be drawn randomly from the associated Haar-random distribution as above.  However, there are several issues to consider further:

\begin{enumerate}[i)]

 \item {} \label{cor} There is in general some correlation between states on the $I$th and $J$th pair of pants.  For example, if the moduli of these surfaces agree, the states are identical.  It would be interesting to understand the extent to which  such correlations are important for reproducing various properties of the $n$-boundary state.  It would also be of great interest to understand how rapidly such correlations decay as the moduli are deformed away from equality.  One should similarly study how these correlations depend on the Dehn twist parameters $\theta$ along the cuts.

\item {}  The decomposition of $\Sigma$ into pairs of pants is not unique.  In simple cases, in a given region of moduli space, it may be possible to single out the decomposition associated with the tightest possible set of constraints.  But in general we expect different decompositions to restrict the state in different ways. For a given prescription for dealing with point \ref{cor} above, the $A$th decomposition effectively defines some measure $\mu_A$ on the space of $n$-boundary states.  So we must combine these $\mu_A$ into a single measure $\mu_\Sigma$.  While there is no canonical way to do so, one may hope that some universality renders the details unimportant for large $L_a$.   One simple recipe that is then to write each measure as $\mu_A = f_A \mu_{Haar}$ in terms of the Haar measure on $n$ boundaries and, noting that any $\Sigma$ has a finite number of decompositions $A$, to define $\mu_\Sigma = \mathcal{N} \left(\prod_A f_A \right) \mu_{Haar}$ by simply multiplying the functions $f_A$ and choosing the coefficient $\mathcal{N}$ to properly normalize the result.  As desired, this $\mu_\Sigma$ vanishes on any state forbidden by any pair of pants decomposition.

  \item {} When the decomposition involves many pairs of pants, one might begin to question the microcanonical approximation in which the Boltzmann factors of \eqref{Va} are replaced by truncation to a finite-dimensional Hilbert space.  One might like to instead explore properties of states chosen randomly from a Gibbs ensemble with some specified density of states.
\end{enumerate}
A class of such models for certain domains of moduli space will be explored in the follow-up paper~\cite{hayden2014random}.

\section{Conclusions}
\label{conclude}

Our work has focused on the holographic description of CFT states $|\Sigma \rangle$ defined by the Euclidean path integral over
Riemann surfaces $\Sigma$ with $n$ asymptotic boundaries.  Such states may be interpreted as entangled states of $n$ separate CFTs each defined on the cylinder $S^1 \times \mathbb{R}$; i.e., as states of ${\rm CFT}{}_1 \otimes \dots \otimes {\rm CFT}{}_n$.  This study was motivated by the fact that -- at least in appropriate regions of moduli space -- the dual bulk states
describe time-symmetric black hole geometries where the $t=0$ surface is precisely the surface $\Sigma$.  Our main goal was to understand the relation of these geometries to entanglement in $|\Sigma \rangle$ as a function of both external and internal moduli ($L_a$ and
$\tau_\alpha$). We found a number of interesting features.

First, we noted in section \ref{phases} that the multiboundary black hole is not always the dominant saddle-point in the bulk semi-classical approximation.
As the moduli of $\Sigma$ vary, there can be phase transitions between this maximally connected phase and phases where only certain subsets of boundaries are connected in the bulk.  As with any first-order phase transition, these are associated with discontinuities in the entropy and energy of each asymptotic region (and thus of each CFT factor).  But by studying a factorization limit, we found in section \ref{factorization} that any notion of temperature one might assign is also discontinuous across the phase boundary.  A particular striking example occurs when
a pair of boundaries splits off from the rest but remain connected to each other.  The two-boundary factor is then (conformal to) a thermofield double state, but with inverse temperature given by the {\rm sum} of those for the corresponding black holes in the maximally connected phase. The disconnection of an $m > 2$ boundary component  -- perhaps the remainder of the geometry from which the above pair of boundaries was severed -- will also feature a discontinuity in all temperatures, though a milder one that vanishes in the puncture limit of small moduli $L_a$. When a single boundary disconnects, it is unclear whether a meaningful notion of temperature can be associated with the resulting factor at all.

While unfamiliar, this phenomenon should be expected since our path integrals do not generally describe thermal equilibrium.  There is thus no canonical assignment of temperatures to a given point in moduli space.  Instead, any approximate notion of temperature must emerge from the physics in the same sense that we may expect generic pure states to equilibrate to thermal states at late times in sufficiently ergodic systems with many degrees of freedom.  This places the notion of temperature in our context on a footing similar to those of energy and entropy so that such discontinuities are natural.  Our understanding of the above phase transitions was based on CFT calculations and indirect arguments from an HRT calculation; it would be very interesting to understand the structure better
in the bulk.

Another interesting result was the strong moduli-dependence of the multipartite nature of $|\Sigma \rangle$.  Consider for example the three-boundary case.  In the limit $L_3 \to \infty$ we found CFT${}_1$ and CFT${}_2$ to be essentially unentangled -- and indeed, uncorrelated; see section \ref{sec:entang-measures} for further discussion of correlation vs. entanglement.  Holographically we found that the mutual information between 1 and 2 vanished at leading order in the central charge whenever $L_3 > L_1 + L_2$. Thus the separate entanglement of systems 1 and 2 with system 3 appears sufficient to produce a geometric connection between 1 and 2; multiparty entanglement  is unimportant in this regime.

One might have expected this bipartite entanglement to produce two separate wormholes, one linking 1 to 3 and one linking 2 to 3, but in AdS$_3$ it is not possible to have two separate black hole horizons in a single asymptotic region; the two must lie inside a single horizon \cite{Steif:1995pq}.  In higher dimensions one expects\footnote{Though see comments in section 3 of \cite{Wall:2012uf}.} to be able to construct states having only separate 1,3 and 2,3 bipartite entanglement for which the holographic dual does indeed have two initially-separate wormholes which link 3 to 1 and 2 respectively. But at least in the absence of angular momentum, gravitational attraction will eventually cause the two black holes to coalesce and form a connected multiboundary wormhole with properties similar to those considered here.  Whether or not this possibility is realized in a given state may depend on the spatial separation of the degrees of freedom in 3 that are entangled with 1 and 2.  Similar extreme regions of moduli space with purely bipartite entanglement (where $n-1$ smaller systems are each entangled only with one big system at leading order in $c$) exist for any number of boundaries $n$.   In contrast, for $n=4$ or more boundaries there are complementary regions of moduli space where all bipartite mutual informations between pairs of boundaries vanish.

For typical parameter values the degree of multipartite entanglement appears to increase with the number of boundaries. In particular, for all horizon lengths equal (and when the internal moduli impose no constraints) the entanglement includes at least some $n$-partite entanglement when $n$ is even and $(n-1)$-partite entanglement when $n$ is odd.  Thus, multipartite entanglement is increasingly important, but the dominant form of entanglement may involve only a subset of the boundaries connected geometrically in the bulk. The techniques developed here to reach those conclusions, namely bounding the entanglement entropy of $k$-producible quantum states, may be of independent interest.

Intriguingly, all of the above entanglement results in both bipartite and multipartite domains agree precisely with the predictions of a simple random-state model.  As discussed in section \ref{random}, extensions of this model to cases with important internal constraints are more subtle.  These cases remain to be explored in detail, though some initial results will appear in \cite{hayden2014random}.

Our studies were facilitated by writing the state $|\Sigma \rangle$ as a sum of products of  CFT $n$-point functions of operators $O_i$ and the states made by the action of certain operators $V_a$, one for each boundary, on the state made by the action of $O_i $ on the 
vacuum (see Sec.~\ref{3ptf}).
The $V_a$ are universal, in the sense that they are (complexified) conformal transformations dictated only by the moduli of $\Sigma$ and independent of any details of the CFT.  Each $V_a$ in fact takes the form $U_a e^{-\frac{1}{2} \hat \beta_a H_a} \tilde U_a$, where $U_a, \tilde U_a$ are unitary conformal transformations acting on $CFT_a$ and $H_a$ is the associated Hamiltonian.  Although the moduli-dependence of $U_a, \tilde U_a$ and the inverse temperature $\hat \beta_a$ may be complicated in general, it simplifies in the puncture limit $L_a \rightarrow 0$ where one finds $U_a, \tilde U_a \rightarrow \mathds{1}$ and -- up to a constant offset -- $\hat \beta$ is just the inverse temperature of the corresponding bulk black hole in the fully-connected single-wormhole bulk geometry.  Thus we obtain an explicit formula in this limit.

We also remind the reader of the interesting phase structure we found in the 3-boundary case when we increased $L_3$ with $L_1,L_2$ held fixed.  Due to the transition of the minimal surface discussed in section \ref{p3}, for  $L_3 \ge L_1+L_2$ the mutual information $I(B_1:B_2)$  is only order 1 (as opposed to ${\cal O}(c)$).  Thus for most probes of CFT${}_1$ and CFT${}_2$ the reduced density matrix $\rho_{12}$ obtained by tracing out $B_3$ is well approximated by the product $\rho_1 \otimes \rho_2$. But as noted in section \ref{reduced}, the bulk geometry in this region is still the three-boundary wormhole. Although particularly dramatic here, this structure is similar to that associated with other Ryu-Takayanagi transitions where one finds related-but-distinct transitions for various Renyi entropies $S_n$ though the bulk geometry for the state changes continuously \cite{Faulkner:2013yia}.

Despite the above results, many features of $|\Sigma \rangle$ remain to be understood.  Interesting avenues for further work include
developing better measures of multipartite entanglement -- which could diagnose more clearly the extent of mutlipartiteness in our states -- and using them to explore the moduli dependence of the entanglement in more detail.  Even restricting attention to simple measures,
it seems likely that one can make further use of HRT to explore the structure of \eqref{L3large} by sewing two identical pairs of pants together along boundary 3 with a general Dehn twist $\theta$.

One would also like to gain some further insight into details of the CFT states.  For example, one might study model systems for which the $n$-point functions are explicitly known.  In another direction, numerical computations may be of use in understanding both the conformal transformations $U, \tilde U$ and the inverse temperature $\hat \beta_a$ in \eqref{Va} beyond the puncture limit.
More generally, it would be desirable to obtain a precise characterization of those CFT states whose bulk duals are described by a single bulk geometry.  It may be that such states exhibit specific information theoretic features -- such as the monogamy of mutual informations -- that distinguish them from generic CFT states.

\section*{Acknowledgements}
We thank David Berenstein, Steve Giddings, Tom Hartman, Sean Hartnoll, Juan Maldacena, John Parker, Joe Polchinski, Grant Salton, Volkher Scholz, Kostas Skenderis, Lenny Susskind, Nate Thomas, and David Tong for useful conversations.   V.B. was supported in part by DOE grant DE-FG02-05ER-41367. V.B. also thanks the Fondation Pierre-Gilles de Gennes for support, and enjoyed the hospitality of the LPT at the Ecole Normale Sup\'{e}rieure and the LPTHE at the Universit\'{e} Pierre et Marie Curie in the early stages of this project.   P.H. was supported in part by CIFAR, FQXi, NSERC and the Templeton foundation. A.M. was supported in part by NSERC, the Simons Foundation, FQXi and the Templeton foundation.  D.M.  was supported in part by the National Science Foundation under Grant No PHY11-25915, by FQXi grant FRP3-1338, by funds from the University of California, and as a Visiting Fellow Commoner at Trinity College during much of this work.  In addition, he would like to thank DAMTP, Cambridge U. for their hospitality throughout much of this work.  S.F.R. is supported by STFC. D.M. and S.F.R. also thank the Isaac Newton Institute for its hospitality during critical stages of the project.
\appendix

\section{Details on Bulk Phases and the Mapping Class Group}
\label{bulkdetails}
\subsection{Mapping Class Group}
 \label{mcg}

The phase structure of section \ref{phases}  can be understood a bit more formally as follows.  We will begin with the torus case  (for which see \cite{Maloney:2007ud, Maldacena:1998bw, Dijkgraaf:2000fq} for details).  The Euclidean geometry is, topologically, a solid donut whose boundary is a torus. This torus has two non-contractible cycles -- the $\phi$ circle and the Euclidean time $t_E$ circle -- which we will call $a$ and $b$, respectively, with intersection number $a \cap b=1$.
One can then consider the action of the mapping class group ($MCG$) -- the group of diffeomorhpisms, modulo those which are continuously connected to the trivial diffeomorphism -- on the torus. This mapping class group can be most easily visualized by considering its action on $H^1(T^2)$, i.e. on the cycles $a$ and $b$ of the boundary torus. It must act linearly on these cycles, and it must preserve the intersection number $a \cap b=1$.  Therefore an element of the $MCG$ can be regarded as an element of $SP(2,\ZZ)\cong SL(2,\ZZ)$.  This is the usual modular  group of the torus.

The action of the $MCG$ on the bulk solid donut is a diffeomorphism, which reduces to the former diffeomorphism on the boundary torus. As this diffeomorphism is non-trivial at the boundary, it is not a gauge transformation of the bulk theory, and the action of the MCG is non-trivial in the bulk; these should be thought of as distinct bulk saddles.  For example, it takes the geometry where the $a$ cycle is contractible in the bulk (thermal AdS) to a geometry where a linear combination of $a$ and $b$ are contractible. Thus we end up with a family of geometries labelled by the elements of $SL(2,\ZZ)$. Typically, one of these will dominate, but at certain points in moduli space two of these saddles might have equal action.  Since saddles are related by the action of the $MCG$, this will  happen only when the boundary torus is a fixed point of the $MCG$.  At this point the boundary torus has a nontrivial discrete symmetry (an automorphism), since it is invariant under the action of an element of the $MCG$. The inverse-temperature $\beta$ can be interpreted as the imaginary part of the conformal structure modulus of the torus, via ${\rm Im}~\tau = 2 \pi \beta$.  The modulus $\tau$ is acted on by $SL(2,\ZZ)$ in the usual way, and has a fixed point at $\tau=i$. This is precisely the Hawking-Page transition.  It is important to note that the two different Euclidean saddle points -- thermal AdS and the Euclidean BTZ black hole --  have the same metric, but that they are related by a diffeomorphism which acts nontrivially on the boundary, so should be regarded as giving independent contributions to the path integral of the theory.

For Riemann surfaces of higher genus, the story is similar.  For a genus $g$ Riemann surface $M$, the non-contractible cycles in $H^1(M)$  are similarly divided into $a$ cycles and $b$ cycles,  $a_i$ and $b^i$, with intersections $a_i\cap b^j = \delta_{i}^j$. The action of the $MCG$ on the cycles is determined by the modular group $Sp(2g,\ZZ)$, which acts linearly on the $a$ and $b$ cycles and preserves the intersection number. For higher genus, some elements of the $MCG$ act trivially on the $a$ and $b$ cycles; mathematically, this is the statement that the Torelli group (the quotient of the mapping class group by the modular group) is nontrivial. However, our interest is on the action on the cycles, as this is what affects the interpretation of the bulk handlebody.

 In the bulk handlebody, half of these cycles will become contractible. There is a choice of division where the cycles that are contractible in the bulk are all $a$ cycles. The action of the $MCG$ will then map this to different combinations of $a$ and $b$ cycles being contractible in the bulk. Again, this is a diffeomorphism of the bulk handlebody, but these should be thought of as distinct bulk saddles, and will have different physical interpretations.

 Focusing on the case of genus two, the modular group is $Sp(4,\ZZ)$. There are four fundamental cycles, $a_1$, $a_2$, $b_1$, $b_2$. Choices for which cycles become contractible in the bulk correspond to the different phases discussed in section \ref{phases} and shown in Figure \ref{genus2}. The cases of interest, which have moments of time reversal symmetry, are:
 \begin{itemize}
 \item
 $(a_1, a_2)$ contractible.  This is the naive handlebody obtained by filling in the genus 2 Riemann surface when we embed it in $\RR^3$.  This is depicted on the left hand side of Figure \ref{genus2}.  This analytically continues in Lorentzian signature to three disconnected AdS's.
\item
  $(b_1, b_2)$ contractible.  This is the "dual" of the naive handlebody described above; it is the exterior of the Riemann surface, if we think of the Riemann surface as embedded in $\RR^3$ along with a point at infinity.
  This is depicted in the middle of Figure \ref{genus2}. This geometry analytically continues to the connected wormhole.
  \item
 $(a_1, b_2), (a_2, b_1)$ or $(a_1 + a_2, b_1 + b_2)$ contractible.  These three cases  correspond to three different AdS + BTZ geometries; the different choices determine which pair of  horizons is connected by a BTZ wormhole.
 The case $(a_1,b_2)$ contractible is shown on the right hand side of Figure \ref{genus2}.
  \end{itemize}

 The different bulk saddles are related by the action of $Sp(4,\ZZ)$, and the phase transitions where bulk saddles exchange dominance must occur at fixed points of $Sp(4,\ZZ)$. In appendix \ref{hp} below, we use this to identify the location of the Hawking-Page like phase transition between the naive handlebody and its dual in a particularly simple subspace of the moduli space of genus two surfaces.

\subsection{Spin structures}
\label{spins}

Section \ref{phases} discussed the phase structure without worrying about boundary conditions for fermions. But explicit realizations of the AdS$_3$/CFT$_2$ duality contain fermions in both the CFT and the dual bulk system. These fermions require a choice of boundary condiitons.  Each copy of the CFT lives on a circle on which we can have either periodic or antiperiodic boundary conditions corresponding to the Ramond (R) or Neveu-Schwarz (NS) sectors.

This choice of boundary conditions makes an important difference to the field theory dynamics.  In particular, as we now review, defining a CFT state on $n$ cylinders by a path integral over $\Sigma$ leads to a vanishing result unless the spin structures chosen on the $n$ circles satisfy certain constraints.  From the bulk point of view, the choice of boundary conditions influences which bulk saddles can contribute as they must admit a consistent spin structure for bulk fermions.

Thinking of $\Sigma$ as a sphere with $n$ holes shows that the discussion is equivalent to that associated with computing $n$-point functions.  The conformal transformation that maps the CFT from the cylinder to the plane flips the fermion periodicity, so a boundary with periodic (R sector) boundary conditions corresponds to the insertion of an R sector operator, which has the property that fermionic operators on the worldsheet pick up a minus sign on transport around a circle enclosing the R sector operator. Thus, an R sector operator has an associated  branch cut for fermions.  Since a branch cut on the sphere must have two ends, the nonzero $n$-point functions must involve an even number of R sector operators.  In the puncture limit this is a familiar statement, but it is clearly true more generally: the CFT path integral defining our state is nonzero only if we have an even number of boundaries with R boundary conditions. So, for example, in the two-boundary case we can either have NS on both boundaries or R on both boundaries.  This is not surprising; two boundaries connected by a cylinder must have the same behavior. For the three-boundary case, the possible fermion boundary conditions are NS on all three boundaries, or NS on one boundary and R on the others.

For the CFTs of interest in string theory, the NS sector has a unique vacuum state, so the behaviour in the limits $L_3 \to 0, \infty$ is as discussed in the main text. But the behavior in the R case is more subtle.  There is a degenerate space of R vacuum states, so if we choose a periodic spin structure on the degenerating cycle, the R vacuum states would propagate along this channel, and the density matrix would not exactly factorize as above. However, the space of R vacuum states is parametrically smaller at large $c$ than the space of states at any finite temperature, so the picture remains qualitatively similar: In the $L_3 \to 0$ limit the entropy of $\rho$ would drop from that of a macroscopic black hole to the smaller entropy obtained by counting the R ground states.

Another important effect, however, is that the logarithmic density of states (measured by the entropy) near the R vacuum is of order $c$ times a coefficient that vanishes at precisely zero energy.  Thus the limiting $L_3 \rightarrow 0$ behavior becomes visible only at extremely small moduli with $L_3/\ell \lesssim e^{-A c}$ for some $A$ of order $1$. Since the bulk semi-classical approximation breaks down in this regime, holography provides no tools for studying this limit.\footnote{The problem is closely related to distinguishing black hole microstates using bulk methods.} In particular, there is no reason to expect semi-classical bulk physics to exhibit an associated phase transition.

Indeed, the choice of fermion boundary conditions influences which of the possible bulk saddle-points discussed in section \ref{phases} can contribute, as the bulk saddle must admit fermion fields consistent with the choice of boundary conditions on each boundary. In differential geometry, the choice of boundary conditions for fermion fields on a non-contractible cycle in spacetime is referred to as a choice of spin structure.  Since fermions must have antiperiodic boundary conditions on a contractible cycle, consistency requires fermions to pick up a minus sign under a $2\pi$ rotation.

For the two-boundary case, the possible bulk saddles were BTZ or two copies of global AdS$_3$.  BTZ has spin structures consistent with either boundary condition (both NS or both R), but in double-global-AdS$_3$ saddle both spatial circles shrink to zero in the interior and so require the NS boundary condition.  We see that the Hawing-Page like phase transition between these geometries occurs only for the both-NS case. For the R boundary conditions the BTZ black hole dominates at all temperatures \cite{Maldacena:1998bw}.

For the three-boundary case, imposing NS boundary conditions on all boundaries similarly allows all of the bulk saddles discussed in section \ref{phases}. But for one NS and two R boundaries we are not allowed to fill in the R boundaries with global AdS${}_3$.  Thus there are only two possible saddles given by the  three-boundary connected wormhole and the BTZ-plus-global-AdS${}_3$ saddle with the AdS${}_3$ factor having the NS boundary.  There should be a single phase transition between them as one varies the length of the geodesic associated with the NS boundary.

The story is similar for $n > 3$ boundaries: the connected $n$-boundary black hole is an allowed bulk saddle for all consistent choices of boundary conditions, and the remaining saddles describe disconnected Lorentzian bulk geometries with the $i$th component connecting $m_i$ boundaries of which an even number have R boundary conditions.  So long as $\Sigma$ has genus zero, the choice of fermion boundary conditions on the asymptotic circles in fact fully determines the spin structure on $\Sigma$.  This can be seen from the fact that any minimal geodesic cuts $\Sigma$ into two disconnected pieces.  Since the total number of R asymptotic boundaries is even, both pieces will have the same number of R asymptotic boundaries counted modulo 2. But we obtain a nonzero path integral on each piece only if the total number of R boundaries on each side is even.  So an even number of R asymptotic boundaries on one side requires an NS cut, and an odd number requires an R cut.  This affects the factorization limits of section \ref{factorization} in direct parallel with the discussion of phase transitions above, with factorization associated with degenerating R cycles being invisible to semi-classical bulk physics.

However, we note that there is a special case immune from such concerns. This arose in our discussion of the $\rho_{12}$ density matrix obtained from the 3-boundary state by tracing over boundary $3$ in the limit $L_3 \rightarrow \infty$.  There the shrinking cycle cut $\Sigma$ into two pieces for which {\em each} piece was separately invariant under a $\mathbb{Z}_2$ symmetry that exchanged boundaries in pairs.  Thus each piece always has an even number of R boundaries, and the cut must be NS.  It follows that there is always an associated phase transition at some finite-but-large value of $L_3$.

\subsection{Hawking-Page like phase transition for three boundaries}
\label{hp}

Section \ref{phases} argued that varying the moduli $L_a$ of our 3-boundary state should lead to certain phase transitions in the dual bulk geometries.
The simplest transition to analyze occurs when all moduli are equal ($L_1 = L_2 = L_3 =L)$ and involves the exchange of relative dominance of the 3-global-AdS${}_3$ phase and the connected 3-boundary wormhole phase.  If the BTZ + global AdS${}_3$ phase were to dominate, it would constitute spontaneous breaking of the $S_3$ permutation symmetries on such boundaries.  This seems unlikely to occur, and we will neglect this phase below.

From the Euclidean point of view, this phase transition can be viewed as a change in the preferred way of filling in the genus two surface, which is the doubled version of our path integral over the sphere with three holes. In the two-boundary case, the analogous surface was a torus, and we know that the transition took place at a point of enhanced symmetry, where the torus was square, with both cycles of the same length. Therefore here rather than trying to explicitly calculate the Euclidean action for the two different saddles to determine the transition point, we will simply look for the point of enhanced symmetry as we vary $L$.

To determine the location of this enhanced symmetry point, it is more convenient to work with the representation of the surface $\Sigma$ where it is formed from identifications on a finite region in the hyperbolic plane. Our geometry on the pair of pants is depicted in figure \ref{pants}, and is determined by the lengths $L_1, L_2, L_3$ of the three geodesics. This alternative representation rewrites $\Sigma$ as the region in figure \ref{pants} bounded by the minimal geodesics, for some other values $L_1', L_2', L_3'$. Unfortunately the map $L_a'(L_a)$ is not known explicitly. But in the symmetric case $L_1 = L_2 = L_3 = L$, we will have $L_1' = L_2' = L_3' = L'$.

\begin{figure}
\centering
\includegraphics[keepaspectratio,width=0.4\linewidth]{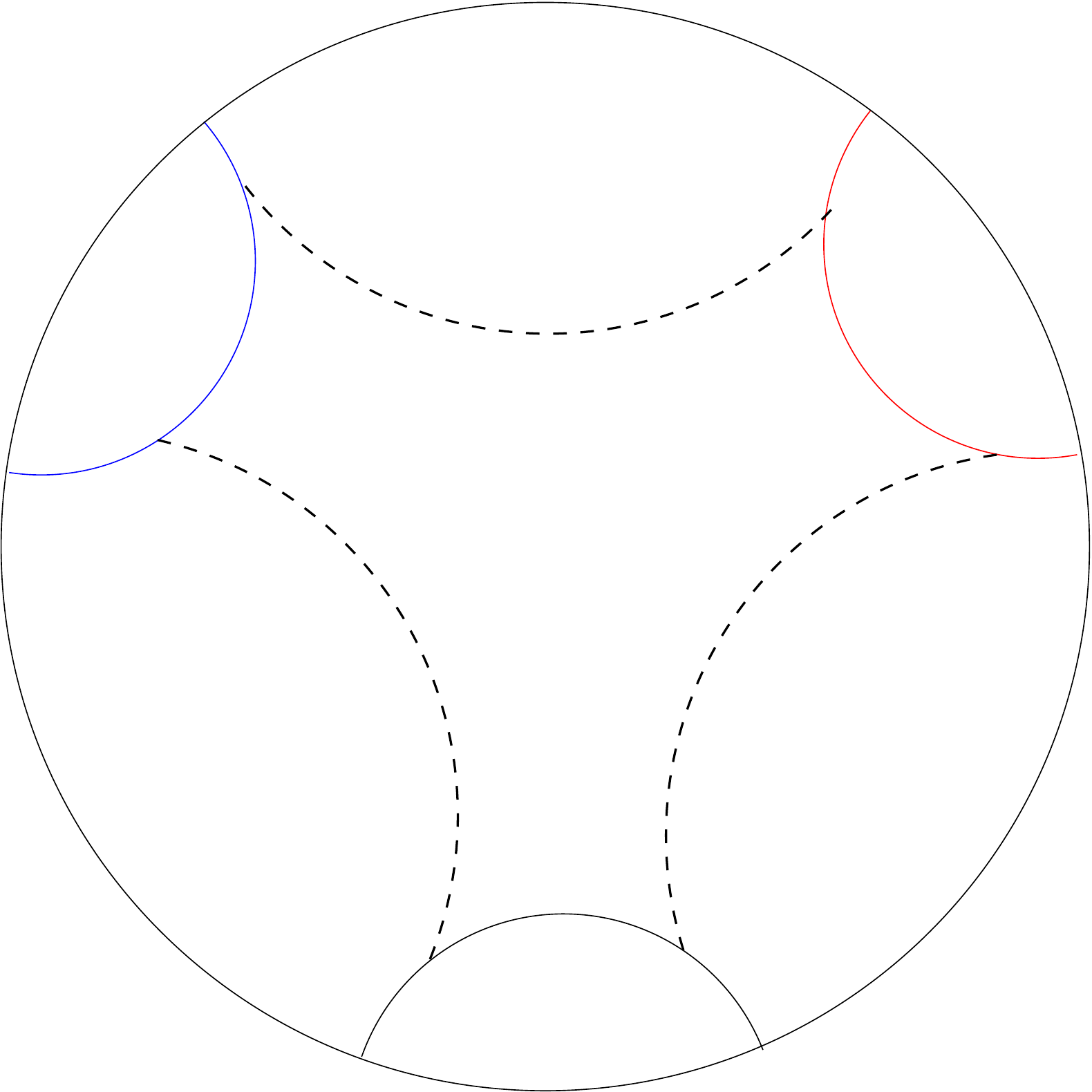}
\caption{The symmetric representation of the pair of pants geometry; the geometry is obtained by taking two copies of the Poincar\'e disc and identifying corresponding geodesics in each copy. For the symmetric configuration where the horizon lengths are equal, this figure has a symmetry under rotations by $2\pi/3$. The minimal closed geodesics are made up of the dashed lines in the two copies. The symmetric point is where the length of the minimal geodesics is equal to that of the identified lines between the minimal geodesics, when the figure becomes a hyperbolic regular hexagon.}
\label{symm}
\end{figure}

There are three parameters $\psi_1$, $\psi_2$, $\alpha_2$ labeling this identification, which determine the lengths $L_1, L_2, L_3$ of the three geodesics. To consider the symmetric case, we need to set these three lengths to be the same; we need to determine the choice of $\psi_1$, $\psi_2$, $\alpha_2$ to which this corresponds. A relatively simple way to do so is to split the fundamental region in figure \ref{pants} along the straight line $\phi =0, \pi$.  We may then think of our pair of pants as being built by gluing together two regions in the hyperbolic plane, each bounded by three geodesics;  we will henceforth refer to these bounding geodesics as seams.  The above split has the effect of separating every minimal closed geodesic $H_a$ of figure \ref{pants} into a pair of segments, with each segment running between some pair of seams.  This description is manifestly symmetric under permuting the boundaries when the seams are taken to be the geodesics
\begin{equation}
\tanh \chi \cos (\phi - \alpha_a') = \cos \psi',
\end{equation}
whith $\alpha_1' = \frac{\pi}{6}$, $\alpha_2' = \frac{7\pi}{6}$, and $\alpha_3' = \frac{3\pi}{2}$, see figure \ref{symm}. A series of boosts relates this to the presentation in figure \ref{pants}, from which we learn that
\begin{equation}
\cos^2 \psi = 1- \frac{ \sin^4 \psi'}{(\sin \pi/6 + \cos^2\psi')^2}
\end{equation}
and thus that
\begin{equation}
L = 2 \ell \tanh^{-1} \left[ \left( 1- \frac{ \sin^4 \psi'}{(\sin \pi/6 + \cos^2\psi')^2 }\right)^{1/2} \right]
\end{equation}
The single parameter $\psi$ can clearly run from $0$ (where $L \to \infty$) to $\pi/3$ (where $L \to 0$). For the representation of $\Sigma$ as the interior region, there is a similar expression for $L'$ as a function of the corresponding angle.

Gluing together two of these pairs of pants along the seams gives us the genus two surface corresponding to the case with all $L_a$ equal. In the representation as a finite region, it is clear that the interesting enhanced symmetry point is when the pair of pants has a symmetry under interchanging the identified and minimal geodesics, that is when the length along the identified geodesic between two minimal geodesics in figure \ref{symm} is the same as $L'$. At this symmetric point, the genus two surface will be the Bolza surface, the most symmetric genus two surface (see e.g. \cite{katz07} for further discussion of symmetric genus two surfaces).

This is most conveniently worked out in the coordinates of figure \ref{pants}, where the length along the identified geodesic $\phi = \pi$ between the two minimal geodesics is simply $L_{id} =  \ell \chi_{min}$, where $\chi_{min}$ is the minimum value of $\chi$ along the minimal geodesic. Boosting to the coordinates of figure \ref{symm}, we find this is
\begin{equation}
L_{id} =  2 \ell \tanh^{-1} \left[  \frac{\cos \pi/6}{(1 + \sin \pi/6)} \tan \psi' \right],
\end{equation}
so the value of $\psi'$ at which $L_{id} = L'$ is determined by solving
\begin{equation}
\frac{\cos^2 \pi/6}{(1+\sin^2 \pi/6)^2} \tan^2 \psi = 1- \frac{\sin^4 \psi}{(\sin \pi/6+ \cos^2 \psi)^2},
\end{equation}
which gives $\psi = 0.848906$, corresponding to a length $L' = 1.701$. As expected, the transition occurs at parameter values of order one. The corresponding value of $L$ can't be explicitly determined, but it should also be order one.

\section{Derivation of the $k$-producible bound}
\label{bound}

We now derive the bound \eqref{eqn:avbound} on entropies associated with a $k$-producible pure $n$-party state.
We define $\delta S(B_X) := S_{\rm{sum}}(B_X) - S(B_X)$ and $X^c := [n] \setminus X$, with all other notation as in section \ref{intrinsic}.

The first step will be to prove the following statement: let $X, Y \in \choice{[n]}{j}$ label collections of systems
and $K \subseteq [n]$ be a subset of size $k \leq n-1$, then
\be
\max\left[ \delta S(B_X^K), \delta S(B_Y^K) \right]
	\geq \Sadd( B_{X \cap Y }^K ) - \Sadd( B_{X^c \cap Y^c}^K ). \label{eqn:sadd1}
\ee
To get a feel for the inequality, consider the simplest case, when $X \cup Y = K$, and suppose for the sake of argument that $\delta S(B_X^K) = \delta S(B_Y^K) = 0$. Then
$S(B_Y^K) = S(B_{Y^c}^K)$ since the state on $B_{Y \cup Y^c}^K$ is pure, so
\beq
S( B_Y^K )
=
S(B_{Y^c}^K)
&=&
\Sadd (B_{Y^c}^K) \\
&=&
\Sadd (B_X^K) - \Sadd (B_{X \cap Y}^K) \\
&=&
\Sadd(B_{X^c}^K) - \Sadd (B_{X \cap Y}^K) \\
&=&
\Sadd( B_Y^K ) - 2 \Sadd( B_{X \cap Y}^K).
\eeq
Comparing the left and right hand sides of the equation, we conclude that $\Sadd( B_{X \cap Y}^K)=0$. Increasing $\delta S(B_X^K)$ or $\delta S(B_Y^K)$ allows $\Sadd( B_{X \cap Y}^K)$ to also increase in a controlled way, which is the content of \eqref{eqn:sadd1}.

The proof of \eqref{eqn:sadd1} hinges on the following easily proved but complicated-looking formula:
\be \label{eqn:entropy-identity}
S(B_Y^K)
= \Sadd( B_Y^K ) - 2 \Sadd( B_{X \cap Y} ) + 2 \Sadd( B^K_{X^c \cap Y^c} )
	+  \delta S( B_X^K ) - \delta S( B_{X^c}^K ) - \delta S( B_{Y^c}^K ).
\ee
The demonstration is just an exercise in set-theoretic manipulations and use of the fact that $S(B_Z^K) = S(B_{Z^c}^K)$.
\beq
S(B_Y^K)
&=& S_{\rm{sum}}( B_{Y^c}^K ) - \delta S(B_{Y^c}^K) \\
&=& S_{\rm{sum}}( B_{Y^c}^K \cap B_X^K ) + S_{\rm{sum}}( B_{Y^c}^K \cap B_{X^c}^K ) - \delta S(B_{Y^c}^K) \\
&=& S_{\rm{sum}}( B_X^K ) - S_{\rm{sum}}( B_{X \cap Y}) + S_{\rm{sum}}( B_{X^c \cap Y^c} ) - \delta S(B_{Y^c}^K) \\
&=& S( B_X^K ) - S_{\rm{sum}}( B_{X \cap Y}) + S_{\rm{sum}}( B_{X^c \cap Y^c} ) - \delta S(B_X^K) - \delta S(B_{Y^c}^K) \\
&=& S( B_{X^c}^K) - S_{\rm{sum}}( B_{X \cap Y}) + S_{\rm{sum}}( B_{X^c \cap Y^c} ) - \delta S(B_X^K)
	- \delta S(B_{Y^c}^K) \\
&=&
S_{\rm{sum}}( B_{X^c}^K) - S_{\rm{sum}}( B_{X \cap Y}) + S_{\rm{sum}}( B_{X^c \cap Y^c} ) \nonumber \\
&\,& \quad \quad \quad	 - \delta S(B_X^K) - \delta S(B_{X^c}^K) - \delta S(B_{Y^c}^K).
 \label{eqn:entropy-identity1}
\eeq
But
\beq
S_{\rm{sum}}( B_{X^c}^K )
&=& S_{\rm{sum}}( B_{X^c \cap Y}^K ) + S_{\rm{sum}}( B_{X^c \cap Y^c}^K ) \\
&=& S_{\rm{sum}}( B_Y^K ) - S_{\rm{sum}}( B_{X\cap Y}^K ) + S_{\rm{sum}}( B_{X^c \cap Y^c}^K ).
\eeq
\eqref{eqn:entropy-identity} then follows by substitution into \eqref{eqn:entropy-identity1}.

We will need to eliminate $\delta S(B_{X^c}^K)$ and $\delta S(B_{Y^c}^K)$ from \eqref{eqn:entropy-identity} to get a state-independent bound.
By the subadditivity of entropy, $\delta S$ is nonnegative. Moreover, the monotonicity of the relative entropy function $S(\rho \| \sigma) = \tr \rho (\log \rho - \log \sigma)$ can be used to conclude that
\beq
\delta S( B_Z^K )
= S( \psi^K_Z \| \otimes_{z \in Z} \psi_z^K )
\geq S( \psi^K_{Z'}  \| \otimes_{z \in Z'} \psi_z^K )
= \delta S( B_{Z'}^K ),
\eeq
provided $Z' \subseteq Z$. Therefore,
\beq
&\,& 2 \Sadd( B^K_{X^c \cap Y^c} ) - \delta S( B_{X^c}^K ) - \delta S( B_{Y^c}^K ) \\
&\leq& 2 \Sadd( B^K_{X^c \cap Y} ) - 2 \delta S( B_{X^c \cap Y^c}^K ) \\
&=& 2 S( B^K_{X^c \cap Y^c} ).  \\
\eeq
Substituting back into \eqref{eqn:entropy-identity} leads to the inequality
\be
\delta S(B_Y^K) \geq 2 \Sadd( B_{X \cap Y} ) - 2 \Sadd( B^K_{X^c \cap Y^c} ) + \delta S( B^K_X ),
\ee
which means that
\beq
\delta S(B_X^K) + \delta S(B_Y^K)
&\geq& \max \left[ \delta S(B_X^K), \delta S(B_Y^K) \right] \\
&\geq& \max \left[
	\delta S(B_X^K), 2 \Sadd( B_{X \cap Y}^K ) - 2 \Sadd( B^K_{X^c \cap Y^c} )
		- \delta S( B^K_X ) \right] \\
&\geq&  \Sadd( B_{X \cap Y}^K ) - \Sadd( B^K_{X^c \cap Y^c} ), \label{eqn:sadd2}
\eeq
which subsumes inequality \eqref{eqn:sadd1}. The third line is just a consequence of optimizing the unknown nonnegative quantity $\delta S( B_X^K )$.

\begin{figure}
\centering
\includegraphics[keepaspectratio,width=0.5\linewidth]{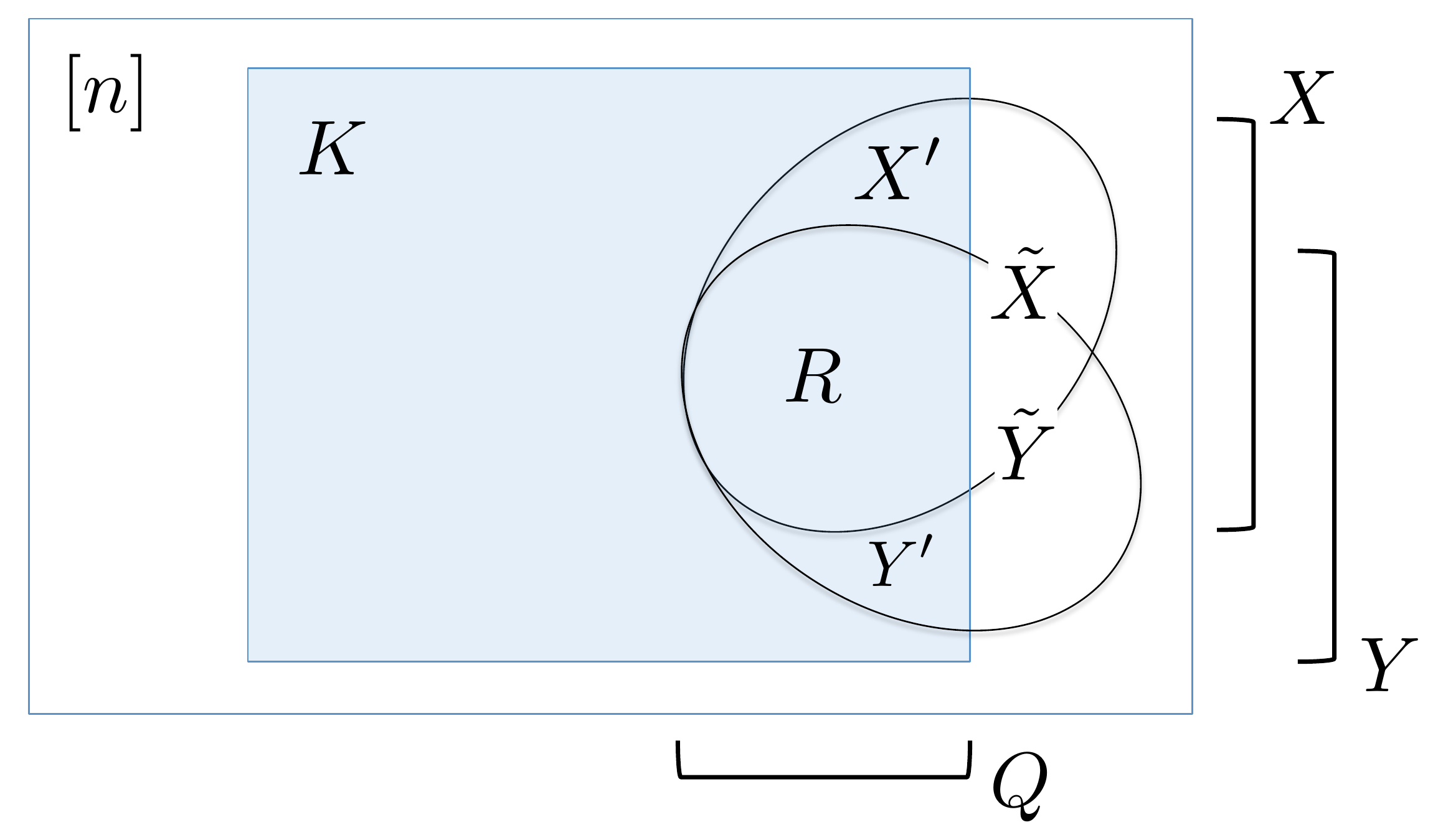}
\caption{Decomposition of the sets $X$ and $Y$ used in the evaluation of \eqref{eqn:rhs-kpartite1}.}
\label{fig:combipic}
\end{figure}

Up to now, we have been studying individual states that are pure on $B^K$. To get a bound applicable to all $k$-producible pure states, it will be necessary to remove the dependence on $K$, which we will do by summing \eqref{eqn:sadd2} over $X, Y$ and $K$.
On the left hand side, using that $B_X$ is isometrically related to $\otimes_{K \in \choice{[n]}{k}} \otimes_{x \in X} B_x^K$, we can calculate
\beq
 \sum_{X,Y \in \choice{[n]}{j}} \sum_{K \in \choice{[n]}{k}}
	\left[ \delta S(B_X^K) + \delta S(B_Y^K) \right]
&=& \sum_{X,Y \in \choice{[n]}{j}} \left[ \delta S(B_X) + \delta S(B_Y) \right] \\
&=& 2 \binom{n}{j} \sum_{X \in \choice{[n]}{j}} \delta S(B_X). \label{eqn:lhs-kpartite}
\eeq
The evaluation of the right hand side of \eqref{eqn:sadd2} is significantly more involved. As a first observation, recall that for generic states, $\delta S(B_X)$ is zero only if $|X| \leq n/2$ so we are only concerned with $j = |X| = |Y| \leq n/2$, in which case $|X \cap Y| \leq |X^c \cap Y^c|$. Summing \eqref{eqn:sadd2} over all $K$, however, will typically give a trivial bound under those circumstances because of negative contributions. Instead, since $\delta S(B^K_X) + \delta S(B^K_Y)$ is nonnegative, \eqref{eqn:lhs-kpartite} can more fruitfully be bounded below by

\beq
&\,& \sum_{X,Y \in \choice{[n]}{j}} \sum_{K \in \choice{[n]}{k}}
	I[ (X,Y,K) \in \mathcal{G} ]
	\left( \Sadd( B_{X \cap Y }^K ) - \Sadd( B^K_{X^c \cap Y^c} ) \right), \label{eqn:rhs-kpartite1}
\eeq
where $I[\cdot]$ denotes the indicator function ($I[{\cdot}] =1$ when the condition is satisfied, and is zero otherwise) and we have the freedom to choose $\mathcal{G}$ as we please in order to get the best bound possible. Let $Q= (X \cup Y) \cap K$ and $R = X \cap Y \cap K$. We will take $\mathcal{G}$ to be the set such that $|Q| \in \mathcal{Q}$ and $|R| \in \mathcal{R}$ for choices of $\mathcal{Q}$ and $\mathcal{R}$ to be determined. To facilitate the calculation, break $X$ and $Y$ each into three parts, namely their intersection $R$, their portions in $K$ but not in $R$, and the rest, as depicted in figure \ref{fig:combipic}.
In terms of those definitions, \eqref{eqn:rhs-kpartite1} can be rewritten as
\beq
&\,& \sum_{(X,Y,K)\in \mathcal{G}}
	\left[ \Sadd( B_{X \cap Y }^K ) - \Sadd( B^K_{X^c \cap Y^c} ) \right] \\
&=& \sum_{K \in \choice{[n]}{k}} \sum_{Q \in \choice{K}{q}} \sum_{R \in \choice{Q}{r}}
	\sum_{X' \in \mathcal{D}} \sum_{\tilde{X} \in \choice{[n] \setminus K}{j-r-|X'|}}
	\sum_{\tilde{Y} \in \choice{[n] \setminus K}{j-r-q+|X'|}}
	\left[ \Sadd( B_{X \cap Y }^K ) - \Sadd( B^K_{X^c \cap Y^c} ) \right], \label{eqn:horrible}
\eeq
where $X$ and $Y$ are the disjoint unions $R \cup X' \cup \tilde{X}$ and $R \cup Y' \cup \tilde{Y}$, respectively, while $\mathcal{D} = \{ X' \subseteq Q \setminus R : \max(0,q-j) \leq |X'| \leq \min (j-r,q-r) \}$. The constraints on the size of $X'$ arise from the requirement that $|X'| + |Y'| + r = q$ but that $|X'|, |Y'| \leq j - r$.

Since the summand is independent of $X', \tilde{X}$ and $\tilde{Y}$, the three rightmost sums can be evaluated, yielding a multiplicative factor of
\be
C_{jknqr}
= \sum_{l = {\max(0,q-j)}}^{\min(q-r,j-r)}
	\binom{q-r}{l}  \binom{n-k}{j-r-l} \binom{n-k}{j-r-q+l},
\ee
with $l$ playing the role of $|X'|$. Next, we compute the sum of $\Sadd(B^K_{X\cap Y})$:
\beq
\sum_{K \in \choice{[n]}{k}} \sum_{Q \in \choice{K}{q}} \sum_{R \in \choice{Q}{r}}
	\Sadd(B^K_{X \cap Y})
&=&
\sum_{K \in \choice{[n]}{k}} \sum_{R \in \choice{K}{r}} \sum_{Q' \in \choice{K\setminus R}{q-r}}
	\Sadd(B^K_R) \\
&=&
\binom{k-r}{q-r} \sum_{K \in \choice{[n]}{k}} \sum_{R \in \choice{K}{r}}
	\Sadd(B^K_R) \\
&=&
\binom{k-r}{q-r}  \sum_{x \in [n]} \sum_{K' \in \choice{[n] \setminus \{x\}}{k-1}}
	\sum_{R' \in \choice{K'}{r-1}} S(B_x^K) \\
&=&
\binom{k-r}{q-r} \binom{k-1}{r-1} \sum_{x \in [n]} S(B_x) \\
&=&
\binom{k-1}{k-q, q-r, r-1} \sum_{x \in [n]} S(B_x),
\eeq
In simplifying the sums, we have used the fact that $S(B_x) = \sum_K S(B_x^K)$.
The contribution to \eqref{eqn:horrible} from the $\Sadd(B^K_{X^c \cap Y^c})$ terms can then be disposed of with a similar calculation.
\beq
\sum_{K \in \choice{[n]}{k}} \sum_{Q \in \choice{K}{q}} \sum_{R \in \choice{Q}{r}}
	\Sadd(B^K_{Q^c})
&=&
\binom{q}{r} \sum_{K \in \choice{[n]}{k}} \sum_{Q^c \in \choice{K}{k-q}}
		\Sadd(B^K_{Q^c}) \\
&=&
\binom{q}{r} \sum_{x \in [n]} \sum_{K' \in \choice{[n]\setminus\{x\}}{k-1}}
	\sum_{\tilde{Q}^c \in \choice{K'}{k-q-1}}
	S( B_x^{\{ x \} \cup K'} ) \\
&=&
\binom{q}{r} \binom{k-1}{k-q-1} \sum_{x \in [n]} S(B_x) \\
&=&
\binom{k-1}{k-q-1, q-r, r}  \sum_{x \in [n]} S(B_x).
\eeq
Substituting leads to the conclusion that \eqref{eqn:horrible} is equal to
\be
C_{jknqr} \binom{k-1}{k-q, q-r, r-1} \left( 1 - \frac{k-q}{r} \right) \sum_{x \in [n]} S(B_x).
\label{eqn:multi-lb2}
\ee

While that completes the proof of the bound, the conclusion can be expressed slightly more conveniently in terms of
\be
\tilde{C}_{jknqr}
= \sum_{l = {\max(0,q-j)}}^{\min(q-r,j-r)}
	\binom{k-1}{k-q,q-r-l,l,r-1}  \binom{n-k}{j-r-l} \binom{n-k}{j-r-q+l}.
\ee
We have shown that
\be
 2 \binom{n}{j} \sum_{X \in \choice{[n]}{j}} \delta S(B_X)
\geq
\sum_{q \in \mathcal{Q}} \sum_{r \in \mathcal{R}}
\tilde{C}_{jknqr} \left( 1 - \frac{k-q}{r} \right) \sum_{x \in [n]} S(B_x).
\label{eqn:multi-lb}
\ee
Note that the formula only makes sense when $q$ and $r$ are chosen to be consistent with $j, n$ and $k$, which requires that $r \leq q \leq k$ and $q \leq 2j-r$. This is a general inequality that must be satisfied for every $k$-producible state on $n$ factors. The strongest bound occurs when $\mathcal{Q}$, $\mathcal{R}$ are selected to contain precisely those $q,r$ making positive contributions, which is the choice made to produce the plots in the main text.

\bibliographystyle{JHEP}
\bibliography{draft}

\end{document}